\documentclass[prd,preprintnumbers,twocolumn,eqsecnum,floatfix,a4paper,nofootinbib,superscriptaddress]{revtex4-2}

\usepackage{color}
\usepackage{amsmath,amssymb,graphicx,mathtools}
\usepackage{bm,lipsum}
\usepackage{times}
\usepackage{microtype}
\usepackage[utf8]{inputenc}
\usepackage{booktabs}
\usepackage[normalem]{ulem}
\usepackage[varg]{txfonts}
\usepackage{bm,graphics,graphicx,epsfig,xcolor,ulem}
\usepackage[colorlinks, pdfborder={0 0 0}]{hyperref}
\usepackage{multirow}
\usepackage{pifont}
\usepackage{array,threeparttable}
\usepackage{diagbox}
\usepackage{boldline}
\usepackage{gensymb}

\definecolor{LinkColor}{rgb}{0.75, 0, 0}
\definecolor{CiteColor}{rgb}{0, 0.5, 0.5}
\definecolor{UrlColor}{rgb}{0, 0, 0.75}
\hypersetup{linkcolor=LinkColor}
\hypersetup{citecolor=CiteColor}
\hypersetup{urlcolor=UrlColor}
\allowdisplaybreaks
\begin{document}
\title{Cosmography with bright and Love sirens}


\newcommand{\penncosmos}{\affiliation{Institute for Gravitation and the Cosmos, Department of Physics, Pennsylvania State University, University Park, PA, 16802, USA}}
\newcommand{\pennastro}{\affiliation{Department of Astronomy \& Astrophysics, Pennsylvania State University, University Park, PA, 16802, USA}}
\newcommand{\cardiff}{\affiliation{School of Physics and Astronomy, Cardiff University, Cardiff, UK, CF24 3AA}}
\newcommand{\olemiss}{\affiliation{Department of Physics and Astronomy, The University of Mississippi, University, Mississippi 38677, USA}}

\author{Arnab Dhani}
\email{aud371@psu.edu}
\penncosmos

\author{Ssohrab Borhanian}
\affiliation{Theoretisch-Physikalisches Institut, Friedrich-Schiller-Universit\"at Jena, 07743, Jena, Germany}
\penncosmos

\author{Anuradha Gupta}
\olemiss

\author{B.S. Sathyaprakash}
\penncosmos
\pennastro
\cardiff

\begin{abstract}
    Precision cosmology is crucial to understand the different energy components in the Universe and their evolution through cosmic time. Gravitational wave sources are standard sirens that can accurately map out distances in the Universe. Together with the source redshift information, we can then probe the expansion history of the Universe. We explore the capabilities of various gravitational-wave detector networks to constrain different cosmological models while employing separate waveform models for inspiral and post-merger part of the gravitational wave signal from equal mass binary neutron stars. We consider two different avenues to measure the redshift of a gravitational-wave source: first, we examine an electromagnetic measurement of the redshift via either a kilonova or a gamma ray burst detection following a binary neutron star merger (the electromagnetic counterpart method); second, we estimate the redshift from the gravitational-wave signal itself from the adiabatic tides between the component stars characterized by the tidal Love number, to provide a second mass-scale and break the mass-redshift degeneracy (the counterpart-less method). We find that the electromagnetic counterpart method is better suited to measure the Hubble constant while the counterpart-less method places more stringent bounds on other cosmological parameters. In the era of next-generation gravitational-wave detector networks, both methods achieve sub-percent measurement of the Hubble constant $H_0$ after one year of observations. The dark matter energy density parameter $\Omega_{\rm M}$ in the $\Lambda$CDM model can be measured at percent-level precision using the counterpart method, whereas the counterpart-less method achieves sub-percent precision. We, however, do not find the postmerger signal to contribute significantly to these precision measurements.
\end{abstract}

\maketitle

\section{Introduction}
\label{sec:introduction}
The coalescence of binary black holes (BBHs), binary neutron stars (BNSs), and black hole-neutron star binaries (BHNSs) are the most abundant sources of gravitational waves (GWs) for ground-based detectors. To date, the Advanced Laser Interferometer Observatory (LIGO) and Advanced Virgo detector network~\cite{LIGOScientific:2014pky,VIRGO:2014yos,KAGRA:2018plz} has observed around 100 of such events, with the farthest observed source GW200208\_222617 at a distance of $\sim$2.23 Gpc~\cite{LIGOScientific:2021djp}. Planned upgrades to the current network of detectors and future detector networks are expected to reach farther distances, having the ability to observe BBH mergers occurring at redshifts up to $z\sim100$ and BNS mergers at $z\sim5$~\cite{Hall:2019xmm}.

Gravitational waves from compact binary mergers are standard sirens, meaning the observed strain amplitude and the rate of change of the signal's frequency can be used to infer both the apparent and intrinsic luminosities of the source and, in turn, the luminosity distance~\cite{Schutz:1986gp}. Thus, compact binary coalescences can be used as a cosmological distance measure. In conjunction with the redshift of the source, we have an independent means to map the expansion history of the Universe~\cite{Schutz:1986gp,Holz:2005df}. The SH0ES team accomplishes this by utilising Type Ia supernovae (SNeIa) as standard candles~\cite{Riess:2021jrx}. Even though these sources have a constant intrinsic luminosity, their value cannot be inferred from terrestrial observations alone and, therefore, the nearby SNeIa are standardised using a different distance measure, such as the cepheid variables~\cite{Sandage1996CepheidCO}. Compact binaries have the added benefit of not requiring any external calibration. 

Independent measurements to map the evolutionary history of the Universe are crucial since the measurement of the Hubble parameter by the Planck collaboration~\cite{Planck:2018vyg} is in conflict with those made by the SH0ES team with the former's median estimate lying outside the 5$\sigma$ confidence interval of the latter~\cite{Riess:2021jrx}. The Planck experiment infers the Hubble parameter from an early Universe model ($z\sim1100$), while the redshift of the most distant supernova in the catalog of sources used by the SH0ES team is at a redshift of $z\sim1.5$. Given this, a variety of models have been proposed that change the physics in some way in the intermediate redshift range to resolve the tension. An in depth review of this tension and proposed solutions can be found in~\textcite{DiValentino:2021izs}. Gravitational waves provide both an independent measurement of the cosmological parameters and also fill the gap in redshift observing events at higher redshifts than the apropos electromagnetic probe. 

The first detection of gravitational waves from a BNS merger~\cite{LIGOScientific:2017vwq} and the subsequent observation of short gamma ray bursts (GRBs) and kilonova~\cite{LIGOScientific:2017ync} from the same source led to the first measurement of the Hubble constant using gravitational waves~\cite{LIGOScientific:2017adf}. A crucial aspect of this measurement was the detection of the electromagnetic counterpart, which allowed for the localisation of the galaxy where the event occurred and, thereby, the redshift measurement. Gravitational wave antennas have a very large field of view and, hence, are poor at source localisation. \textcite{Borhanian:2020vyr,Nishizawa:2016ood,Yu:2020vyy} looked at probable populations of sources with an accurate localisation using gravitational-wave detectors that can be observed with planned upgrades to the current detectors or with future detector networks. The authors forecast a  measurement accuracy of better than 2\% for the Hubble constant without an electromagnetic counterpart. One can also do a statistical measurement of the redshift based on galaxy catalogs in a 3D volume localised by gravitational-wave detectors~\cite{Schutz:1986gp,DelPozzo:2011vcw}. This method was used to measure the Hubble constant from LIGO-Virgo's catalog of BBH mergers and will require hundreds of BBHs to achieve 2\% in Hubble constant~\cite{DES:2019ccw,LIGOScientific:2019zcs,DES:2020nay}. Another technique known as the \emph{GW-galaxy cross-correlation} explores the spatial clustering of the GW sources with galaxies to infer the clustering redshift of the gravitational-wave sources~\cite{Oguri:2016dgk,Mukherjee:2020hyn} which is demonstrated on the third LIGO-Virgo Gravitational-Wave Transient Catalog, \textit{GWTC-3}~\cite{Mukherjee:2022afz,LIGOScientific:2021djp}. In the future, this technique will provide a 2\% measurement of the Hubble constant in synergy with galaxy surveys such as DESI~\cite{DESI:2016fyo} and SPHEREx~\cite{Diaz:2021pem,Dore:2014cca}.

A multitude of other methods have also been used in the literature to make cosmological inference in the absence of an electromagnetic counterpart. \emph{Mass function cosmology}, or \emph{spectral sirens}, uses various aspects of the mass spectrum of neutron stars (NSs) and black holes (BHs), such as the upper mass gap in the BH mass spectrum induced by \textit{pair instability supernova}, the lower mass gap demarcating NSs and BHs, and the narrowness of the NS mass distribution to imprint an additional mass scale to break the mass-redshift degeneracy~\cite{Chernoff:1993th,Taylor:2012db,Farr:2019twy,Ezquiaga:2022zkx}. Furthermore, cosmological constraints using the redshift distribution of compact binary sources have been proposed in~\textcite{Mukherjee:2021rtw, Karathanasis:2022rtr,Ding:2018zrk,Leandro:2021qlc,Mastrogiovanni:2022hil} showing a hint towards the redshift evolution of the GW mass distribution. A direct measurement of the redshift utilising the tidal information in BNS mergers was proposed in Refs.~\cite{Messenger:2011gi,Messenger:2013fya} and explored in Refs.~\cite{Li:2013via,Wang:2020xwn}.

Although compact binary coalescences are calibration-free standard candles, a number of unknown model systematics can bias the measurement of cosmological parameters. For example, in the case of kilonova, which are necessary to identify the host and measure its redshift, we do not know if the emission is isotropic nor what the dependence is on the co-latitude $\theta_J$ with respect to the total angular momentum of the system. Assuming that the emission is isotropic when it is not could mean the observed sample is incomplete and not all binary orientations are observed, leading to biases in the inferred parameters~\cite{Chen:2020dyt}.  Likewise, to make use of tidal deformability to measure the redshift of a BNS merger it is necessary to know what the equation of state is and a  wrong equation of state could bias the measurement of redshift. It should be possible to relax model assumptions, include additional model parameters and use gravitational wave observations to simultaneously measure both the model as well as cosmological parameters or marginalize over the model parameters if they cannot be inferred from gravitational-wave observations. This procedure will require a greater number of observations or an alternative approach to determining the model parameters. For example,~\textcite{Chen:2020dyt} estimates that an unknown kilonova emission mechanism would require twice as many BNS observations to accomplish the same accuracy as if the model is known precisely.  We do not model all possible systematics that could bias the results as it is our intention to explore the power of this method. Application of the methods explored in this study to real data would require an in-depth analysis of all the systematics which we hope to pursue in a follow-up paper.

In this paper, we assess the potential of an astrophysical population of BNS to constrain cosmological models using proposed upgrades to the current detector network and future observatories. To that end, we will compare two ways of cosmological inference using BNSs. The first procedure is the electromagnetic counterpart method where a coincident detection of a GRB or a kilonova is used to infer the source redshift. This part of our work is along the lines of \textcite{Belgacem:2019tbw}. Our analysis relaxes some of the assumptions in their analysis and we do a broader comparison of several future detector network configurations. We contrast the results in two distinct cases: when the inclination angle of the binary is (i) known from the GRB counterpart via modeling of the GRB afterglow and (ii) estimated from the gravitational-wave signal alone. We, thereby, make the case for an accurate modeling of the GRB jet profile for the precise measurement of the inclination angle. \textcite{CalderonBustillo:2020kcg} explores the effect of higher harmonics in breaking the luminosity distance -- inclination angle degeneracy and obtaining a percent level precision in the measurement of the Hubble constant. \textcite{Chen:2020zoq} forecasts cosmological constraints in selected detector configurations of upcoming and future gravitational-wave observatories and makes the case for a target-of-opportunity in various current and future electromagnetic observatories. We refer the reader to~\textcite{Bulla:2022ppy} for a review on multi-messenger constraints on Hubble constant using kilonovae and GRBs.

The second method uses the tidal information in the waveform to break the mass-redshift degeneracy and infer the redshift directly from the gravitational-wave signal. \textcite{Chatterjee:2021xrm} performs a Bayesian estimation of the Hubble constant while fixing the functional form of the dependence of tidal deformability on source-frame mass using binary Love relations~\cite{Yagi:2013sva}. \textcite{DelPozzo:2015bna} carry out a simultaneous Bayesian inference of multiple cosmological parameters for a set of BNSs observed by the Einstein Telescope using the inspiral part of the signal to determine the redshift and assuming a specific equation of state (EoS). \textcite{Wang:2020xwn} and \textcite{Jin:2022qnj} used an astrophysical population of BNSs to estimate the constraints on cosmological parameters using Fisher matrix formalism, again utilising the inspiral signal to break the mass-redshift degeneracy and measure the redshift to the source. The former determines the EoS of NSs using the nearby observations and farther sources to determine the dark energy parameters. The latter assumes the EoS to be known and calculates constraints on all the parameters of a given cosmological model. \textcite{Ghosh:2022muc} shows that it is possible to precisely measure the EoS and the Hubble constant simultaneously using next-generation detectors assuming realistic constraints on the EoS at the time of operation of these detectors. We include information not only from the inspiral phase but also from the post-merger phase of the signal to assess the contribution, if any, of the post-merger signal towards cosmological inference. Additionally, as mentioned earlier, we consider a wider array of future observatories and detector configurations that would help in their cost-benefit analysis. 

The summary of our results is as follows. The electromagnetic counterpart method can determine the Hubble constant $H_0$ at a sub-10\% accuracy for the $\Lambda$CDM model, while the \emph{counterpart-less} method also yields an accuracy of sub-10\% only with the A+ network (see Tab.~\ref{tab:detector_networks} for a description of the various networks considered in this study). This set of detectors is uninformative on the current fraction of dark matter energy density $\Omega_M$ if measured jointly with $H_0$. This is because of the low reach of these detectors and high correlation of $\Omega_M$ with $H_0$. However, if $H_0$ is known a priori, the counterpart method can achieve an accuracy of $\sim$100\% for $\Omega_M$. The counterpart-less method performs poorly because of redshift errors. The best network, ECS, consisting of three third-generation detectors, on the other hand, can determine $H_0$ to an accuracy of $\sim0.20\%$ using the counterpart method and 50\% better with the counterpart-less method, while $\Omega_M$ is measured at $\sim4\%$ level using electromagnetic counterparts and sub-percent level in the counterpart-less method. In the case of a non-trivial but constant dark energy EoS, $w_0$ [one of the dark energy EoS parameters; see Eq.~(\ref{eq:de_eos})] can be measured to an accuracy of $\sim0.1$ ($\sim0.02$) with (without) electromagnetic counterparts. If $H_0$ is known from other cosmological probes, these constraints improve to $\sim0.08$ and $\sim0.01$, respectively. For an interacting dark energy model, the marginalised bounds on the other dark energy EoS parameter $w_a$ are $\sim2$ and $\sim0.3$, respectively, with and without the use of electromagnetic counterparts. As earlier, with a priori knowledge of $H_0$, these improve to $\sim1$ and $\sim0.3$. A cosmological model that modifies not just the background evolution with respect to $\Lambda$CDM but also the tensor perturbations has differing luminosity distances for gravitational and electromagnetic waves. In such a model, the $\Xi_0$ parameter parameterizes the ratio of the two luminosity distances and can be constrained at $\sim140\%$ ($\sim90\%$) level using electromagnetic counterparts if $H_0$ is unknown (known). With the counterpart-less method, it can be constrained at the level of $\sim20\%$ ($\sim10\%$).

The rest of the paper is organised as follows. In Sec.~\ref{sec:network_and_population}, we describe the various detector networks that are considered in this study and their parameter estimation capability with respect to the population of BNS mergers accessible to them. Then in Sec.~\ref{sec:em_counterpart_method} we report on the constraining power of each network for different types of cosmological models in the presence of an electromagnetic counterpart. In Sec.~\ref{sec:cosmology}, we repeat the same but utilise gravitational waves alone for cosmological inference. Finally, we conclude and discuss our results in Sec.~\ref{sec:conclusion}.

\section{Characterization of an astrophysical source population in different detector networks}
\label{sec:network_and_population}
In this section, we will introduce the various detector networks considered in this study and the BNS source population accessible to them. Thereafter, we will describe the waveform model used to simulate the gravitational waves from a BNS merger and the parameter estimation capabilities of the different detector networks with a focus on the parameters necessary for cosmological inference.

\subsection{Detector networks}
\label{subsec:network}
\begin{table*}[ht]
    \centering
    \begin{tabular}{c|c|c}
        \toprule
        \toprule
        Network Name & Detectors & Detection Rate [$\rm yr^{-1}$] \\
        \midrule
        A+   & LIGO (HLI+), Virgo+, KAGRA+ & 192 \\
        A+V  & LIGO (HLI-Voy), Virgo+, KAGRA+ & 1,969 \\
        A+E  & ET, LIGO (HLI+), KAGRA+ & 43,161 \\
        A+C  & CE, Virgo+, KAGRA+, LIGO-I+ & 113,309 \\
        A+EC & ET, CE, KAGRA+, LIGO-I+ & 180,340 \\
        ECS  & ET, CE, CE-South & 281,131 \\
        \toprule
        \toprule
    \end{tabular}
    \caption{This table shows the acronyms for different detector networks in this study along with their detector configurations. In addition, we quote the number of expected detections per year assuming a network signal-to-noise ratio cutoff of 10.}
    \label{tab:detector_networks}
\end{table*}

We start by briefly describing the various detector configurations used in this study. They are abbreviated and, together with the technologies used, listed in Tab.~\ref{tab:detector_networks}. For an in-depth discussion of the various network configurations and technologies used in a particular detector, we refer the interested reader to~\textcite{Evans:2021gyd,Borhanian:2022czq}.

Our reference network (A+) is the upgraded network of currently operating and planned detectors consisting of LIGO-Hanford, LIGO-Livingston, and LIGO-Aundh~\cite{aLIGO_ref}, the Virgo detector, and the KAGRA detector operating at their targeted sensitivities of A+~\cite{LIGOScientific:2014pky}, Virgo+~\cite{VIRGO:2014yos}, and KAGRA+~\cite{KAGRA:2018plz}, respectively. The proposed improvement to the LIGO detectors without building a new facility consists of its upgrade to `Voyager' technology~\cite{LIGO:2020xsf}, which involves cryogenic cooling of silicon test masses to mitigate quantum noise and mirror thermal fluctuations. Voyager can be designed to either optimize the low frequency sensitivity, termed compact binary optimized (CBO), or the high frequency sensitivity, known as post-merger optimized (PMO) configurations. In this study, we only consider the CBO configuration of the Voyager network with the Virgo and KAGRA detectors continuing at their `plus' sensitivities. The network is abbreviated as A+V.

We, then, look at various possible next-generation detector networks. The proposed triangle-shaped interferometer Einstein Telescope (ET)~\cite{Punturo:2010zza} is an European detector with 10 km arms. We imagine a scenario where this will be the only next-generation detector that will be built. In this case, we consider a background of second-generation detectors consisting of 3 LIGO detectors and the KAGRA detector operating in conjunction with the ET to form the A+E network. Here we do not deem the Virgo detector to be still commissioned.

\begin{figure}[h]
    \centering
    \includegraphics[width=\columnwidth]{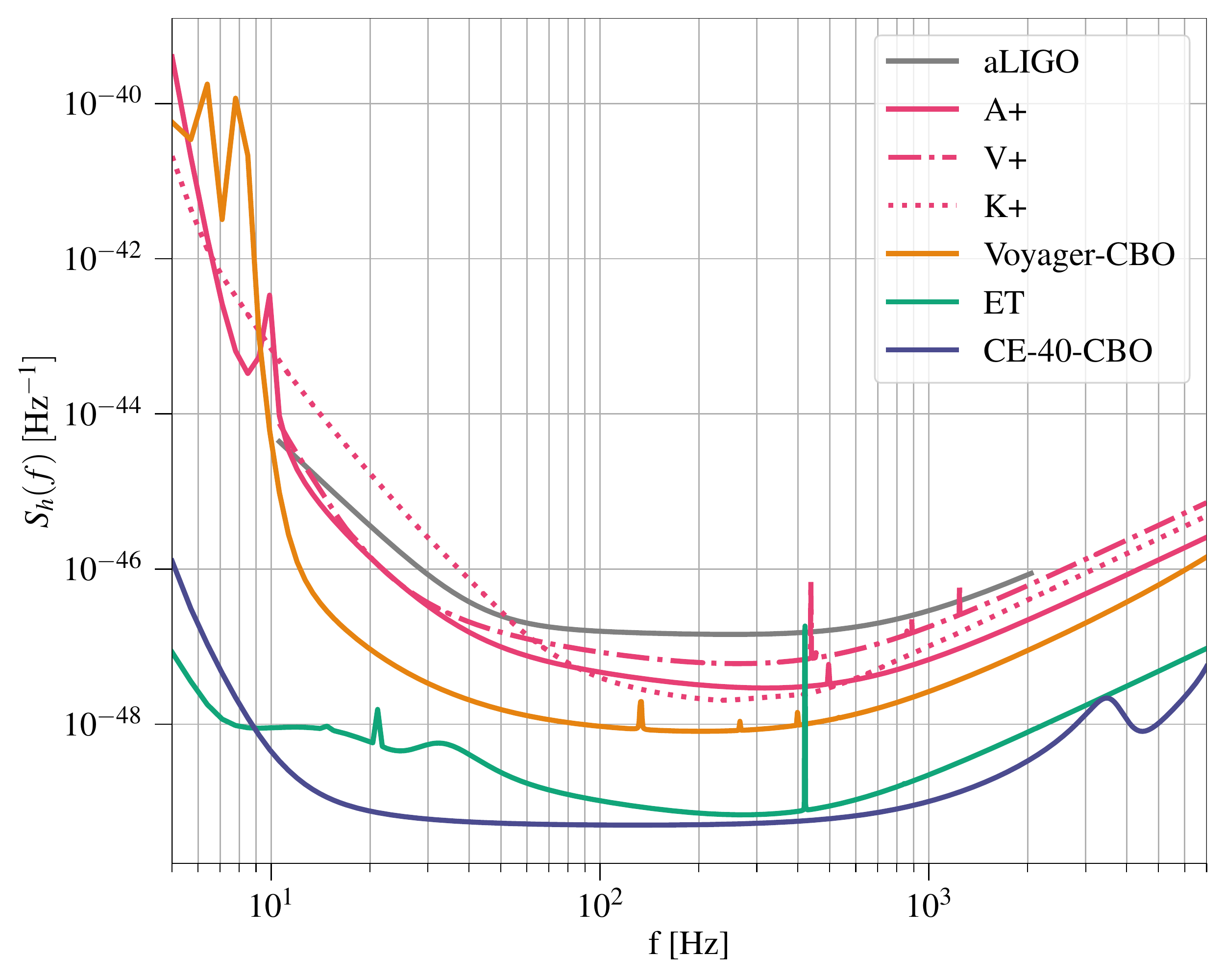}
    \caption{The estimated noise power spectral densities (PSDs) of the current, planned, and future ground-based gravitational-wave detectors.}
    \label{fig:psd}
\end{figure}

For completeness, we contrast this scenario with the one where the only next-generation detector is a single Cosmic Explorer detector (CE)~\cite{Reitze:2019iox}, which is an US undertaking. In this case, the second-generation background would consist of the Virgo, KAGRA, and LIGO-Aundh detectors. As before, we assume that LIGO-Hanford and LIGO-Livingston would be decommissioned once CE is operational. CE can also be preferentially optimized for either low-frequency (CBO) or high-frequency (PMO) and their sensitivities can be seen in Fig.~\ref{fig:psd}. Again, we only consider the CBO setting with a 40 km CE and the network is named A+C. Notice that the post-merger optimized 20 km detector has a greater high-frequency sensitivity than the corresponding 40 km detector~\cite{Martynov:2019gvu}. This is because the free spectral range of a 20 km detector is larger and, hence, at frequencies in the range 1-2 kHz, the antenna response could be designed to be better than a 40 km detector. However, because of the poorer low frequency sensitivity, the detectable population is greatly reduced. 

Next, we consider a third-generation detector network with one CE and one ET in a background of KAGRA and LIGO-Aundh and name it A+EC. The most advanced network configuration that we envision has three next-generation detectors consisting of two 40 km CEs -- one in the US and one in Australia -- and an ET in Europe forming the ECS network. We do not consider a background of second-generation detectors for this most advanced network.

\subsection{Source population}
\label{subsec:population}
We simulate a population of BNS mergers up to a redshift of $z=10$. The redshift distribution of the population is given by
\begin{equation}
    p(z) = \frac{R_z(z)}{\int_0^{10} R_z(z) dz}\,,
\end{equation}
where $R_z(z)$ is the merger rate density in the observer frame and can be expressed as
\begin{equation}
    R_z(z) = \frac{R_m(z)}{1+z}\frac{dV(z)}{dz}.
\end{equation}
Here $dV(z)/dz$ is the comoving volume element and $R_m(z)$ is the merger rate per comoving volume in the source frame which, in turn, is assumed to be proportional to the star formation rate (SFR), $R_f(t)$, and takes the form,
\begin{equation}
    R_m(z) = \int_{t_{\rm min}}^{t_{\rm max}} R_f(t(z) + t_d) \,P(t_d) \,dt_d\,,
\end{equation}
where $t(z)$ is the lookback time given by 
\begin{equation}
    t(z) = \frac{1}{H_0}\int_{z}^{z_f} \frac{dz'}{(1+z')E(z')}.
\end{equation}
This equation signifies that the binaries that form at time $t(z_f)=t(z)+t_d$ merge at lookback time $t(z)$ (i.e., redshift $z$) after a delay time $t_d$. Here, we choose the cosmic SFR to follow~\textcite{Vangioni:2014axa}. The probability distribution for a binary to coalesce after merger is taken to be $P(t_d) \propto 1/t_d$ with $t_{\rm min} = 20 \;\rm Myr$ and $t_{\rm max} = 1/H_0$ ($z\approx30$) in geometric units. The local merger rate $R_m(z=0)$, taken from the second LIGO-Virgo Gravitational-Wave Transient Catalog, \textit{GWTC-2}~\cite{LIGOScientific:2020kqk}, is 
\begin{equation}
    R_m(z=0) = 320 \; \rm Gpc^{-3} yr^{-1}.
\end{equation}
This is an overall normalization factor for the detection rate. If subsequent observing runs of LIGO-Virgo-KAGRA infer this rate to be substantially different, all the bounds on the cosmological parameters would scale as $1/\sqrt{N}$ where $N$ is the ratio of the new rate to the \emph{GWTC-2}~\cite{LIGOScientific:2020kqk} rate used here.

The parameters of the binaries in our population are distributed as follows. We consider non-spinning equal-mass binaries for our population with total mass uniformly distributed between 2.4$\rm M_{\odot}$ and 3.1$\rm M_{\odot}$. The EoS for a NS is taken to be MPA1~\cite{Muther:1987xaa}. The choice for the mass distribution and EoS will be elucidated in Sec.~\ref{subsec:parameter_estimation}. The luminosity distance to a binary is calculated from its redshift using the cosmological parameters inferred by \emph{Planck18} results~\cite{Planck:2018vyg}. The remaining extrinsic parameters, cosine of the inclination angle $\cos\iota$, location of the source on the plane of the sky (cosine of the declination angle $\cos\delta$ and right ascension $\alpha$), polarization angle $\psi$, and the phase of coalescence $\phi_c$, of the fiducial BNS population are drawn from a uniform distribution across their domains.

The detection rate per year and the maximum redshift up to which a given network could detect BNS mergers in our population, assuming a network signal-to-noise (SNR) cutoff of 10, for each of the networks studied here is given in Tab.~\ref{tab:detector_networks}.

\subsection{Parameter estimation}
\label{subsec:parameter_estimation}
\begin{figure}[h]
    \centering
    \includegraphics[width=\columnwidth]{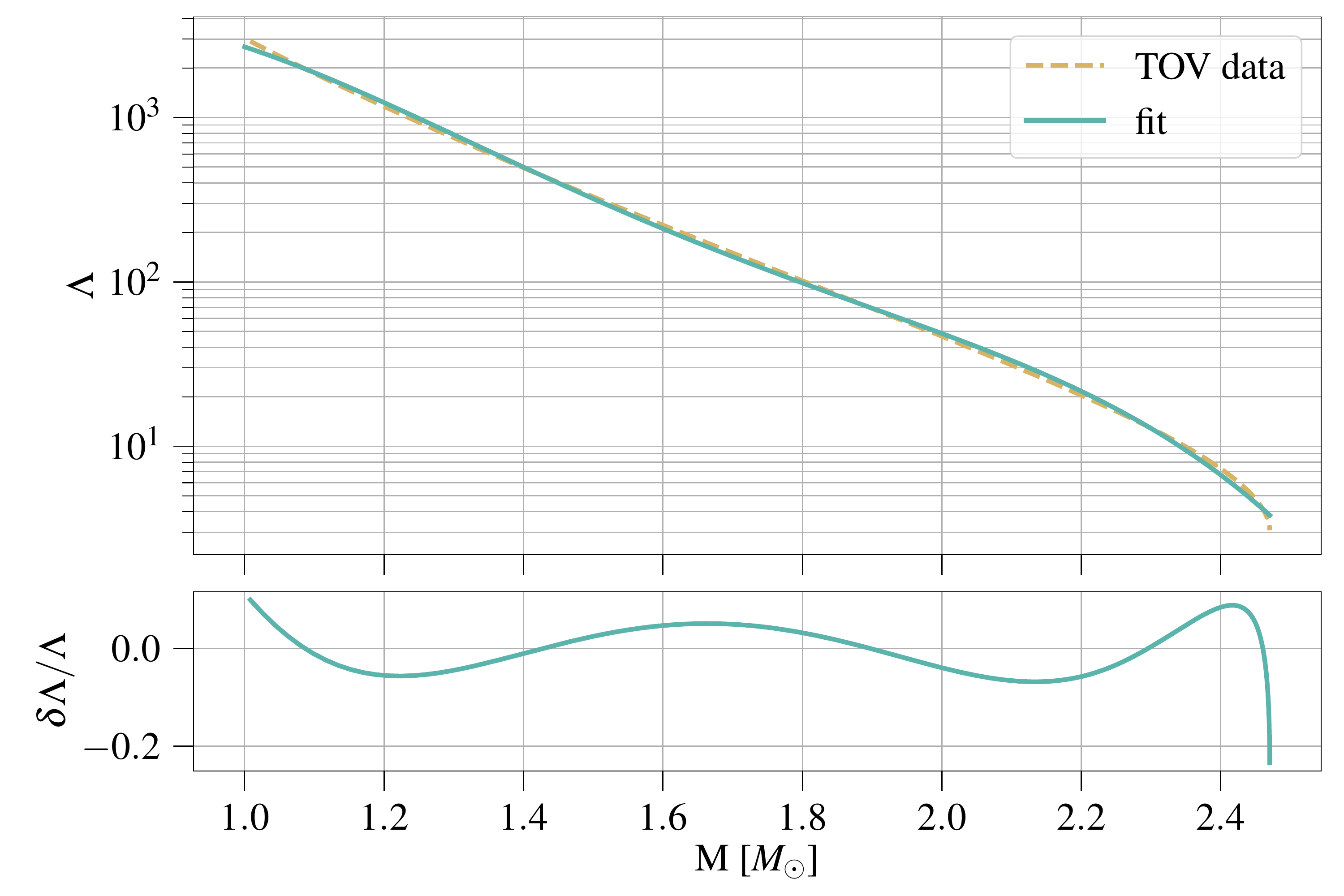}
    \caption{Top: The dimensionless tidal deformability versus NS mass curve and its polynomial fit for the chosen equation of state, MPA1. Bottom: The fractional error between the fit function and the true solution. The true solution for $\Lambda$ for each mass $M$ is obtained by solving the TOV equations. We use the fit to break the mass-redshift degeneracy in the inspiral part of the gravitational-wave signal.}
    \label{fig:lam_M}
\end{figure}

Gravitational waves emitted from BNS mergers are simulated using two separate analytic waveform models for the inspiral and the postmerger parts of the signal. For the inspiral part, the frequency domain \textit{TaylorF2} waveform model is used with added tidal effects incorporated at the 5th post-Newtonian (PN) order~\cite{Flanagan:2007ix,Favata:2013rwa}. The point-particle terms in the waveform model are included up to the 3.5PN order~\cite{Buonanno:2009zt}. For the post-merger part, we use a simplified model proposed in~\textcite{Soultanis:2021oia} consisting of only damped sinusoids. This would be a variation of their model sAc (see Table 1 of Ref.~\cite{Soultanis:2021oia}), without the phenomenological Tukey window. This simplified model allows us to compute the analytical Fourier transform of exponentially damped sinusoids in terms of Lorentzian functions as done in~\textcite{Berti:2005ys}. We give a brief overview of their model in Appendix~\ref{sec:pm_model} for the benefit of the reader. This is an equal-mass model with total masses between 2.4$\rm M_{\odot}$ and 3.1$\rm M_{\odot}$. Additionally, the numerical simulations used to construct this model used the MPA1 EoS for NSs. These are the reasons for our choice of mass distribution and EoS.

The measurement errors in the parameters of the gravitational-wave signal are calculated using the Fisher approximation to the likelihood of a signal using the publicly available code \textit{\sc gwbench}~\cite{Borhanian:2020ypi}. For a set of parameters $\vec{\theta}$, the Fisher matrix is given by
\begin{equation}
    \mathcal{F}_{ij} = \left<\frac{\partial h(f)}{\partial \theta^i}, \frac{\partial h(f)}{\partial \theta^j} \right>\,,
\end{equation}
and, consequently, the covariance matrix $\mathcal{C}_{ij} = \mathcal{F}_{ij}^{-1}$, where the inner product is defined as 
\begin{equation}
    \left<a(f),b(f)\right> = 2\int_{f_{\rm low}}^{f_{\rm high}} \frac{a(f)^*b(f)+a(f)b(f)^*}{S_h(f)} df\,,
\end{equation}
$h(f)$ is the frequency-domain gravitational-wave waveform, $f_{\rm low}$ is 10 Hz for the current detectors and their proposed upgrades and 5 Hz for next-generation detectors, and $f_{\rm high}=4096\, \rm Hz$. The parameter space of the Fisher matrix is $\vec{\theta}=\{\mathcal{M}_c, D_L, z, \cos\iota, \alpha, \cos\delta, \psi, (\phi_c, t_c)\}$ where $\mathcal{M}_c$, $\phi_c$, and $t_c$ are the chirp mass, the phase of the signal at coalescence, and the time of coalescence, respectively. The parenthesis on $\phi_c$ and $t_c$ indicates that the errors on these parameters are estimated from the inspiral signal only. 

The astute reader will notice that mass ratio is not included in the Fisher parameter space. This is because \emph{\sc gwbench} cannot handle numerical derivatives with respect to the mass ratio for \emph{exactly} equal mass systems. Nonetheless, we do not expect it to change our results in any material way since the mass ratio is well-determined from the early inspiral. This can be seen from Fig. 3 of~\textcite{Borhanian:2022czq} which shows that the symmetric mass ratio can be measured better than a few times $10^{-4}$ for $\sim50\%$ of the sources.

The inspiral and the post-merger is joined at the merger frequency $f_{\rm merg}$~\cite{Dietrich:2020eud} defined by
\begin{equation}
    f_{\rm merg} = 2\pi\omega_0\sqrt{\frac{1}{q}}\frac{1+n_1\kappa_{\rm eff}^T+n_2(\kappa_{\rm eff}^T)^2}{1+d_1\kappa_{\rm eff}^T+d_2(\kappa_{\rm eff}^T)^2} \,,
\end{equation}
with $n_1=3.354\times10^{-2}$, $n_2=4.315\times10^{-5}$, $d_1=7.542\times10^{-2}$, $d_2=2.236\times10^{-4}$, $\omega_0=0.3586$, and $\kappa_{\rm eff}^T=(3/16)\Tilde{\Lambda}$. $\Tilde{\Lambda}$ is a certain mass-weighted linear combination of the individual tidal deformabilities, $\Lambda_i$, of the companion stars called \emph{reduced tidal deformability}. For equal-mass binaries, as is our case, the reduced tidal deformability coincides with the individual tidal deformabilities of the two NSs.
Since we are only interested in the Fisher errors, which do not get contributions from an overall phase as is clear from the definition of the inner product, we do not try to match the inspiral and post-merger waveform phases at the merger frequency. Instead, our method can be thought of as independent measurements of the inspiral and post-merger signals giving independent constraints on the parameters, which can then be combined to get the joint constraints as done in multi-band studies of gravitational-wave signals using space-based and ground-based detectors \cite{Gupta:2020lxa,Datta:2020vcj}. Nevertheless, abrupt cutoffs of waveforms can affect the parameter estimation accuracy as discussed in~\textcite{Mandel:2014tca} due to the information present in the termination. However, they find that the effect is inconsequential for a BNS in a detector with sensitivity comparable to that of Initial LIGO~\cite{Damour:2000zb}. The same is assumed to hold true in this paper though this requires further investigation.

Another point of contention would be our use of the inspiral waveform up to the merger frequency which is beyond the frequency of the \emph{innermost stable circular orbit} (ISCO) for heavier masses. We will later see that the postmerger signal makes minimal contribution to the main conclusions of the paper and, by extension, we do not expect a high frequency termination of the waveform at the merger frequency instead of the ISCO frequency to make any significant differences to our results. We note that the postmerger for signals from higher redshifts ($z>3$) occur at the heart of next-generation detector sensitivity but these signals are weak with a BNS detection efficiency of 20\% at z=4 and merger rates at these redshifts are low.

In Sec.~\ref{sec:em_counterpart_method}, we assume two types of electromagnetic counterparts for cosmological inference -- kilonovae and short GRBs (see that section for more details on the two populations). Kilonovae can be very accurately localised in the sky and the host galaxy can be identified. Therefore, for the kilonova population we recompute the Fisher errors assuming that the sky position of the source is known exactly. Similarly, GRB pulse modeling can lead to an exact measurement of the inclination angle $\iota$ of the binary~\cite{Belgacem:2019tbw,Sathyaprakash:2009xt,Arun:2014ysa} and its follow-up in other electromagnetic bands can give an accurate sky localisation. Therefore, for the GRB population, we recompute the Fisher errors assuming that the sky position and inclination angles are known. We have also computed the distance errors assuming a Gaussian prior of width $10^{\circ}$ on the inclination angle and the results are similar to assuming a delta function prior. We will later see that an independent measurement of the GRB inclination angle from electromagnetic observations is crucial for the use of this population in cosmological inference. 

We get the errors on the redshift by utilizing the tidal terms in the PN expansion, which depends on the source-frame NS mass, to break the mass-redshift degeneracy. We do a polynomial fit for the log of the dimensionless tidal deformability parameter of a NS $\Lambda$ in terms of its source-frame mass $M$ in solar mass units. The best-fit polynomial is given by
\begin{equation}
    \log_{10}(\Lambda) = -1.21 M^4 + 7.80 M^3 - 18.2 M^2 + 16.5 M - 1.46,
\end{equation}
and  we show the fit in Fig.~\ref{fig:lam_M}.
The true solution for $\Lambda$ as a function of $M$ is given by solving the \emph{Tolman–Oppenheimer–Volkoff}~\cite{Tolman:1939jz,Oppenheimer:1939ne} (TOV) equations for a given nuclear EoS but this fit suffices our purposes. We also verify that the fit matches the slope of the $\Lambda(M)$ curve since the contribution to the Fisher matrix is the derivative of the curve.

For the post-merger signal, the amplitudes, frequencies, damping times, and the phase differences of the various damped sinusoids are given in terms of the total source-frame mass of the binary (see Appendix~\ref{sec:pm_model}) and these break the mass-redshift degeneracy. 

The errors on the cosmological parameters are determined using the luminosity distance-redshift relation, $D_L \equiv D_L(z;\vec{\phi})$, and depends on a set of cosmological parameters, $\vec{\phi}$. The specific parameters depend on the type of cosmology assumed and are described in detail in Secs.~\ref{sec:em_counterpart_method} and~\ref{sec:cosmology}. Given independent measurements of the redshift and the luminosity distance, the errors can be propagated from the $D_L$--$z$ space that was calculated above to that of the cosmological parameters using another Fisher matrix~\cite{Heavens:2014xba} given by
\begin{equation}
\label{eq:cosmo_fisher}
    \mathcal{G}_{ij} = \sum_{k=1}^N \frac{1}{\sigma_{D_L,k}^2} \frac{\partial D_L^k(z)}{\partial \phi^i} \frac{\partial D_L^k(z)}{\partial \phi^j} \,,
\end{equation}
where $N$ is the number of events and $\sigma_{D_L,k}$ is the total error on the luminosity distance $D_L^k$ of the $k$-th event. The total luminosity distance error has two contributions. One is the error on the gravitational-wave strain amplitude, $\sigma_{D_L}^{h}$ and the second is the error in the measurement of the redshift via the relation
\begin{equation}
\label{eq:errz_dl}
    \sigma_{D_L}^{z} = \left|\frac{\partial D_L}{\partial z}\right| \sigma_{z}.
\end{equation}
Therefore, the total luminosity distance error is given by
\begin{equation}
    (\sigma_{D_L})^2 = (\sigma_{D_L}^{h})^2 + (\sigma_{D_L}^{z})^2.
\end{equation}
For the electromagnetic counterpart method, we assume the redshift to be known accurately and, therefore, the total error on the luminosity distance is given solely by the error in the gravitational-wave strain amplitude. This assumption is justified because once a galaxy is identified, photometric and spectroscopic methods can estimate the redshift very precisely. On the other hand, when we estimate the redshift from gravitational-wave itself using the tidal effects to break the mass-redshift degeneracy, the redshift error too contributes to the total luminosity distance error.

\section{Cosmological inference using kilonova/GRB counterpart}
\label{sec:em_counterpart_method}
In this section, we will determine how well a set of cosmological parameters can be measured if a gravitational-wave event is followed by an observed kilonova or a GRB counterpart. For that, we will first describe our calculation to determine the possibility of a kilonova/GRB counterpart observation. We will then look at the luminosity distance errors for this population. As stated in the previous section, we will assume the redshift to be known accurately for this set of events. Finally, we will propagate the errors on the luminosity distance to the set of cosmological parameters.

\subsection{Rates of electromagnetic counterparts}
\label{subsec:em_rates}
\begin{table}[h]
    \centering
    \begin{tabular}{c|c|c}
        \toprule
        \toprule
        \multirow{2}{*}{Network name} & \multicolumn{2}{c}{Detection Rate [$\rm yr^{-1}$]} \\
        \cmidrule{2-3}
        & $z\leq0.5$ & GW+GRB \\
        \midrule
        A+ & 192 & 11 \\
        A+V & 1,945 & 23 \\
        A+E & 9,282 & 48 \\
        A+C & 9,976 & 55 \\
        A+EC & 11,201 & 69 \\
        ECS & 11,426 & 60 \\
        \bottomrule
        \bottomrule
    \end{tabular}
    \caption{This table shows the number of expected gravitational-wave events within a redshift of $z=0.5$ and the number of coincident GRBs that cross the detection threshold in a given year for various detector networks considered in this study. We assume 10\% of the events detected within a redshift of $z=0.5$ to be followed up electromagnetically to detect kilonova emission and a quarter of the coincident GRBs to be in the direction scanned by GRB detectors.}
    \label{tab:em_detection_rate}
\end{table}

BNS mergers leave behind a relativistic ejecta that produces radiation across the entire electromagnetic spectrum \cite{LIGOScientific:2017ync} ranging from high energy short GRBs that are generated seconds after the merger \cite{LIGOScientific:2017zic}, to optical and infrared emissions produced hours and weeks after merger \cite{Coulter:2017wya, Cowperthwaite:2017dyu, Kasen:2017sxr, DES:2017kbs, Valenti:2017ngx, Arcavi:2017xiz, Tanvir:2017pws, Lipunov:2017dwd, Evans:2017mmy}, X-rays emitted days and years after merger \cite{Margutti:2017cjl, Hajela:2021faz} and radio afterglows \cite{Hallinan:2017woc} that are observed over weeks to months or even years after the merger \cite{Balasubramanian:2022sie}. Short GRB observatories, like Swift and Fermi Space Observatory \cite{Swift:2022}, are sensitive to about a quarter of the sky and can provide accurate sky position for follow-up observations by optical and infrared telescopes to identify the host galaxy and its redshift. Dedicated follow-up initiatives are also required to observe the prompt thermal kilonova \cite{Metzger:2019zeh}, which can then be used to identify the host galaxy and its redshift.

\subsubsection{Kilonova counterpart}
\label{subsubsec:kilonova_rates}
Electromagnetic telescopes, other than GRB observatories, cannot observe binary mergers to large distances accessible to future gravitational-wave detector networks. Among the current and upcoming telescopes, Vera Rubin Observatory has the greatest coverage and is expected to be able to observe kilonovae up to a redshift of about $z=0.5$ (see, e.g., Table 2.2 in Ref.~\cite{Kalogera:2021bya}). We take this to be the largest redshift from which a kilonova counterpart can be observed in following up a gravitational-wave event. Furthermore, we assume that only 10\% of the observed BNS mergers will be followed-up electromagnetically. This is because for the next 10 years Vera Rubin observatory is likely to focus on its primary science mission and follow-up only the best localized 10\% of all GW events. In the era of next-generation observatories, there will be thousands of BNS mergers accessible to GW each year and it is reasonable to expect only some 10\% will be followed up and randomly sample for this from the full detectable BNS population. 
We give the number of BNS mergers expected to be detected per year within a redshift of $z=0.5$ for the various detector networks used in this study in Tab.~\ref{tab:em_detection_rate}. Our results for cosmological parameter inference based on kilonovae are calculated using 10\% of the events given in that Table. Note that the estimates scale with the number of events $N$ as $1/\sqrt{N}$ and, thus, one can recompute our results for one's preferred number of events. 

\subsubsection{GRB counterpart}
\label{subsubsec:grb_rates}
GRB detectors, on the other hand, have a large sky coverage and would not need a targeted follow-up. It depends only on the jet opening angle relative to the angular momentum direction.
The rate of a coincident GRB detection can be calculated following the procedure outlined in \textcite{Belgacem:2019tbw} and is sketched out here for completeness. We assume a Gaussian structured jet profile~\cite{Howell:2018nhu} for a GRB and the luminosity $L(\theta_V)$ is given by
\begin{equation}
\label{eq:grb_lum}
    L(\theta_V) = L_p \exp\left(-\frac{\theta_V^2}{2 \theta_c^2}\right),
\end{equation}
where $\theta_V$ is the viewing angle and $\theta_c = 4.7\degree$ represents the variation in the GRB jet opening angle. $L_p$ is the peak luminosity of each burst assuming isotropic emission in the rest frame in the $1-10^4$ keV energy range and can be sampled from the probability distribution 
\begin{equation}
    \Phi(L_p) \propto
    \begin{cases}
        (L_p/L_*)^{\alpha},\qquad L_p<L_*, \\
        (L_p/L_*)^{\beta}, \qquad L_p\geq L_*,
\end{cases}
\end{equation}
where the parameters of the broken power-law distribution are $L_{*}=2 \times 10^{52} \,\rm erg/s$, $\alpha=-1.95$, and $\beta=-3$~\cite{Wanderman:2014eza}. A GRB is assumed to be detected if the observed peak flux $F_P(\theta_V) = L(\theta_V)/4\pi D_L^2$, given the GW luminosity distance and inclination angle, is greater than the flux limit of $1.1 \rm \,ph\, s^{-1}\, cm^{-2}$~\cite{Belgacem:2019tbw} in the $50$--300 keV band for Fermi-GBM. The total time-averaged observable sky fraction for the Fermi-GBM is taken to be 0.6~\cite{Burns:2015fol}.

Using this procedure, the number of coincident GRB detections expected for various gravitational-wave networks is tabulated in Tab.~\ref{tab:em_detection_rate}. However, since GRB detectors are sensitive to only a fourth of the sky, we use only a quarter of the coincident events for constraining cosmological parameters. Short GRBs are much more luminous than a kilonova and, hence, can be observed at much farther distance. Nevertheless, the farthest observable GRB that we find in our population has a redshift of $\sim 3$.

\subsection{Luminosity distance errors}
\label{subsec:em_dl_err}
\begin{figure*}[ht]
    \centering
    \includegraphics[width=2\columnwidth]{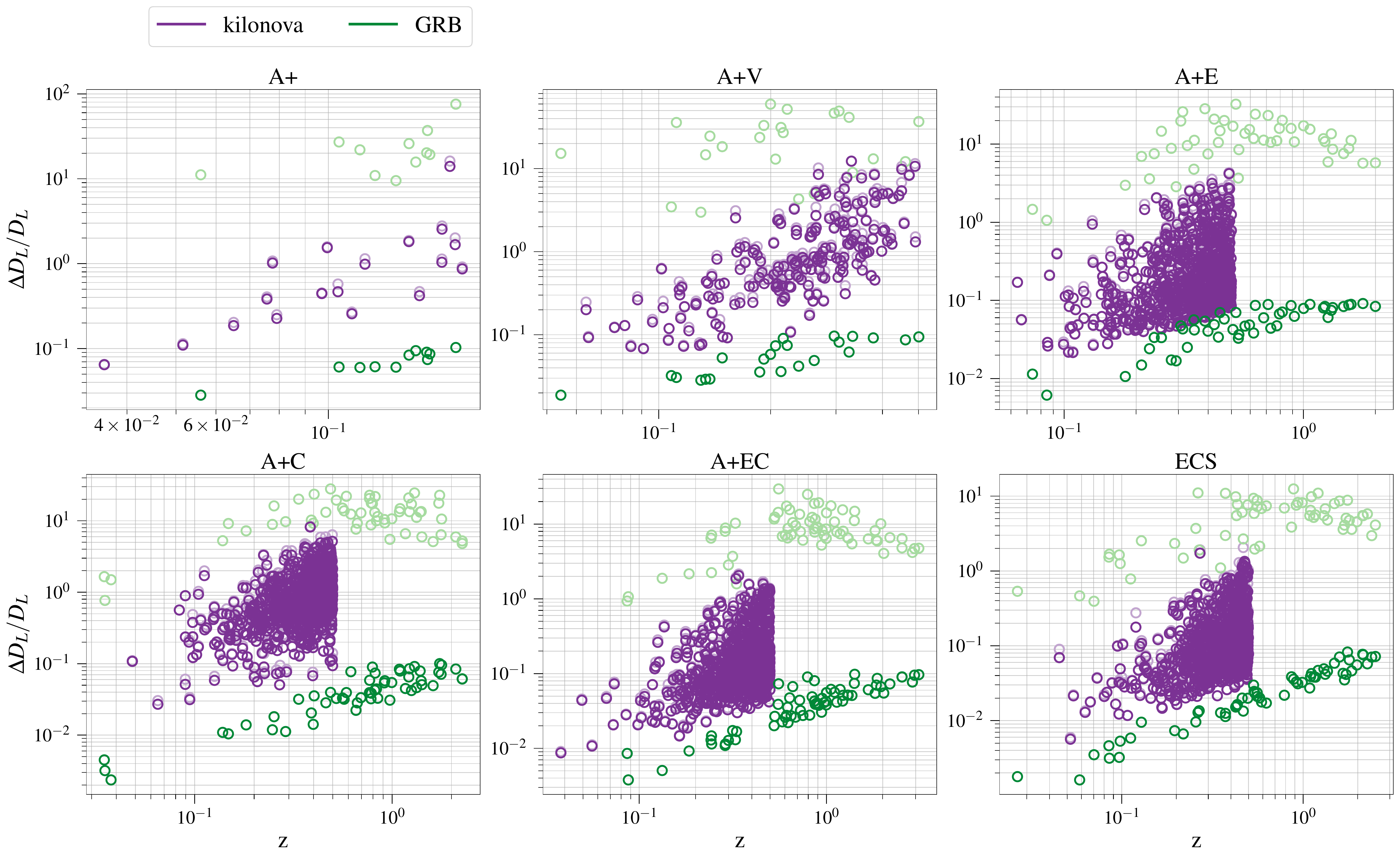}
    \caption{Scatter plot of the fractional luminosity distance errors as a function of redshift for the kilonova and GRB populations for the different detector networks. Lighter shades show the distance errors for the full parameter space while the darker shades illustrate the case where the sky position is known for the kilonova population and for the GRB population inclination angle is also known.}
    \label{fig:em_dl_err}
\end{figure*}

Let us now look at the luminosity distance errors for the population of kilonovae and GRBs selected using the method outlined in the previous section. This is the population that is used for cosmological inference. 

In Fig.~\ref{fig:em_dl_err}, we show a scatter plot of the fractional errors in the luminosity distance as a function of redshift for the expected kilonova and GRB populations for all the networks considered in this study. The kilonova population is shown in purple while the GRBs are in green. In each figure, the lighter colors show the luminosity distance errors assuming unknown position in the sky and inclination angle while the darker colors show the errors with known sky position for the kilonova population and known inclination angle, in addition to sky position, for the GRB population.

The luminosity distance errors for the kilonova population is generically smaller than that of the GRB population by significantly more than an order of magnitude, if the errors are computed for the full parameter space. This can be understood as follows. The GRB population is composed of nearly-face-on systems because of the small opening angle of the GRB and a Gaussian pulse profile, which falls off very sharply [see Eq.~(\ref{eq:grb_lum})]. These are systems that have the strongest $D_L$--$\iota$ degeneracy and, hence, the poorest luminosity distance measurements. On the other hand, the kilonova population is sampled randomly from the full population where the edge-on systems are more probable than face-on systems. The $D_L$--$\iota$ degeneracy breaks as a system becomes more edge-on. Therefore, this population has a markedly better luminosity distance estimate. Note also that the GRB population can be observed from farther out in redshift and, consequently, are weaker signals. 

If the position of the source on the sky is known, there is marginal improvement in the distance estimates for the kilonova population. On the other hand, if the inclination angle is known too, the distance estimates improve by over two orders of magnitude for most of the sources in the GRB population and comfortably over an order of magnitude for all. We stress that the utility of the GRB population to do cosmology relies heavily on the modeling of the inclination angle and its inference independently of from gravitational waves and wrong models could result in large systematic biases in cosmological parameters. We find that similar results are obtained by imposing a prior of width $10^{\circ}$ on the inclination angle.

A further observation from Fig.~\ref{fig:em_dl_err} is that the distance estimates for the kilonova population in the A+E network is noticeably better than the A+C network, even though both have only a single next-generation detector in a background of second-generation detectors. Note also that this is despite A+C having a better detection rate and redshift reach (see Tab.~\ref{tab:detector_networks}). We, however, point out that the same does not hold for the GRB population.

We repeat the same calculation where the A+V, A+C, A+EC, and one CE in the ECS networks are in the PMO configuration. Modulo the detection rate, the PMO configurations that we considered gave similar results.

\subsection{Cosmological models}
\label{subsec:cosmological_models}
In this section, we look at various cosmological models with dynamical and non-dynamical dark energy components and determine the constraints that can be set on various parameters of these models using different detector networks.

A cosmological model, for our purposes here, looks at the background evolution of different energy components that make up the Universe. At late times, this is made up of two components: matter consisting of non-relativistic ordinary matter and cold dark matter (CDM), and dark energy. Two independent observables inform us of this evolution. One is the redshift $z$ of an electromagnetic or gravitational-wave signal from a source which is affected by the evolution of the different energy components of the Universe along the path of the wave. The other is the luminosity distance $D_L$ to the source which, for a source with a known luminosity, gives us a measure of the distance to the source based on how luminous they appear to us. A relationship between the two, for late times and assuming a flat universe with no contribution from the curvature term ($k=0$), is generically given as~\cite{Maggiore:2018sht},
\begin{equation}
\label{eq:dl_z}
    D_L(z) = \frac{1+z}{H_0} \int_0^z \frac{dz'}{\sqrt{\Omega_M (1+z')^3 + \Omega_{\Lambda}(z')}},
\end{equation}
where $H_0$ and $\Omega_M$ are the Hubble parameter and dark matter energy density at the current epoch. $\Omega_{\Lambda}(z)$, which describes the redshift evolution of the dark energy sector, takes the common form
\begin{equation}
    \Omega_{\Lambda}(z) = \Omega_{\Lambda} \; \text{exp}\bigg\{ 3\int_0^z \frac{dz'}{1+z'}[1+w_{\Lambda}(z)] \bigg\},
\end{equation}
for different dark energy models where $w_{\Lambda}(z)$ is the dark energy EoS parameterized in terms of $w_0$ and $w_a$ as 
\begin{equation}
\label{eq:de_eos}
    w_{\Lambda}(z) = w_0 + w_a \frac{z}{1+z}.
\end{equation}
For a spatially flat Universe, $\Omega_{\Lambda}=1-\Omega_M$ at late times. Note that Eq.~(\ref{eq:dl_z}) is valid for the gravitational-wave luminosity distance in general relativity (GR) and in a cosmological model which do not change the tensor perturbations with respect to GR.

We consider three types of dark energy models of increasing generality. First, we examine the standard cosmological model, $\Lambda$CDM, which consists of a constant dark energy density. This corresponds to $w_0=-1$ and $w_a=0$. Next, we allow for a non-trivial, but constant, EoS of dark energy. This is equivalent to a $w_0\neq-1$ but $w_a=0$ and is named $w$CDM. Finally, we let the dark energy EoS evolve with redshift according to Eq.~(\ref{eq:de_eos}). This model goes by $w_0w_a$CDM. 

We, then, turn our attention to theories of gravity that have modified tensor perturbations with respect to GR. The equations of motion for tensor perturbations in a homogeneous and isotropic background in GR is given by~\cite{Maggiore:2018sht}
\begin{equation}
    \Tilde{h}_{+/\times}'' + 2\mathcal{H}\Tilde{h}_{+/\times}' + k^2\Tilde{h}_{+/\times} = 0,
\end{equation}
where $\Tilde{h}_{+/\times}$ are the two tensor polarization modes in Fourier space, $\mathcal{H}=a'/a$ with $a(\eta)$ the scale factor, and the primes are derivatives with respect to the conformal time $\eta$. In a theory that modifies the luminosity distance, the above equation changes to~\cite{Maggiore:2018sht}
\begin{equation}
    \Tilde{h}_{+/\times}'' + 2\mathcal{H}[1-\delta(\eta)]\Tilde{h}_{+/\times}' + k^2\Tilde{h}_{+/\times} = 0,
\end{equation}
where $\delta(\eta)$ is a function of the extra degree(s) of freedom in the modified theory. The luminosity distance, in this theory is given by
\begin{equation}
    \overline{D}_L(z) = D_L(z) \exp\left\{-\int_0^z \frac{dz'}{1+z'}\delta(z')\right\},
\end{equation}
where $D_L(z)$ is the luminosity distance given in Eq.~(\ref{eq:dl_z}). The usual parameterization of the term in exponential is~\cite{Maggiore:2018sht}
\begin{equation}
    \frac{\overline{D}_L}{D_L} = \Xi_0 + \frac{1-\Xi_0}{(1+z)^n},
\end{equation}
where $n$ determines the rate at which the ratio asymptotes to $\Xi_0$ at large redshifts.

\subsubsection{$\Lambda$CDM cosmology}
\label{subsubsec:lcdm}
\begin{figure}[ht]
    \centering
    \includegraphics[width=\columnwidth]{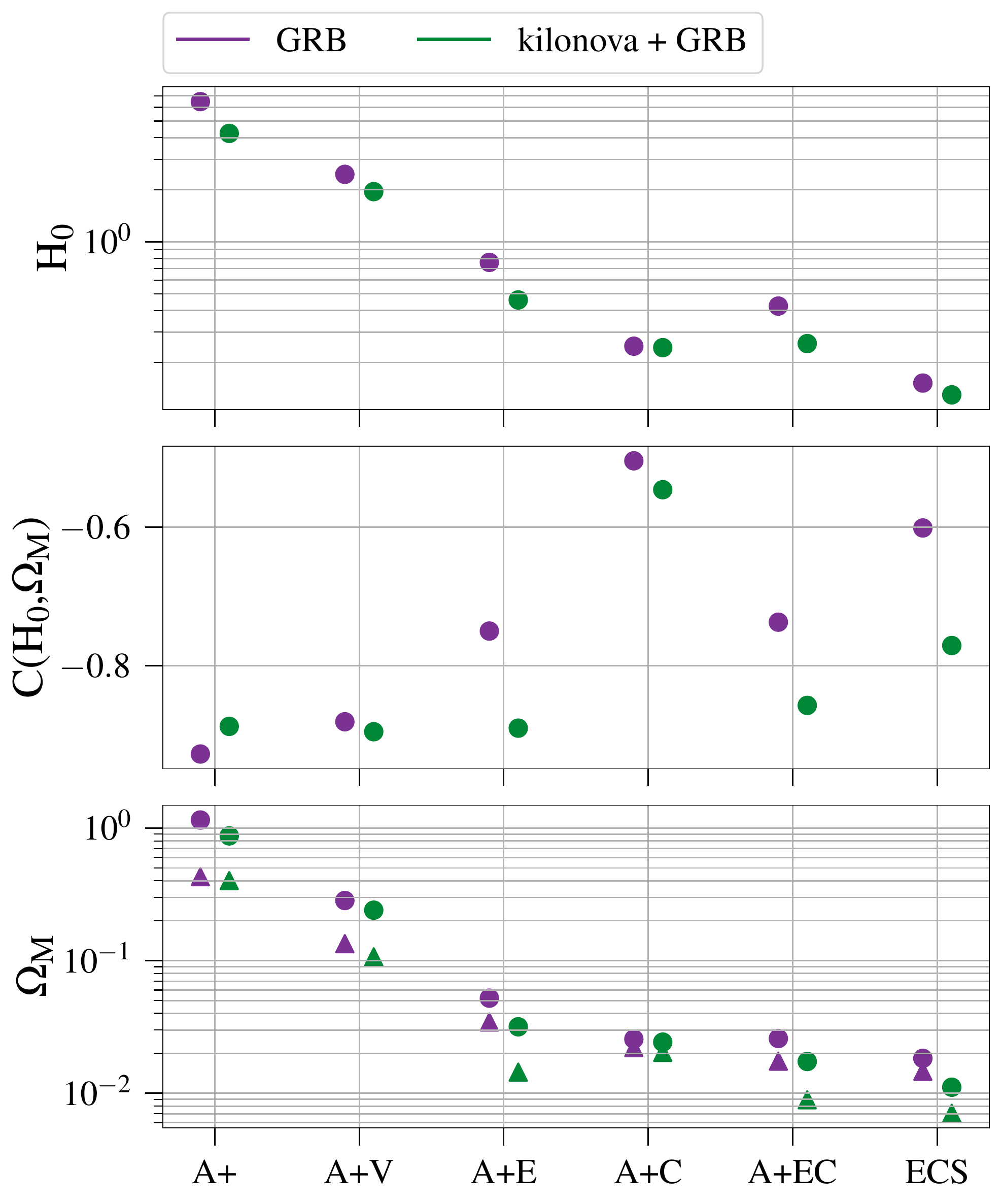}
    \caption{1-$\sigma$ errors on and the correlation coefficients between the parameters of the $\Lambda$CDM cosmological model for the various networks. The triangles on the bottom panel show the 1-$\sigma$ errors on $\Omega_M$ if $H_0$ is assumed to be known. We show the results for the GRB population in purple and the combined population of GRBs and kilonovae is shown in green.}
    \label{fig:lcdm_em}
\end{figure}

The standard model of cosmology, which is consistent with the current observations by the Planck experiment~\cite{Planck:2018vyg}, has a non-dynamical dark energy sector. The luminosity distance in this case simplifies from Eq.~(\ref{eq:dl_z}) to take the form
\begin{equation}
    D_L(z) = \frac{1+z}{H_0} \int_0^z \frac{dz'}{\sqrt{\Omega_M (1+z')^3 + \Omega_{\Lambda}}},
\end{equation}
where all the parameters are as defined above and $\Omega_{\Lambda}=1-\Omega_M$. 

In this section, we will determine the accuracy to which $H_0$ and $\Omega_M$ can be measured using a kilonova and a GRB counterpart to a BNS merger. We will use Eq.~(\ref{eq:cosmo_fisher}) to calculate the errors on these quantities. 
In Fig.~\ref{fig:lcdm_em}, we show these errors for different networks, where the error bars correspond to 1-$\sigma$ errors on the parameters. The triangles in the bottom panel show the errors on $\Omega_M$ if the value of $H_0$ is known, which we take to be the one determined in~\textcite{Planck:2018vyg}. The GRB population is shown in purple, while the green color show the errors for the combined population consisting of both GRBs and kilonovae. In the middle panel, the correlation coefficients between $H_0$ and $\Omega_M$ are plotted. The values for errors are tabulated in Tab.~\ref{tab:em_cosmology}.

The main takeaways from Fig.~\ref{fig:lcdm_em} and Tab.~\ref{tab:em_cosmology} are as follows. 
\begin{itemize}
    \item In general, the GRB population, even though fewer in number, has better error estimates on cosmological parameters. This is because of the $1/\sqrt{N}$ dependence of the errors on the number of events. Recall that the luminosity distance errors for the GRB population is, on average, an order of magnitude lower than the kilonova population. To compensate for the poorer distance estimate, the kilonova population has to be two orders of magnitude larger in size.
    \item $H_0$ and $\Omega_M$ are significantly correlated. Therefore, if $H_0$ is assumed to be known from other cosmological probes, the constraints on $\Omega_M$ improves remarkably.
    \item The second-generation A+ network can achieve a sub-10\% estimate on the Hubble constant $H_0$ while the best network is the ECS network consisting of three third-generation detectors, which can measure $H_0$ to a 0.2\% accuracy. Note that this is the case when $\Omega_M$ is measured in conjunction as well. A measurement of just the Hubble constant assuming the dark matter energy density is known would provide an even better constraint.
    \item We, again, point out that the A+E network performs better than A+C for the kilonova population due to its superior distance measurement despite having a smaller redshift reach and a resulting lower detection rate. The same is not true for the GRB population where the A+C network is significantly better than A+E. However, this difference is possibly because of the handful of extremely close events in our population for the A+C network, as can be seen in Fig.~\ref{fig:em_dl_err}; the distance estimates and the detection rates for the two networks are similar and, therefore, we expect similar cosmological inference. 
\end{itemize}

We  repeat the calculations where the A+V, A+C, A+EC,
and one CE in the ECS networks are in the PMO configuration. We find that the CBO network settings can constrain the parameters to a higher precision. This is primarily due to the greater detection rates for the CBO configuration due to a better low frequency sensitivity.

\subsubsection{wCDM}
\label{subsubsec:wcdm_em}
\begin{figure*}[ht]
    \centering
    \includegraphics[width=2\columnwidth]{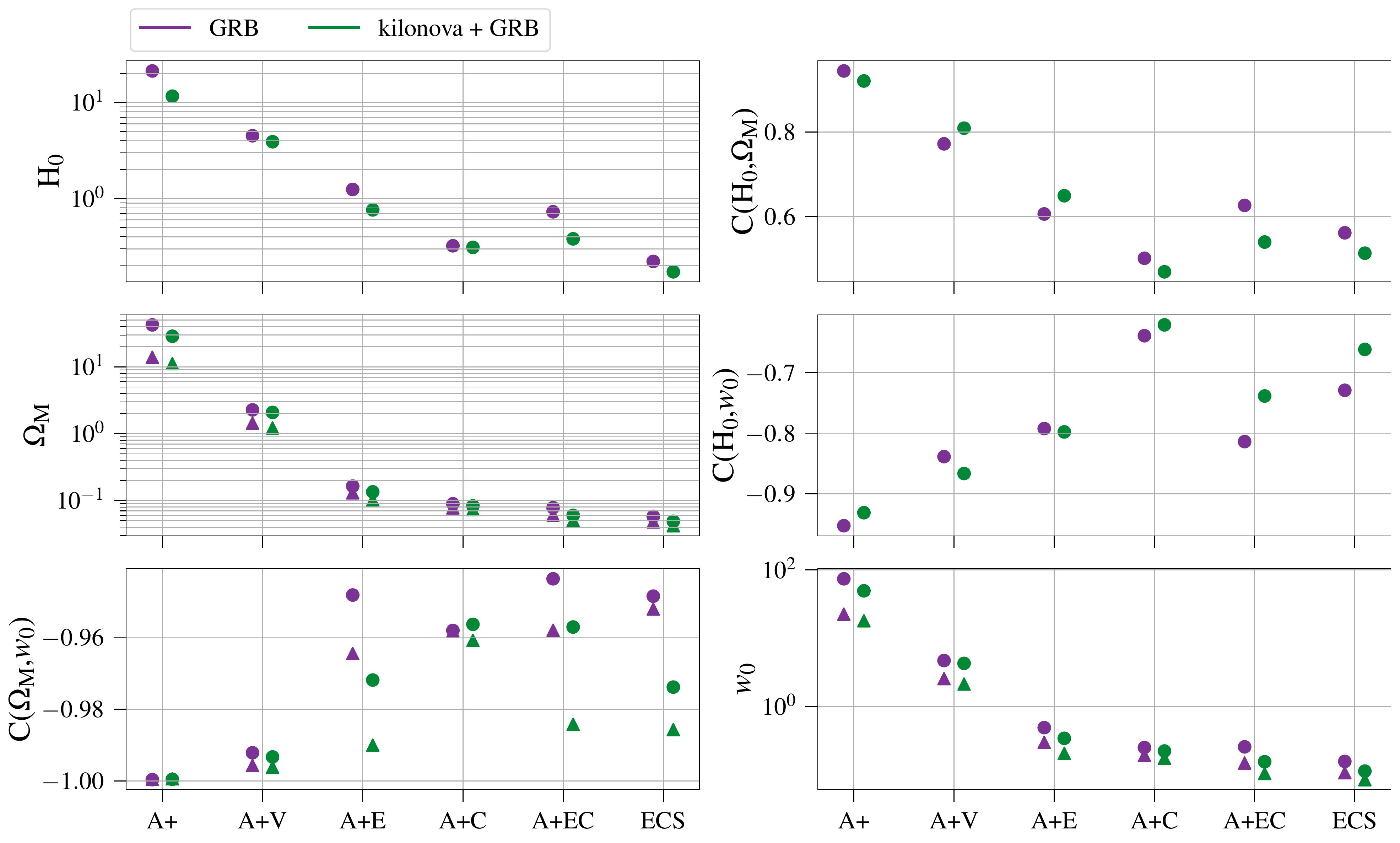}
    \caption{1-$\sigma$ errors on and the correlation coefficients between the parameters of the $w$CDM cosmological model for the various networks. The triangles show the respective quantities if $H_0$ is assumed to be known. We show the results for the GRB population in purple and the combined population of GRBs and kilonovae is shown in green.}
    \label{fig:wcdm_em}
\end{figure*}

\begin{table*}[ht]
    \centering
    \begin{tabular}{cV{3}c|cV{3}c|c|cV{3}*{3}{c|}cV{3}}
        \toprule
        \toprule
        \multirow{2}{*}{\diagbox{Network}{$\sigma(\vec{\phi})$}} & \multicolumn{2}{cV{3}}{$\Lambda$CDM} & \multicolumn{3}{cV{3}}{$w$CDM} & \multicolumn{4}{cV{3}}{$w_0w_a$CDM} \\
        \cmidrule(lr){2-3} \cmidrule(lr){4-6} \cmidrule(lr){7-10}
        & $\sigma(H_0)$ (\%) & $\sigma(\Omega_M)$ (\%) & $\sigma(H_0)$ (\%) & $\sigma(\Omega_M)$ (\%) & $\sigma(w_0)$ & $\sigma(H_0)$ (\%) & $\sigma(\Omega_M)$ (\%) & $\sigma(w_0)$ & $\sigma(w_a)$ \\
        \midrule
        \multicolumn{10}{c}{\textit{GRB}} \\
        \midrule
        \multirow{2}{*}{A+} & 9.6 & 380 & 32 & 14000 & 74 & 180 & 660000 & 2800 & 12000 \\
        & - & 140 & - & 4400 & 22 & - & 130000 & 560 & 2000 \\
        \midrule
        \multirow{2}{*}{A+V} & 3.6 & 92 & 6.6 & 740 & 4.6 & 12 & 7200 & 28 & 180  \\
        & - & 44 & - & 460 & 2.6 & - & 4400 & 18 & 94 \\
        \midrule
        \multirow{2}{*}{A+E} & 1.1 & 17 & 1.8 & 52 & 0.48 & 3.2 & 220 & 0.52 & 9.6 \\
        & - & 11 & - & 42 & 0.30 & - & 150 & 0.52 & 5.6 \\
        \midrule
        \multirow{2}{*}{A+C} & 0.36 & 8.2 & 0.48 & 28 & 0.24 & 0.72 & 100 & 0.26 & 4.4 \\
        & - & 7.2 & - & 24 & 0.19 & - & 78 & 0.28 & 3.0 \\
        \midrule
        \multirow{2}{*}{A+EC} & 0.62 & 8.4 & 1.1 & 26 & 0.26 & 1.8 & 78 & 0.26 & 4.0 \\
        & - & 5.6 & - & 2.0 & 0.15 & - & 60 & 0.20 & 2.4 \\
        \midrule
        \multirow{2}{*}{ECS} & 0.22 & 6.0 & 0.32 & 19 & 0.16 & 0.50 & 60 & 0.18 & 2.6 \\
        & - & 3.6 & - & 15 & 0.11 & - & 48 & 0.18 & 1.7 \\
        \midrule
        
        \multicolumn{10}{c}{\textit{Kilonova}} \\
        \midrule
        \multirow{2}{*}{A+} & 11 & 820 & 27 & 21200 & 110 & 75 & 590000 & 2500 & 10000 \\
        & - & 360 & - & 9200 & 44 & - & 230000 & 1000 & 3700 \\
        \midrule
        \multirow{2}{*}{A+V} & 5.0 & 150 & 13 & 1800 & 11 & 29 & 21000 & 82 & 460 \\
        & - & 57 & - & 830 & 4.4 & - & 11000 & 46 & 210 \\
        \midrule
        \multirow{2}{*}{A+E} & 0.87 & 13 & 2.2 & 130 & 0.96 & 5.8 & 1600 & 5.3 & 42 \\
        & - & 5.1 & - & 61 & 0.37 & - & 710 & 2.9 & 16 \\
        \midrule
        \multirow{2}{*}{A+C} & 2.0 & 34 & 3.9 & 300 & 2.1 & 8.9 & 3600 & 13 & 92 \\
        & - & 17 & - & 180 & 1.1 & - & 1800 & 7.2 & 41 \\
        \midrule
        \multirow{2}{*}{A+EC} & 0.50 & 7.6 & 0.89 & 64 & 0.45 & 1.4 & 680 & 2.5 & 17 \\
        & - & 3.3 & - & 41 & 0.25 & - & 450 & 1.8 & 10 \\
        \midrule
        \multirow{2}{*}{ECS} & 0.37 & 5.8 & 0.70 & 50 & 0.35 & 1.3 & 570 & 2.1 & 15 \\
        & - & 2.6 & - & 31 & 0.19 & - & 330 & 1.4 & 7.7 \\
        \midrule
        
        \multicolumn{10}{c}{\textit{GRB + Kilonova}} \\
        \midrule
        \multirow{2}{*}{A+} & 6.3 & 280 & 17 & 9300 & 50 & 42 & 230000 & 1000 & 4100 \\
        & - & 130 & - & 3600 & 18 & - & 100000 & 450 & 1700 \\
        \midrule
        \multirow{2}{*}{A+V} & 2.9 & 78 & 5.8 & 670 & 4.3 & 11 & 6200 & 24 & 150 \\
        & - & 34 & - & 400 & 2.1 & - & 3700 & 16 & 78 \\
        \midrule
        \multirow{2}{*}{A+E} & 0.68 & 10 & 1.1 & 44 & 0.34 & 2.2 & 150 & 0.38 & 6.2 \\
        & - & 4.6 & - & 33 & 0.20 & - & 100 & 0.38 & 3.2 \\
        \midrule
        \multirow{2}{*}{A+C} & 0.36 & 7.9 & 0.46 & 27 & 0.22 & 0.67 & 96 & 0.26 & 4.1 \\
        & - & 6.6 & - & 24 & 0.17 & - & 76 & 0.26 & 2.8 \\
        \midrule
        \multirow{2}{*}{A+EC} & 0.38 & 5.6 & 0.56 & 19 & 0.15 & 0.87 & 59 & 0.16 & 2.5 \\
        & - & 2.2 & - & 16 & 0.10 & - & 47 & 0.16 & 1.6 \\
        \midrule
        \multirow{2}{*}{ECS} & 0.19 & 3.6 & 0.26 & 16 & 0.11 & 0.38 & 47 & 0.14 & 1.8 \\
        & - & 2.3 & - & 14 & 0.085 & - & 38 & 0.14 & 1.2 \\
        \bottomrule
        \bottomrule
    \end{tabular}
    \caption{1-$\sigma$ errors on the parameters of the various types of cosmologies with differing dark energy models for different detector networks and different populations corresponding to their electromagnetic counterpart. If a parameter in a model is assumed to be known a priori, the corresponding cell has been marked with `-'.}
    \label{tab:em_cosmology}
\end{table*}

The standard model of cosmology is modified to let $w_0$ be a free parameter allowing it to take values different from $-1$. The luminosity distance can then be written in the simplified form
\begin{equation}
    D_L(z) = \frac{1+z}{H_0} \int_0^z \frac{dz'}{\sqrt{\Omega_M (1+z')^3 + \Omega_{\Lambda} (1+z)^{3 (1+w_0)}}}.
\end{equation}

In Fig.~\ref{fig:wcdm_em} we plot the errors in and correlations between the parameters of this model which includes $w_0$ in addition to the ones in $\Lambda$CDM. The plot marker and the plot styles are the same as in Fig.~\ref{fig:lcdm_em}, the additional triangles depicting the correlation between parameters if $H_0$ is assumed to be known, i.e., for this cosmological model it shows the correlation between $\Omega_M$ and $w_0$. The 1-$\sigma$ error values can again be found in Tab.~\ref{tab:em_cosmology}. The primary surmise is the same as for the $\Lambda$CDM model. However, the parameter estimates are poorer than in $\Lambda$CDM, as expected, because of an enlarged parameter space. Moreover, the new parameter $w_0$ is heavily correlated with the other parameters. Specifically, we note that the second-generation detector network, or its proposed upgrade to `Voyager' technology, will not place meaningful constraints on any cosmological parameters other than $H_0$. Despite a lower precision of distance estimates using second-generation detector networks, this is, in part, also because of the hierarchy of the parameters in the cosmological model where $\Omega_M$ and $w_0$ appear at a higher order in redshift compared to $H_0$. They are, therefore, better constrained with higher redshift events that are observable with only the next-generation of detectors.

The most stringent constraints on the dark energy EoS parameter $w_0$ is using the ECS detector network. The measurement uncertainty on $w_0$ is $\sim0.1$ or $\sim0.08$ depending on whether $H_0$ is simultaneously measured with $w_0$ or not. The current best constraints on the parameter is reported by the Planck collaboration~\cite{Planck:2018vyg} which combines the Planck data with \emph{baryon acoustic oscillation} (BAO) and SNeIa data to achieve a measurement of $w_0=-1.03\pm0.03$. The measurement accuracy using the counterpart method is comparable to these estimates and would provide a complimentary measurement.

\subsubsection{$w_0w_a$CDM}
\label{subsubsec:w0wacdm_em}
\begin{figure*}[ht!]
    \centering
    \includegraphics[width=2\columnwidth]{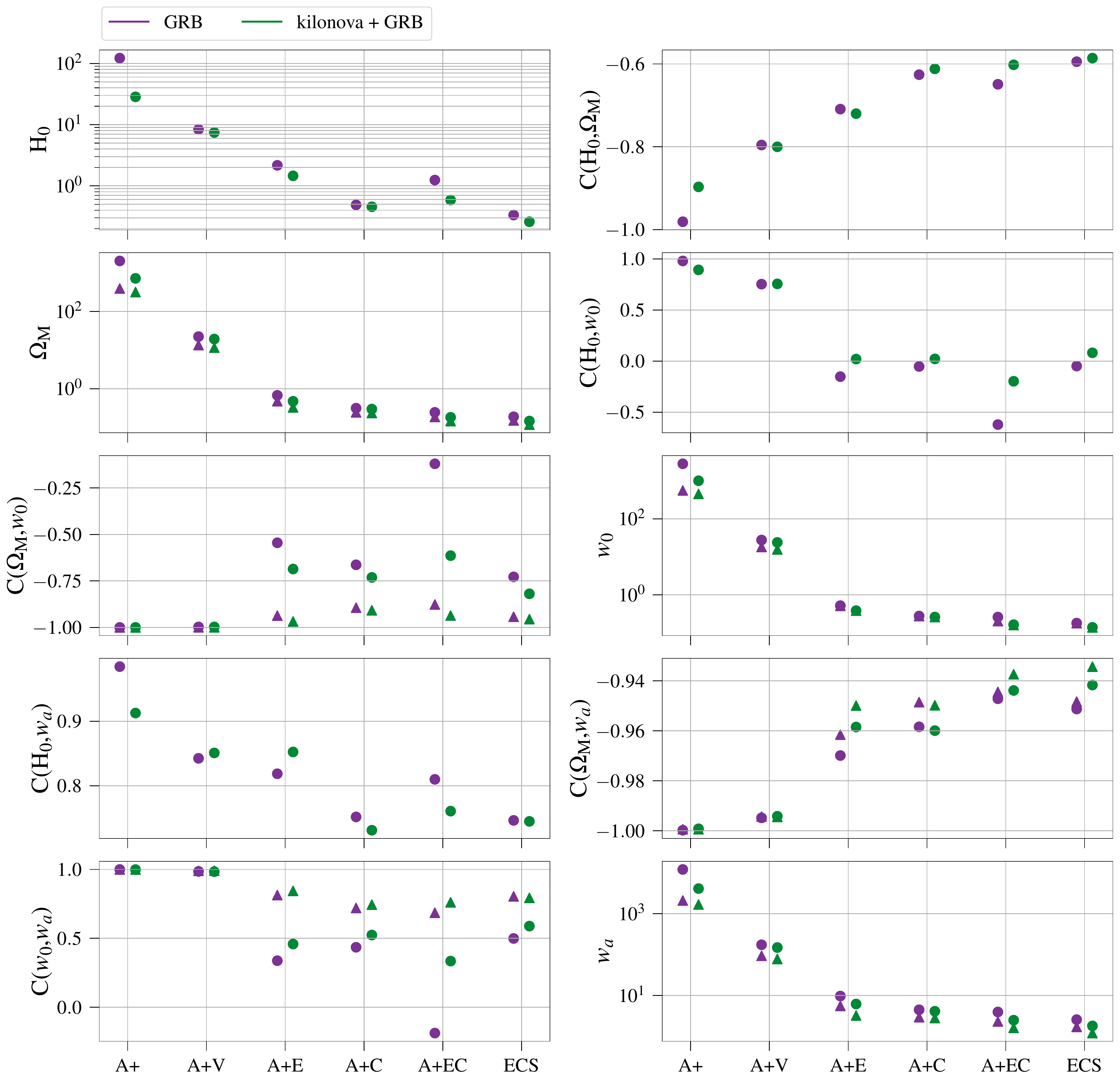}
    \caption{1-$\sigma$ errors on and the correlation coefficients between the parameters of the $w_0w_a$CDM cosmological model for the various networks. The triangles show the respective quantities if $H_0$ is assumed to be known. We show the results for the GRB population in purple and the combined population of GRBs and kilonovae is shown in green.}
    \label{fig:w0wacdm_em}
\end{figure*}

We further extend the late-time cosmological parameter space to account for a time-varying dark energy EoS given by Eq.~(\ref{eq:de_eos}) and parameterised by the additional parameter $w_a$. The errors in and the correlation between the parameters are shown in Fig.~\ref{fig:w0wacdm_em} and the 1-$\sigma$ errors are tabulated in Tab.~\ref{tab:em_cosmology}. We note that we do not expect the current networks or their upgrades to place consequential limits on any cosmological parameters in a joint inference of multiple parameters in lieu of the results of the previous cosmological model but quote the numbers in the Table for completeness. As expected, next-generation detectors drastically improve the estimates.

The most precise measurement of the additional parameter $w_a$ in a joint parameter estimation including and excluding $H_0$ is $\sim1.8$ and $\sim1.2$, respectively. The Planck estimate~\cite{Planck:2018vyg} combining BAO and SNeIa data has an uncertainty of $\sim0.3$. When $w_a$ is also constrained in addition to $w_0$, the current Planck constraint and our best forecast constraint on $w_0$ are both $\sim0.08$.

\begin{figure*}[ht]
    \centering
    \includegraphics[width=2\columnwidth]{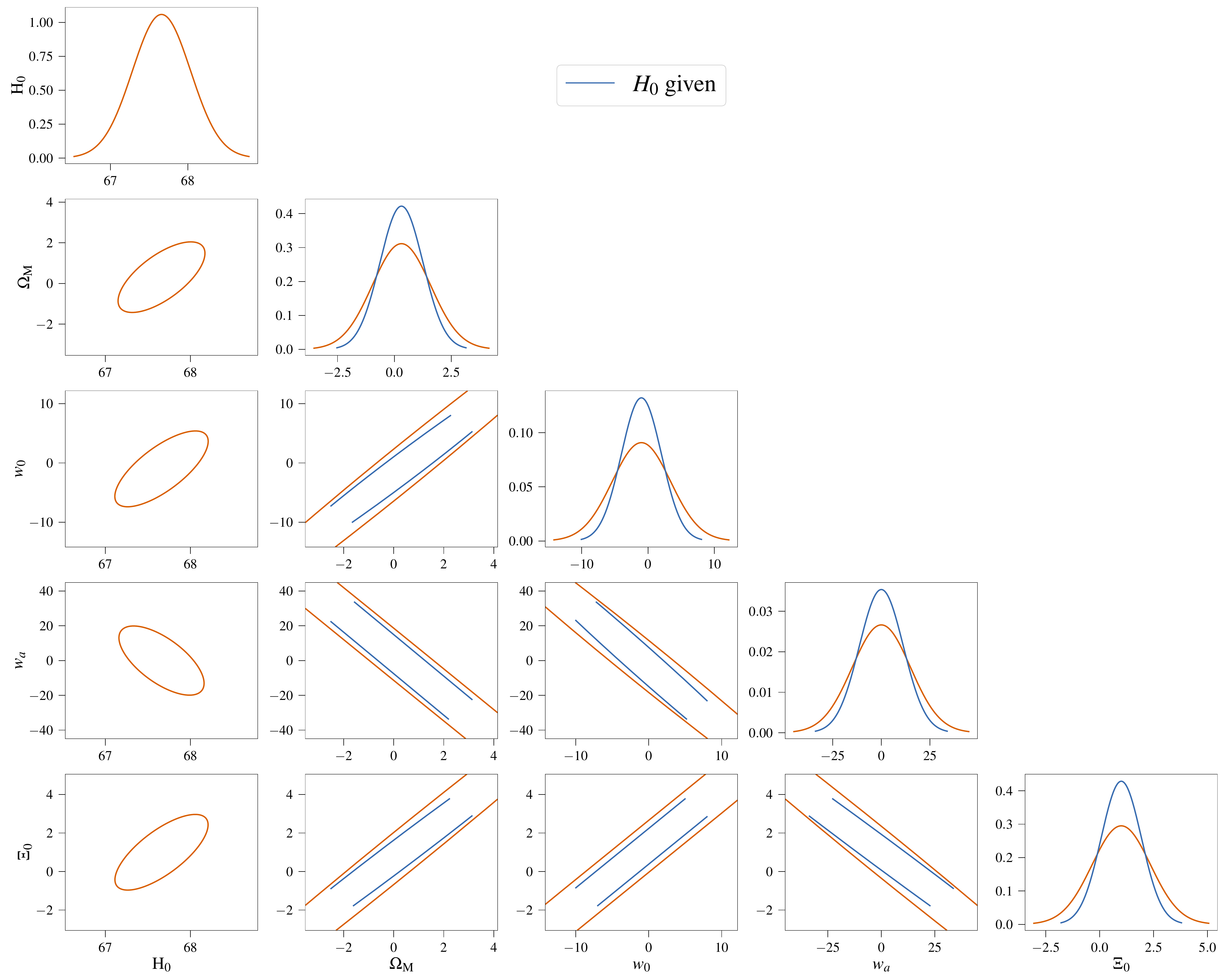}
    \caption{Corner plot for the parameters of the $\Xi_0-w_0w_a$CDM cosmological model using the combined kilonova and GRB population observed by the ECS network. In orange, we show the constraints on the full parameter space whereas for the estimates in blue means $H_0$ is assumed to be known a priori. One can see that the parameters, except $H_0$, are extremely correlated by the almost parallel lines for the error contours.}
    \label{fig:xi-w0wacdm}
\end{figure*}

\begin{table}[ht!]
    \centering
    \begin{tabular}{cV{3}*{4}{c|}cV{3}*{5}{c|}cV{3}}
        \toprule
        \toprule
        \multirow{2}{*}{\diagbox{Network}{$\sigma(\vec{\phi})$}} & \multicolumn{5}{cV{3}}{$\Xi_0-w_0w_a$CDM} \\
        \cmidrule(lr){2-6} 
        & $\sigma(H_0)$ (\%) & $\sigma(\Omega_M)$ (\%) & $\sigma(w_0)$ & $\sigma(w_a)$ & $\sigma(\Xi_0)$ (\%) \\
        \midrule
        \multicolumn{6}{c}{\textit{GRB}} \\
        \midrule
        \multirow{2}{*}{ECS} & 0.66 & 500 & 5.6 & 18 & 170 \\
        & - & 400 & 4.2 & 15 & 130 \\
        \midrule
        
        \multicolumn{6}{c}{\textit{Kilonova}} \\
        \midrule
        \multirow{2}{*}{ECS} & 3.0 & 9400 & 58 & 310 & 2100 \\
        & - & 4600 & 24 & 150 & 940 \\
        \midrule
        
        \multicolumn{6}{c}{\textit{GRB + Kilonova}} \\
        \midrule
        \multirow{2}{*}{ECS} & 0.56 & 410 & 4.4 & 15 & 140 \\
        & - & 300 & 3.0 & 11 & 93 \\
        \bottomrule
        \bottomrule
    \end{tabular}
    \caption{1-$\sigma$ errors on the parameters of the $\Xi_0-w_0w_a$CDM cosmology for the ECS detector network and different populations corresponding to their electromagnetic counterpart. This model has a time-varying dark energy EoS and a modified perturbation of the tensor modes. If a parameter in a model is assumed to be known a priori, the corresponding cell has been marked with `-'. We do not quote the results for the other detector networks because they do not provide any meaningful constraints.}
    \label{tab:xi-w0wacdm}
\end{table}

\subsubsection{$\Xi_0$--$w_0w_a$CDM}
\label{subsubsec:xiw0wacdm}
This is the final cosmological model we will constrain in this section. Here, we add the additional parameter $\Xi_0$ to the parameter space. We do not try to measure the value of $n$, but instead take it to be a fiducial value of 2.5~\cite{Belgacem:2019tbw,Belgacem:2018lbp}. We show the results only for the ECS network because the other networks do not provide anything consequential in a joint estimate of multiple parameters. The corner plot for this model is given in Fig.~\ref{fig:xi-w0wacdm}. The dashed lines show the PMO detector configuration while the solid lines are for the CBO configuration. In orange we show the estimates for the full parameter space whereas the constraints assuming $H_0$ is known is illustrated in blue. The 1-$\sigma$ errors are tabulated in Tab.~\ref{tab:xi-w0wacdm}. We see that the GRB population and, in turn, the combined GRB and kilonova population, can measure $\Xi_0$ at less than 100\% error. 

Of crucial importance is to note that the extra parameter $\Xi_0$ is extremely degenerate with the other cosmological parameters except $H_0$ and, consequently, their estimates worsen by an order of magnitude.

\section{Cosmology using only Gravitational waves}
\label{sec:cosmology}
In this section, we  turn our attention to the potential for cosmological inference using gravitational waves alone. To that effect, we will first ascertain how well gravitational waves can measure the redshift of a BNS merger. Thereafter, we will calculate the contribution of the redshift errors to the total luminosity distance errors. Finally, we will propagate these errors on to the parameters of various cosmological models.

\subsection{Redshift using GWs}
\label{subsec:z_gw}
First, we explore the capabilities of the gravitational-wave detector networks to estimate the redshift to a BNS merger. 

\begin{figure*}[ht]
    \centering
    \includegraphics[width=2\columnwidth]{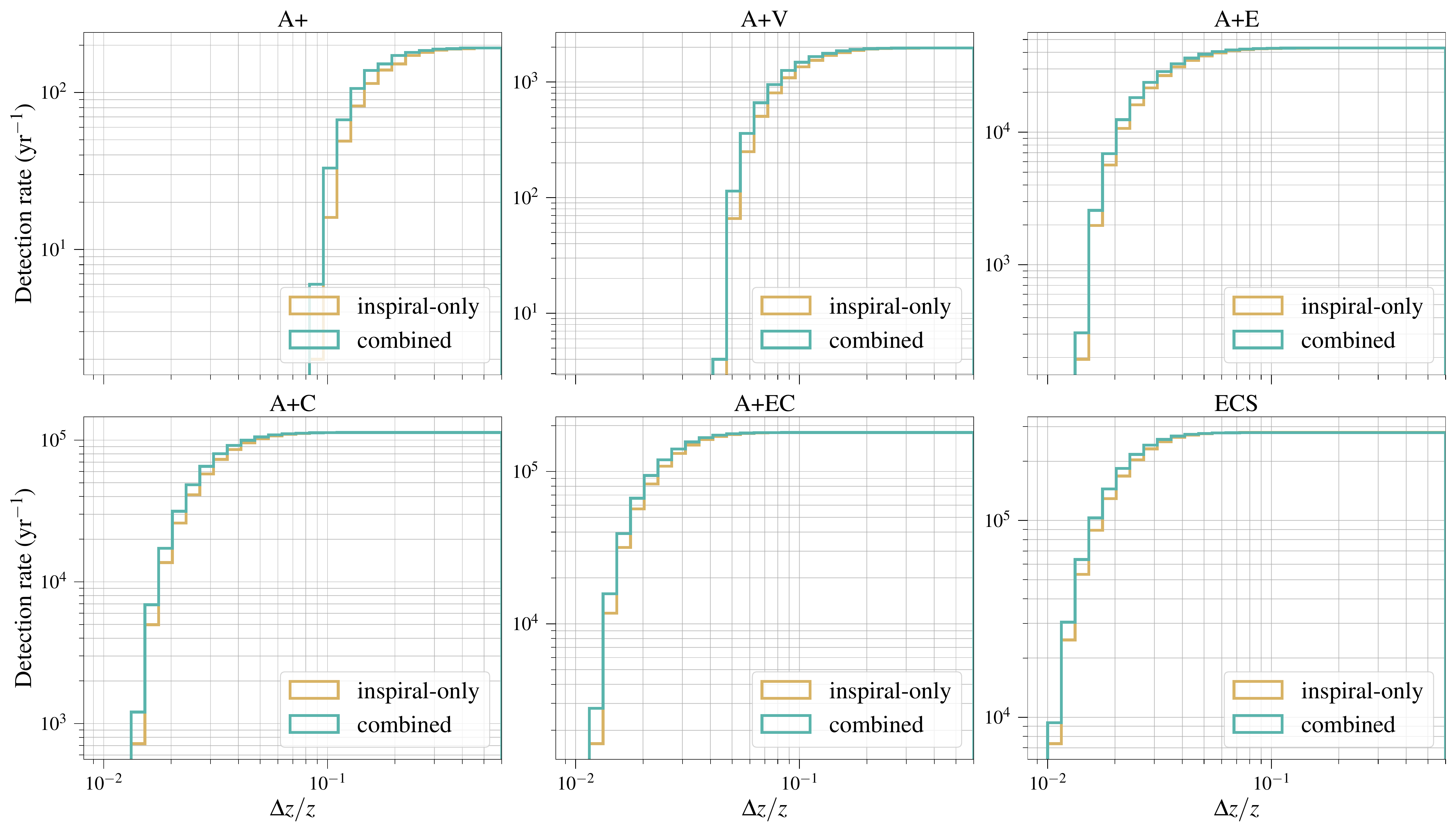}
    \caption{The cumulative distribution functions (CDFs) of the fractional error in redshift measured using gravitational waves for the various networks considered in this study. The CDFs are not normalised and, as a result, the vertical axis depicts the detection rate in a year for the respective network. Each panel additionally shows the redshift estimates using a inspiral-only signal (labeled as `inspiral-only') and a postmerger-included signal (labeled as `combined').}
    \label{fig:frac_errz}
\end{figure*}

In Fig.~\ref{fig:frac_errz}, we show the cumulative distribution functions (CDFs) for the fractional error in the measurement of the redshift from BNS mergers for our populations in different detector networks. Also shown in each panel are the redshifts determined from the inspiral signal alone and the combined inspiral and post-merger signal. We find that the post-merger signal does not improve the redshift estimates appreciably. We note that the postmerger signal does contain significant redshift information for individual sources that are sufficiently close but this sub-population of close-by events does not make a sizeable contribution to the full population. We also found that the redshift measurement using the CBO configuration is better than the corresponding PMO configuration.

\subsection{Total luminosity distance errors}
\label{subsec:dl_errs_gw}
\begin{figure*}[ht]
    \centering
    \includegraphics[width=2\columnwidth]{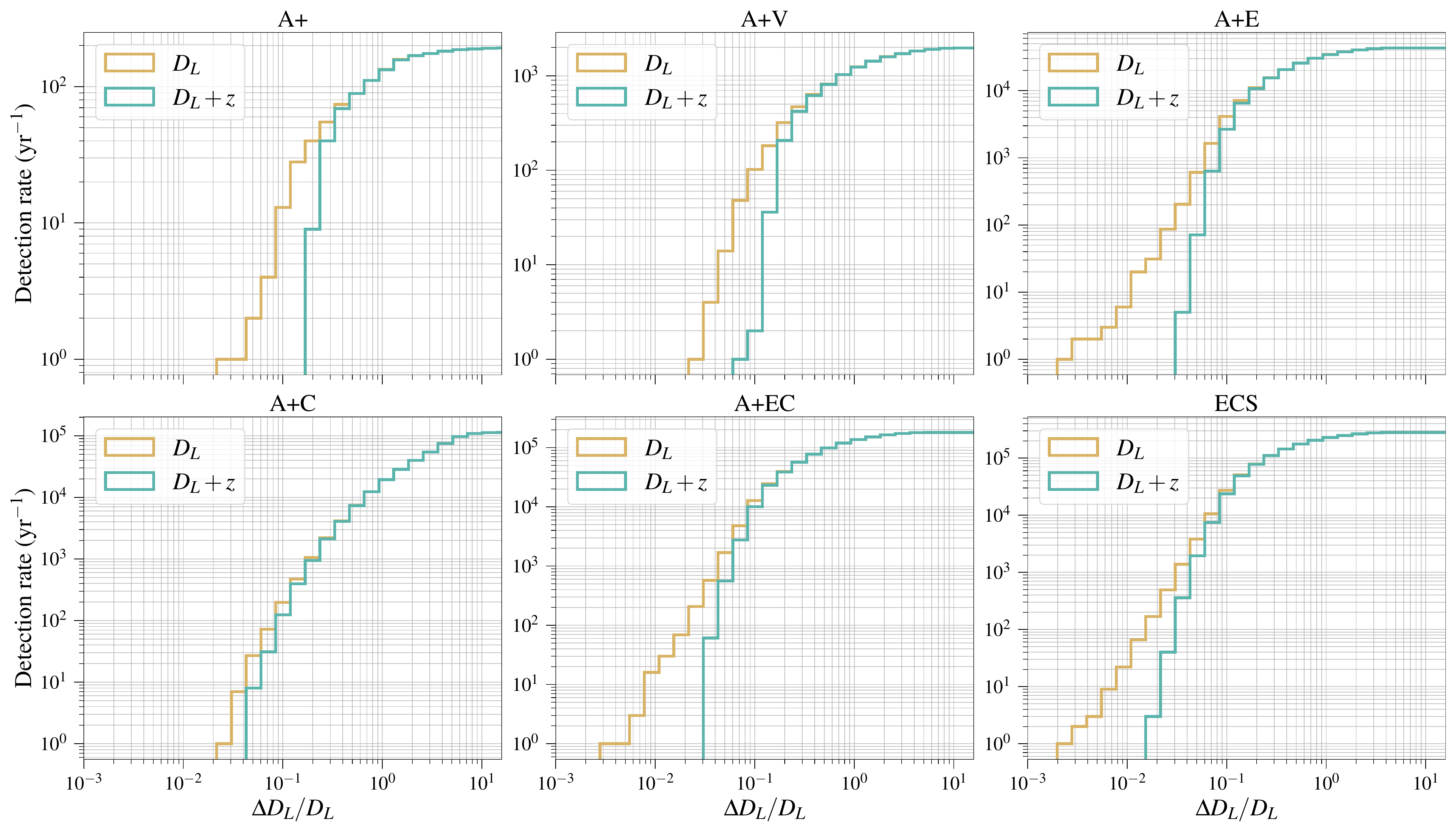}
    \caption{The cumulative distribution functions (CDFs) for the fractional error in the luminosity distance for the various networks considered in this study. The CDFs are not normalised and, as a result, the vertical axis depicts the detection rate in a year for the respective network. The total error in the measurement of the luminosity distance for cosmological parameter estimation has two contributions. One is from the error in the measurement of the gravitational wave strain, which is the standard gravitational-wave luminosity distance error, and the other is the error in the luminosity distance due to the error in redshift measurement. In each of the figures, we show the standard error, which carries the label `$D_L$' on the panels, and the total error, which has the label `$D_L+z$'. We see that the contribution of the error in the redshift measurement to the total error is substantial for only the best measured strain amplitudes and these are the nearby sources. The results shown here are for a postmerger-included signal.}
    \label{fig:frac_errdl}
\end{figure*}

The total luminosity distance error when the redshift of a source is determined from gravitational waves is not just the error in the gravitational-wave strain amplitude but also has a contribution due to the error in the measurement of the redshift as given in Eq.~(\ref{eq:errz_dl}). In Fig.~\ref{fig:frac_errdl} we compare the two sources of errors for different detector networks by plotting the CDFs of the fractional errors. The error due to the strain amplitude is shown in orange while the total error is in green. 

Of note is that the redshift uncertainty contribute significantly to the total error for the best measured strains or, in other words, for nearby sources. We also found that both PMO and CBO configurations can reach similar measurement accuracy, albeit the CBO configuration detects more number of sources. 

\subsection{Cosmological models}
\label{subsec:gw_cosmological_models}
In this section, we consider all the cosmological models examined in Sec.~\ref{sec:em_counterpart_method} and evaluate the performance of each network in the absence of any counterpart observations. The crucial difference with the counterpart method is that the entire population of gravitational-wave observations is available at our disposal.

\subsubsection{$\Lambda$CDM}
\label{subsubsec:lcdm_gw}
\begin{figure}[h]
    \centering
    \includegraphics[width=\columnwidth]{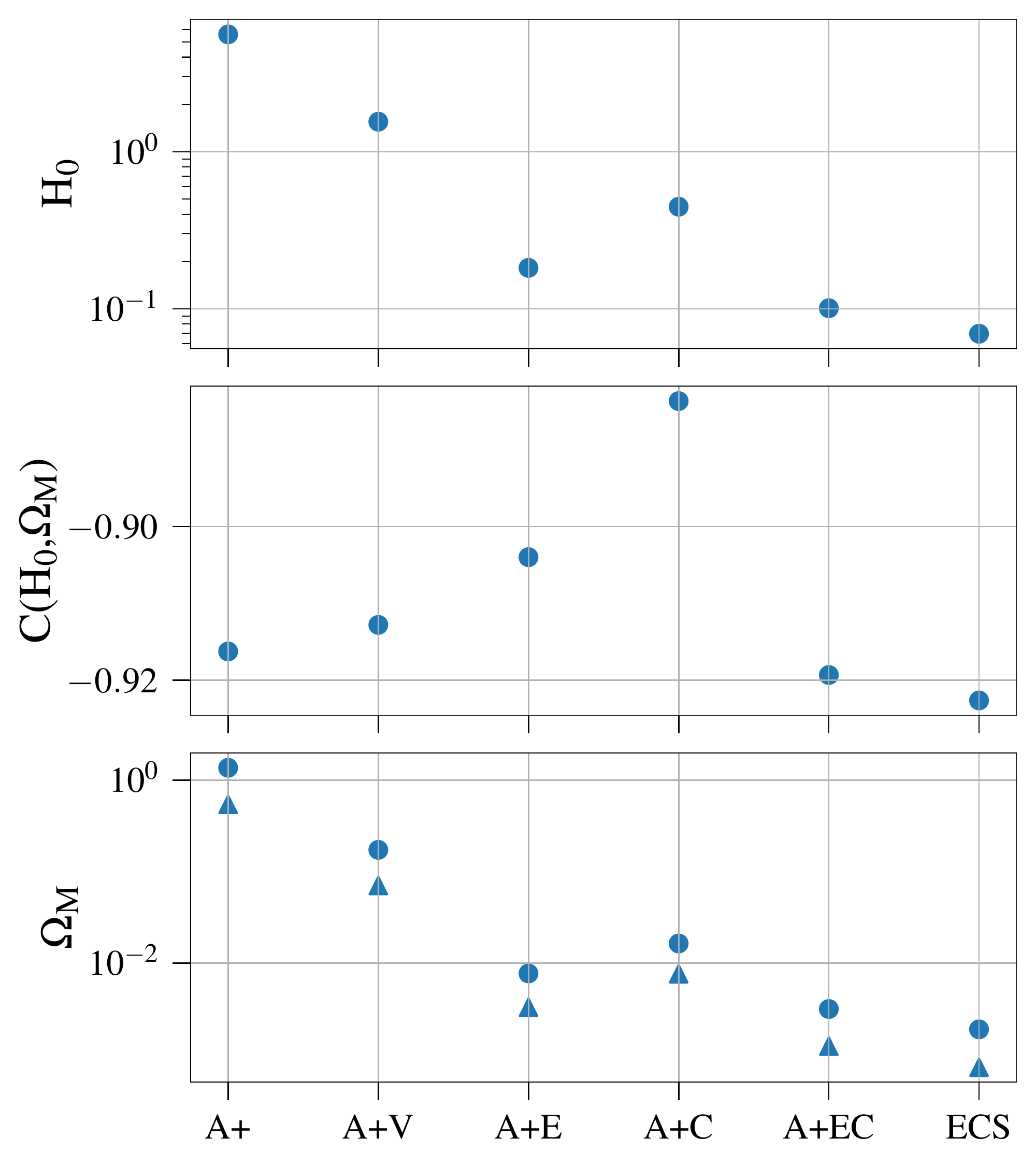}
    \caption{1-$\sigma$ errors on and the correlation coefficients between the parameters of the $\Lambda$CDM cosmological model for the various networks calculated using gravitational-wave observations alone. The triangles on the bottom panel show the 1-$\sigma$ errors on $\Omega_M$ if $H_0$ is assumed to be known.}
    \label{fig:lcdm_gw}
\end{figure}
We estimate the constraints on the parameters of the $\Lambda$CDM cosmological model using gravitational waves alone. As described before, Eq.~(\ref{eq:cosmo_fisher}) is used to calculate the errors on these quantities. Fig.~\ref{fig:lcdm_gw} depicts these 1-$\sigma$ errors for different networks. The error bars correspond to 1-$\sigma$ errors on the parameters. The triangles in the bottom panel show the errors on $\Omega_M$ if the value of $H_0$ is known, which we take to be the one determined in~\textcite{Planck:2018vyg}. In the middle panel, the correlation coefficients between $H_0$ and $\Omega_M$ are plotted. The errors are tabulated in Tab.~\ref{tab:gw_cosmology}.

The main takeaways of the estimates shown in Fig.~\ref{fig:lcdm_gw} and Tab.~\ref{tab:gw_cosmology} are as follows (they are qualitatively the same as for the counterpart method).
\begin{itemize}
    \item The parameters $H_0$ and $\Omega_M$ are highly correlated and knowing $H_0$ significantly improves the bounds on $\Omega_M$.
    \item The A+ network can measure the Hubble constant at $\sim$8\% in a joint estimation of both $H_0$ and $\Omega_M$ cosmological parameters but can place no meaningful bounds on $\Omega_M$ regardless of the knowledge of $H_0$. In contrast, the ECS network can achieve a $0.10\%$ accuracy on $H_0$ and a $0.61\%$ accuracy on $\Omega_M$ after a year of observation. The bound on $\Omega_M$ improves roughly three-fold to $0.23\%$ if the Hubble constant is known a priori.
\end{itemize}

A comparison of the constraints obtained using the  electromagnetic counterpart method and the standalone gravitational-wave inference reveals the following. The bounds on $H_0$ and $\Omega_M$ are stronger using the electromagnetic counterpart method for the A+ network. As pointed out in Sec.~\ref{subsec:dl_errs_gw}, the total luminosity distance error has a significant contribution from the error on the redshift for nearby sources. And since the A+ network can observe only nearby sources, the counterpart method has smaller overall errors and performs better. Of specific note is that the available catalog of sources for the counterpart-less method is about 10 times larger than the kilonova population and 200 times larger than the GRB population wherein it is the GRB population that overwhelmingly dominates the constraints obtained using the counterpart method. This is, as we argued in Sec.~\ref{subsubsec:lcdm}, a consequence of the way errors scale with the number of events. Since the luminosity distance errors for the GRB population is an order of magnitude lower than the kilonova population on average, there has to be two orders of magnitude more kilonova to get the same error estimates on cosmological parameters. The motivation to use gravitational waves alone for cosmological inference was not only because of a ten-fold increase in population size but also due to the superior reach of the gravitational-wave network. But the redshift of the farthest source in our population for the A+ network is 0.2 which is well within the capabilities of electromagnetic telescopes and also not large enough for $\Omega_M$ to start contributing significantly to the luminosity distance--redshift relation. This latter effect becomes clear when comparing the results for the ECS network. We see that the ECS network gives similar constraints on $H_0$ using both methods. But $\Omega_M$ is determined to an accuracy of 4 and 8 times better, respectively, in a joint analysis of the parameters and assuming $H_0$ is known. The other networks show similar trends. The estimates on both $H_0$ and $\Omega_M$ are better with the counterpart method in the absence of any third-generation detectors where the farthest detected gravitational-wave source is within the detection capability of an electromagnetic telescope. With next-generation observatories, the $\Omega_M$ estimates improve drastically in the counterpart-less method with the $H_0$ estimates on par or better than the counterpart method.

\subsubsection{wCDM}
\label{subsubsec:wcdm}
\begin{figure*}[ht]
    \centering
    \includegraphics[width=2\columnwidth]{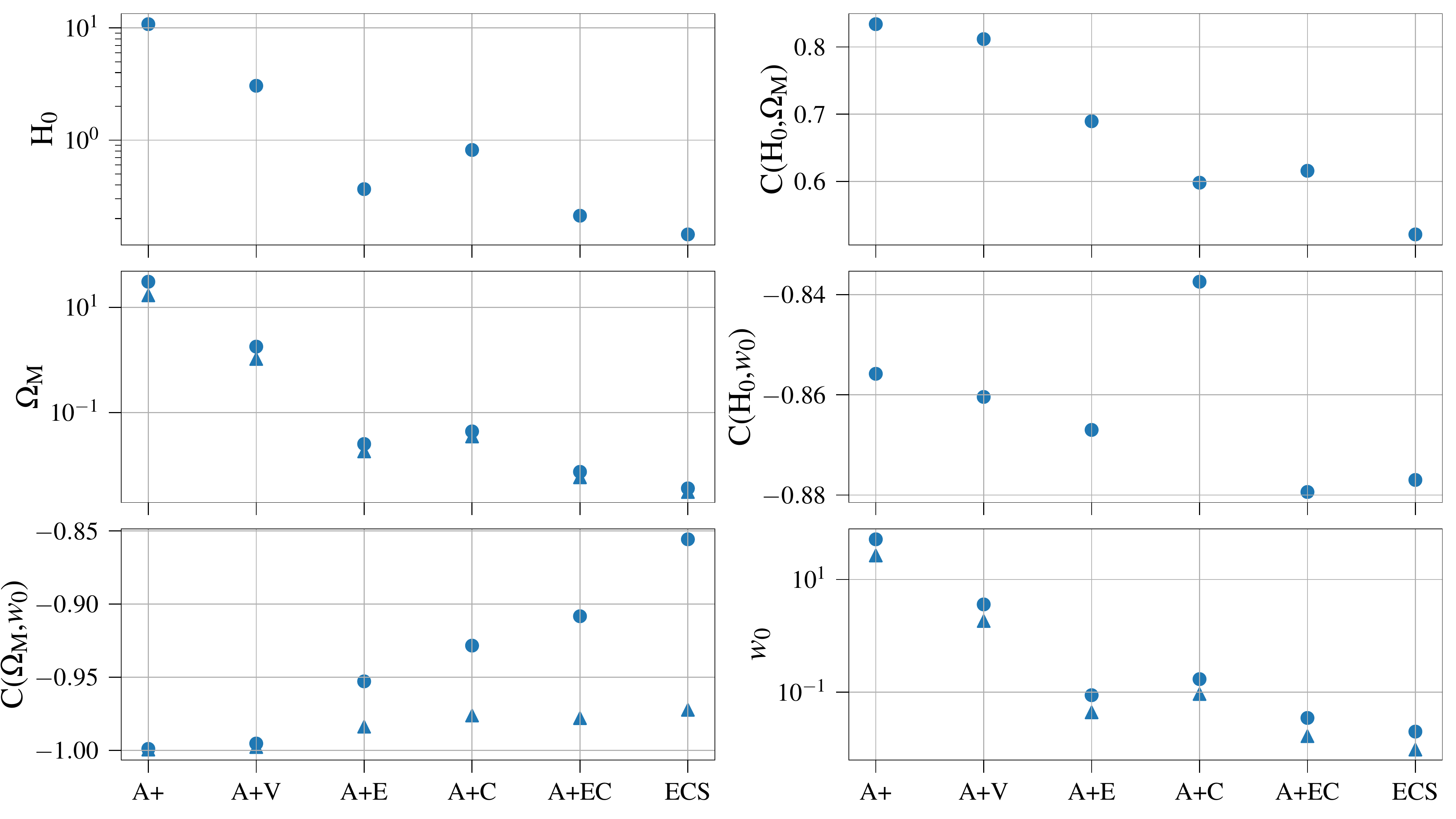}
    \caption{1-$\sigma$ errors on and the correlation coefficients between the parameters of the $w$CDM cosmological model for the various networks calculated using gravitational-wave observations alone. The triangles show the respective quantities if $H_0$ is assumed to be known.}
    \label{fig:wcdm_gw}
\end{figure*}
The $w$CDM cosmological model has an additional parameter $w_0$ for a non-trivial evolution of dark energy. Fig.~\ref{fig:wcdm_gw} shows the correlations between and the error estimates on various cosmological parameters. The plots are the same as in Fig.~\ref{fig:lcdm_gw} other than the triangular markers representing the correlation between parameters other than $H_0$. These are the correlations between the parameters when $H_0$ is known. The 1-$\sigma$ errors are listed in Tab.~\ref{tab:gw_cosmology}.

The difference between the estimates of $\Lambda$CDM and $w$CDM models is due to an enlarged parameter space and the correlations between those parameters in $w$CDM model. Of specific note is that the near degeneracy between $\Omega_M$ and $w_0$ for the second-generation detectors make the estimates in a joint parameter estimation meaningless. This degeneracy partially lifts for most next-generation networks, allowing for a meaningful measurement of the parameters. The best network, ECS, can achieve a measurement error of $\sim0.2\%$ on $H_0$, 1.2\% on $\Omega_M$, and $\sim0.02$ on $w_0$. On assuming that $H_0$ is known, these bounds improve to 1.0\% for $\Omega_M$ and 0.0096 for $w_0$. These bounds are better than those obtained using the electromagnetic counterpart method and also the current Planck~\cite{Planck:2018vyg} measurements.

\subsubsection{$w_0w_a$CDM}
\label{subsubsec:w0wacdm}
\begin{figure*}[ht]
    \centering
    \includegraphics[width=2\columnwidth]{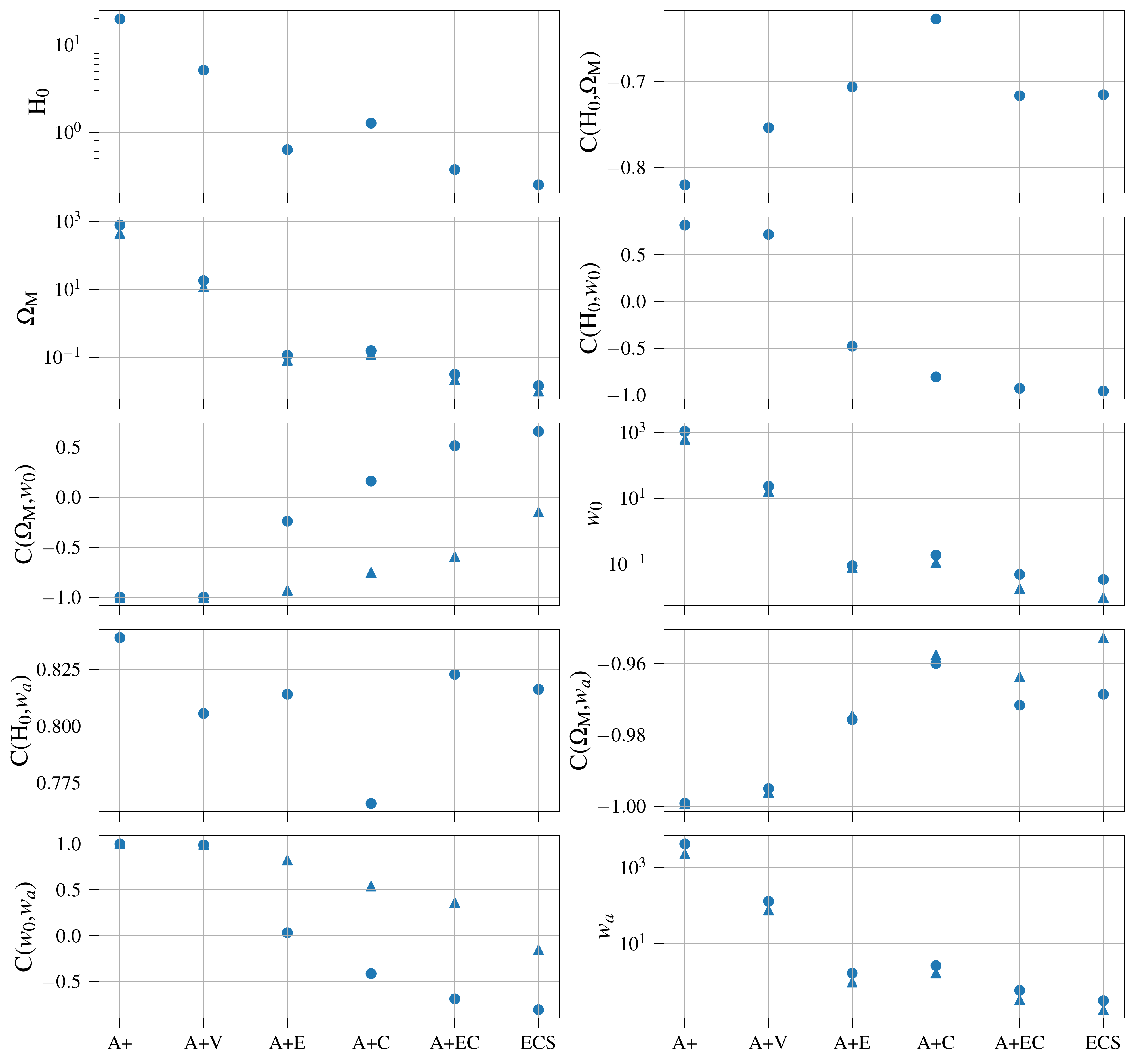}
    \caption{1-$\sigma$ errors on and the correlation coefficients between the parameters of the $w_0w_a$CDM cosmological model for the various networks calculated using gravitational-wave observations alone. The triangles show the respective quantities if $H_0$ is assumed to be known.}
    \label{fig:w0wacdm_gw}
\end{figure*}
The $w_0w_a$CDM model of cosmology allows for a redshift dependent EoS for dark energy with the redshift dependence controlled by the parameter $w_a$. The correlations between and the errors on the model parameters are plotted in Fig.~\ref{fig:w0wacdm_gw} and tabulated in Tab.~\ref{tab:gw_cosmology}.

It can be seen that the new parameter $w_a$ is strongly correlated with both $w_0$ and $\Omega_M$ for the current-generation of networks with almost perfect degeneracy. The situation is different for next-generation networks where the degeneracy of $w_a$ with $\Omega_M$ barely breaks while that with $w_0$ is significantly better. As a result, in a joint parameter estimation, $w_0$ and $w_a$ can be measured to an accuracy of $\sim0.03$ and $\sim0.3$, respectively. If $H_0$ is given, some of the degeneracies break and the two parameters can be measured to an accuracy of $\sim0.01$ and $\sim0.2$. It is noted that a counterpart-less method constrains the dark energy parameters better by factors of a few compared to the electromagnetic counterpart method.

\subsubsection{$\Xi_0$--$w_0w_a$CDM}
\label{subsubsec:xi-w0wacdm}
\begin{figure*}
    \centering
    \includegraphics[width=2\columnwidth]{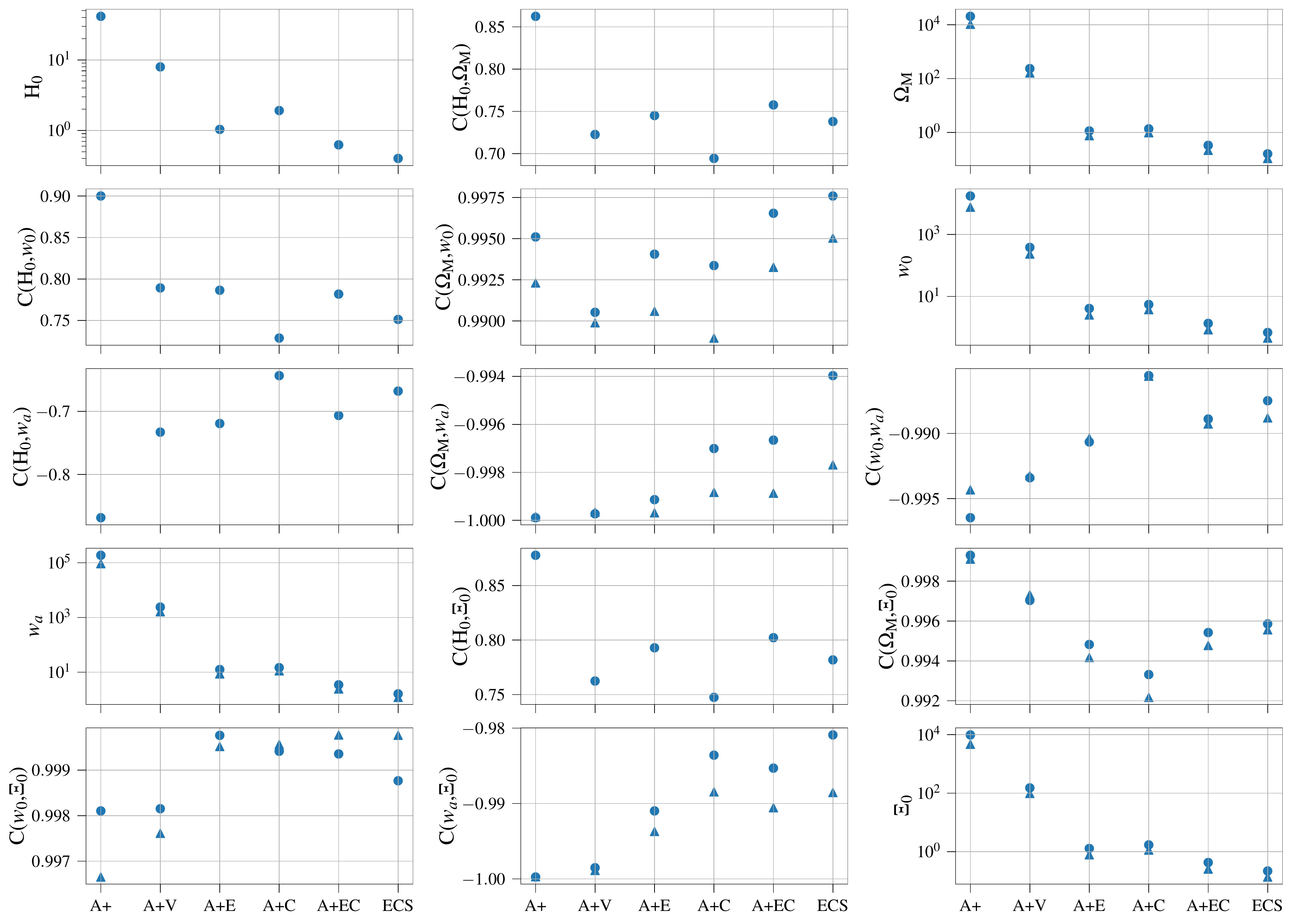}
    \caption{1-$\sigma$ errors on and the correlation coefficients between the parameters of the $\Xi_0$--$w_0w_a$CDM cosmological model for the various networks calculated using gravitational-wave observations alone. The triangles show the respective quantities if $H_0$ is assumed to be known.}
    \label{fig:xi-w0wacdm_gw}
\end{figure*}
Cosmological models that modify the tensor perturbations can have a different damping of the gravitational wave amplitude from the $\Lambda$CDM model. This is modeled by the parameter $\Xi_0$ which is an additional parameter that we include in this model. A fiducial value of $n=2.5$ is also assumed as in the electromagnetic counterpart method. The correlation coefficients between and the errors in the model parameters are shown in Fig.~\ref{fig:xi-w0wacdm_gw} and the values are quoted in Tab.~\ref{tab:xi-w0wacdm_gw}. We quote results only for those networks that have at least one next-generation detector in the network because the current networks do not give any meaningful bounds.

We note that $\Xi_0$ is extremely well correlated with all the parameters except $H_0$. This is why the estimates of all those parameters increase by an order of magnitude or more compared to the $w_0w_a$CDM model. Despite this, we see that $\Xi_0$ can be measured at at $\sim$20\% accuracy, which can rule out models such as that of~\textcite{Belgacem:2019lwx} that predict a large deviation in the gravitational-wave propagation relative to GR. It is again noted that this constraint is better than the counterpart method by a factor of 5.

\begin{table*}[h]
    \centering
    \begin{tabular}{cV{3}c|cV{3}c|c|cV{3}*{3}{c|}cV{3}}
        \toprule
        \toprule
        \multirow{2}{*}{\diagbox{Network}{$\sigma(\vec{\phi})$}} & \multicolumn{2}{cV{3}}{$\Lambda$CDM} & \multicolumn{3}{cV{3}}{$w$CDM} & \multicolumn{4}{cV{3}}{$w_0w_a$CDM} \\
        \cmidrule(lr){2-3} \cmidrule(lr){4-6} \cmidrule(lr){7-10}
        & $\sigma(H_0)$ (\%) & $\sigma(\Omega_M)$ (\%) & $\sigma(H_0)$ (\%) & $\sigma(\Omega_M)$ (\%) & $\sigma(w_0)$ & $\sigma(H_0)$ (\%) & $\sigma(\Omega_M)$ (\%) & $\sigma(w_0)$ & $\sigma(w_a)$ \\
        \midrule
        \multirow{2}{*}{A+} & 8.3 & 440 & 16 & 9900 & 52 & 29 & 250000 & 1100 & 4200 \\
        & - & 180 & - & 5500 & 27 & - & 140000 & 620 & 2300  \\
        \midrule
        \multirow{2}{*}{A+V} & 2.3 & 56 & 4.5 & 580 & 3.6 & 7.6 & 5800 & 23 & 130 \\
        & - & 23 & - & 340 & 1.8 & - & 3800 & 16 & 77 \\
        \midrule
        \multirow{2}{*}{A+E} & 0.27 & 2.5 & 0.54 & 8.2 & 0.089 & 0.93 & 37 & 0.089 & 1.7 \\
        & - & 1.0 & - & 5.9 & 0.044 & - & 26 & 0.078 & 0.97 \\
        \midrule
        \multirow{2}{*}{A+C} & 0.66 & 5.3 & 1.2 & 14 & 0.17 & 1.9 & 51 & 0.19 & 2.6 \\
        & - & 2.5 & - & 11 & 0.093 & - & 39 & 0.11 & 1.7 \\
        \midrule
        \multirow{2}{*}{A+EC} & 0.15 & 1.0 & 0.31 & 2.4 & 0.035 & 0.55 & 10 & 0.048 & 0.59 \\
        & - & 0.40 & - & 1.9 & 0.017 & - & 7.1 & 0.018 & 0.34 \\
        \midrule
        \multirow{2}{*}{ECS} & 0.10 & 0.61 & 0.21 & 1.2 & 0.020 & 0.37 & 4.7 & 0.034 & 0.31 \\
        & - & 0.23 & - & 1.0 & 0.0096 & - & 3.3 & 0.0098 & 0.18 \\
        \bottomrule
        \bottomrule
    \end{tabular}
    \caption{1-$\sigma$ errors on the parameters of the various types of cosmologies with differing dark energy models for different detector networks calculated using gravitational-wave observations alone. If a parameter in a model is assumed to be known a priori, the corresponding cell has been marked with `-'.}
    \label{tab:gw_cosmology}
\end{table*}

\begin{table*}[ht]
    \centering
    \begin{tabular}{cV{3}*{4}{c|}cV{3}}
        \toprule
        \toprule
        \multirow{2}{*}{\diagbox{Network}{$\sigma(\vec{\phi})$}} & \multicolumn{5}{cV{3}}{$\Xi_0-w_0w_a$CDM} \\
        \cmidrule(lr){2-6}
        & $\sigma(H_0)$ (\%) & $\sigma(\Omega_M)$ (\%) & $\sigma(w_0)$ & $\sigma(w_a)$ & $\sigma(\Xi_0)$ (\%) \\
        \midrule
        \multirow{2}{*}{A+E} & 1.5 & 370 & 4.1 & 12 & 130 \\
        & - & 240 & 2.5 & 8.7 & 78 \\
        \midrule
        \multirow{2}{*}{A+C} & 2.8 & 440 & 5.5 & 15 & 170 \\
        & - & 320 & 3.8 & 11 & 110 \\
        \midrule
        \multirow{2}{*}{A+EC} &0.92 & 110 & 1.3 & 3.4 & 42 \\
        & - & 70 & 0.84 & 2.4 & 25 \\
        \midrule
        \multirow{2}{*}{ECS} & 0.59 & 52 & 0.68 & 1.6 & 22 \\
        & - & 35 & 0.45 & 1.2 & 14 \\
        \bottomrule
        \bottomrule
    \end{tabular}
    \caption{1-$\sigma$ errors on the parameters of the $\Xi_0$--$w_0w_a$CDM cosmology for the various detector networks having at least one third-generation detector calculated using gravitational-wave observations alone. This model has a time-varying dark energy EoS and a modified perturbation of the tensor modes. If a parameter in a model is assumed to be known a priori, the corresponding cell has been marked with `-'. We do not quote the results for the other detector networks because they do not provide any meaningful constraints.}
    \label{tab:xi-w0wacdm_gw}
\end{table*}

\section{Conclusion and Discussion}
\label{sec:conclusion}
In this study, we simulate an astrophysical population of equal-mass BNS mergers to evaluate their potential for cosmological inference. We analyse both the electromagnetic counterpart method, where a kilonova or a GRB detection is used to determine the redshift, and the counterpart-less method, where tidal interactions between NSs are used to bring in an additional mass scale to break the mass-redshift degeneracy present in gravitational waveforms. Separate analytical gravitational waveform models are used for the inspiral and post-merger dynamics with the expectation that the post-merger signal would improve the redshift measurement. This is justified because where the mutual tidal interactions break the degeneracy in the inspiral signal, it is the self interactions of the merged NS that provides the mass scale in the post-merger. In the electromagnetic counterpart method, we assume that 10\% of the BNS mergers observed within a redshift of 0.5 will be followed-up electromagnetically while GRB detection is possible for a fifth of the observed binaries due to the large sky coverage of GRB observatories, though only a small fraction is detected due to the viewing angle dependence of the GRB flux. 

We consider a succession of detector networks of increasing sensitivity in this work, from the planned upgrades to the current set of detectors to proposed future observatories, with the goal of understanding the measurement capabilities of different GW networks. Specifically, for the proposed US-based initiatives, we compare the PMO and CBO configurations to assess whether an improved high frequency sensitivity can provide a more accurate redshift measurement with the post-merger signal to compensate for a poorer low frequency sensitivity and, consequently, fewer number of detections.

We draw the following broad conclusions.
\begin{itemize}
    \item The redshift determination from the post-merger signal is ineffective compared to the inspiral signal for the population as a whole though for very nearby events the post-merger signal can have a significant contribution.
    \item The total luminosity distance errors are governed by the errors on the gravitational-wave amplitude and those coming from the redshift determination play a part for only the nearby population.
    \item The luminosity distance errors for the population with a GRB counterpart are smaller by more than an order of magnitude, sometimes two orders, if the inclination angle of the binary can be modeled from the peak luminosity of the GRB. In the absence of such a model, the population becomes uninformative.
    \item In the electromagnetic counterpart method, the GRB counterpart population provides far superior estimates of the cosmological parameters compared to the kilonova population, despite being 20 times fewer in counts. This is if the inclination angle $\iota$ can be measured from the angular modeling of GRB peak luminosity, which breaks the $D_L$--$\iota$ degeneracy, enabling the refinement of distance measurement by two orders of magnitude.
    \item The electromagnetic counterpart method gives a better determination of the Hubble constant because of a finer redshift measurement in the local Universe compared to counterpart-less estimates.
    \item The counterpart-less inference achieves better measurement accuracy in the presence of at least one next-generation detector, for all other cosmological parameters, because a greater number of sources from larger redshifts contribute to the measurement. It is at these redshifts that other cosmological parameters become significant.
    \item For both the methods, the CBO detector configuration performs better than the PMO configuration. This is because in the electromagnetic counterpart method the former observes a greater number of sources. In the counterpart-less method, a better measurement of redshift from the inspiral signal also makes CBO superior.
\end{itemize}

There are some limitations to our work. Perhaps, the most important limitation is that we model the inspiral and post-merger phases separately as there are currently no analytic \emph{inspiral-merger-ringdown} waveform models that include BNS post-merger physics. A phase coherent model across the complete signal can be more informative and should improve redshift measurement. We also point out possible issues in parameter estimation accuracy due to the abrupt termination of the waveforms as discussed in~\textcite{Mandel:2014tca} though they found its effect to be important for heavier binaries.

Next, the NS population that we consider is restricted to equal mass companions in the range 1.25$\rm M_{\odot}$ and 1.55$\rm M_{\odot}$ by the post-merger waveform model we employ. Most EoS of NSs support higher maximum masses and if the Universe contains higher mass NSs, then we would have higher rates and greater signal strengths, leading to better measurements. We also note that the mass ratio, which appears in tandem with the tidal deformability parameter, is not considered in the Fisher parameter space. We do this because of the technical challenges of handling the mass ratio derivatives of \emph{exactly} equal mass systems in the Fisher package \emph{\sc gwbench}. On the surface, this would appear to be a noteworthy limitation as the mass ratio and the tidal deformability terms are both present at the 5PN order and, therefore, would be correlated. However, we note that the mass ratio first appears in early inspiral and, since these are very long-duration signals, its correlation with the tidal deformability is not expected to be significant.

For the counterpart-less method, the EoS has to be known a priori to break the mass-redshift degeneracy. Here, the implicit assumption in our study is that a fraction of nearby events can determine the EoS and cosmological parameters can be inferred with the remaining population. This would primarily affect the Hubble constant estimate but, as we have established in the preceding sections, the Hubble constant is best measured with counterparts which do not require a knowledge of the EoS. It should, nevertheless, be pointed out that the determination of the nuclear EoS from nearby events would be limited by the measurement of the peculiar velocity. An error in its measurement would manifest itself, in our case, as an additional source of error in the redshift estimation via an error in the EoS determination. However, this should not be a serious impediment since, for faraway sources the limitation is due to the measurement of the gravitational-wave amplitude and not the redshift. Furthermore, we have shown that the counterpart method achieves better accuracy for nearby sources. Accordingly, the two methods compliment each other. The loud nearby sources can then be used to constrain the nuclear EoS, in conjunction with the Hubble constant~\cite{Ghosh:2022muc}.

In this paper, the tidal contribution is taken to be purely adiabatic. It would be interesting to estimate the bias in the redshift measurement due to dynamical tides which are known to bias the measurement of the tidal deformability parameter~\cite{Pratten:2021pro}. In the future, we also plan to extend the analysis to a Bayesian framework, folding in the uncertainty on the nuclear EoS, and get more realistic constraints on cosmological parameters.

\acknowledgements
We thank Salvatore Vitale and Philippe Landry for useful discussions. AD is supported by the NSF grant PHY-2012083, SB is supported by the Deutsche Forschungsgemeinschaft, DFG, project MEMI number BE 6301/2-1, AG is  supported by the NSF grant AST-2205920, and BSS is supported in part by NSF grants PHY-1836779, PHY-2012083, and AST-2006384.

\appendix
\section{Analytic post-merger model}
\label{sec:pm_model}
\begin{figure}[h]
    \centering
    \includegraphics[width=\columnwidth]{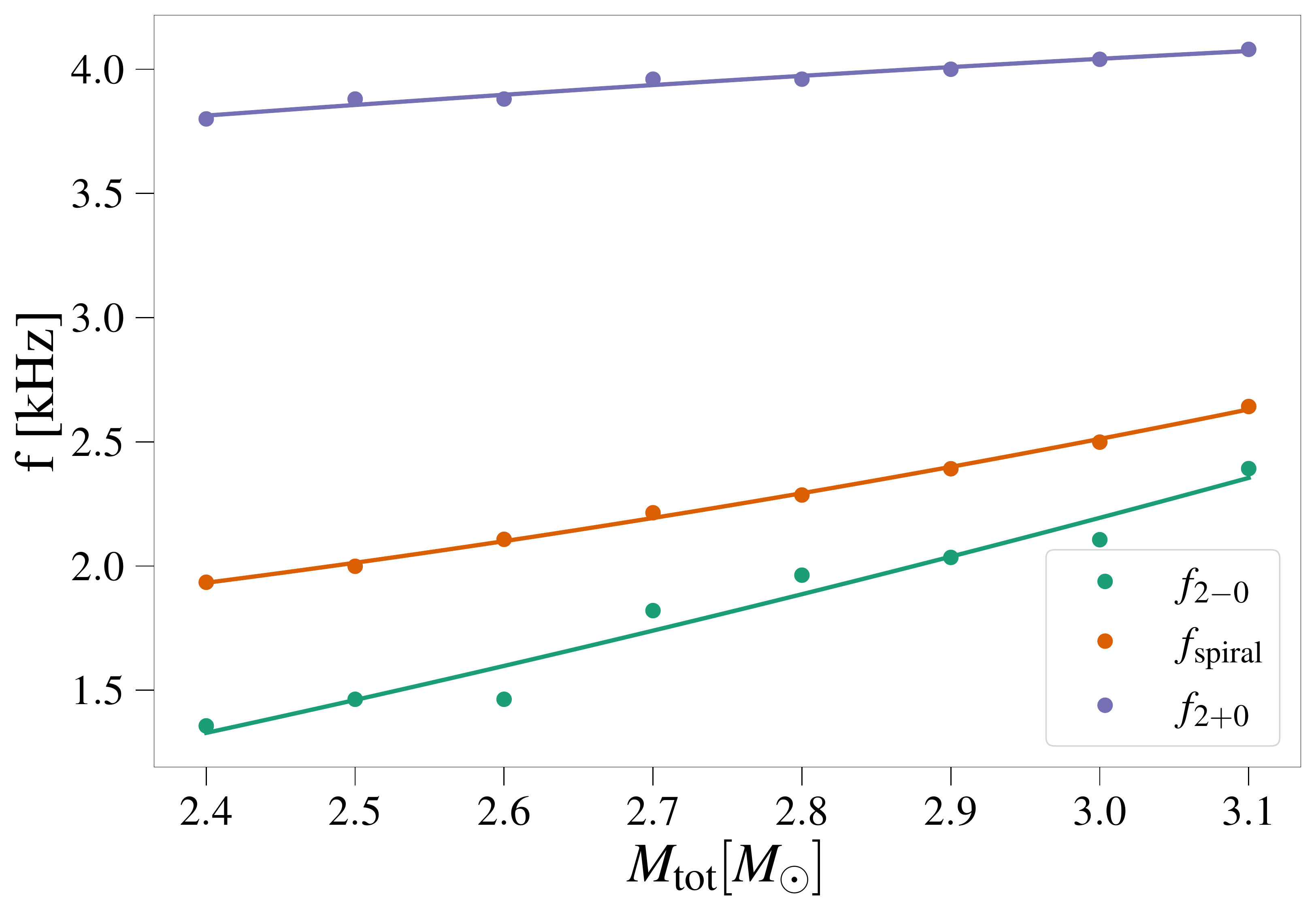}
    \caption{The frequencies of the three non-dominant modes used in~\textcite{Soultanis:2021oia} to model the post-merger signal of a BNS merger as a function of the total mass of the system. Also shown are the quadratic fits to the data whose expressions can be found in Eq.~(\ref{eq:pm_fits}).}
    \label{fig:pm_fits}
\end{figure}

In this appendix, we briefly describe the \textcite{Soultanis:2021oia} model for the post-merger signal of a BNS merger that is used in this study. A set of BNS simulations with equal component masses, no component spins, and MPA1 EoS~\cite{Muther:1987xaa} is taken to construct the post-merger model. Their model consists of a set of four quasi-normal modes that describe various features of the signal. The `plus' polarization of the gravitational waveform takes the form
\begin{equation}
\label{eq:soultanis}
    \begin{split}
        h_{+}(t) &= A_{\rm peak} e^{-t/\tau_{\rm peak}} \sin{(2\pi f_{\rm peak}t + \phi_{\rm peak})} \\
        &+ A_{\rm spiral} e^{-t/\tau_{\rm spiral}} \sin{(2\pi f_{\rm spiral}t + \phi_{\rm spiral})} \\
        &+ A_{\rm 2-0} e^{-t/\tau_{\rm 2-0}} \sin{(2\pi f_{\rm 2-0}t + \phi_{\rm 2-0})} \\
        &+ A_{\rm 2+0} e^{-t/\tau_{\rm 2+0}} \sin{(2\pi f_{\rm 2+0}t + \phi_{\rm 2+0})},
    \end{split}
\end{equation}
with the `cross' polarization obtained by adding a phase of $\pi/2$ to each mode. The various model parameters are as follows. $f_{\rm peak}$ is the dominant feature in the post-merger signal and is attributed to the fundamental quadrupolar oscillations of the merged remnant. A non-linear feature coupling the quadrupolar oscillations to the quasi-radial oscillations $f_0$ is also identifiable. The spectral features corresponding to these are present at frequencies $f_{2\pm0}\approx f_{\rm peak}\pm f_0$. The final spectral feature modeled is $f_{\rm spiral}$ which is attributed to the orbital motion of the tidal antipodal bulges~\cite{Bauswein:2015yca}. We refer the interested reader to their paper for further details. Here, we present fits for the three frequencies not directly reported in their paper. These are the frequencies $f_{\rm spiral}$, $f_{\rm 2-0}$, and $f_{\rm 2+0}$ and their fits are given by
\begin{align}
\label{eq:pm_fits}
    f_{\rm spiral} &= 0.319 M_{\rm tot}^2 - 0.758 M_{\rm tot} + 1.914, \\
    f_{\rm 2-0} &= 0.236 M_{\rm tot}^2 + 0.167 M_{\rm tot} - 0.434, \\
    f_{\rm 2+0} &= -0.095 M_{\rm tot}^2 + 0.895 M_{\rm tot} + 2.213,
\end{align}
where the total mass of the binary $M_{\rm tot}$ is in terms of solar mass and the frequencies have units of kilohertz. For illustrative purposes, we depict the values extracted from the numerical simulations and the fits in Fig.~\ref{fig:pm_fits}.

\bibliographystyle{apsrev4-1}
\bibliography{dark_energy_3g}{}

\begin{thebibliography}{101}%
\makeatletter
\providecommand \@ifxundefined [1]{%
 \@ifx{#1\undefined}
}%
\providecommand \@ifnum [1]{%
 \ifnum #1\expandafter \@firstoftwo
 \else \expandafter \@secondoftwo
 \fi
}%
\providecommand \@ifx [1]{%
 \ifx #1\expandafter \@firstoftwo
 \else \expandafter \@secondoftwo
 \fi
}%
\providecommand \natexlab [1]{#1}%
\providecommand \enquote  [1]{``#1''}%
\providecommand \bibnamefont  [1]{#1}%
\providecommand \bibfnamefont [1]{#1}%
\providecommand \citenamefont [1]{#1}%
\providecommand \href@noop [0]{\@secondoftwo}%
\providecommand \href [0]{\begingroup \@sanitize@url \@href}%
\providecommand \@href[1]{\@@startlink{#1}\@@href}%
\providecommand \@@href[1]{\endgroup#1\@@endlink}%
\providecommand \@sanitize@url [0]{\catcode `\\12\catcode `\$12\catcode
  `\&12\catcode `\#12\catcode `\^12\catcode `\_12\catcode `\%12\relax}%
\providecommand \@@startlink[1]{}%
\providecommand \@@endlink[0]{}%
\providecommand \url  [0]{\begingroup\@sanitize@url \@url }%
\providecommand \@url [1]{\endgroup\@href {#1}{\urlprefix }}%
\providecommand \urlprefix  [0]{URL }%
\providecommand \Eprint [0]{\href }%
\providecommand \doibase [0]{http://dx.doi.org/}%
\providecommand \selectlanguage [0]{\@gobble}%
\providecommand \bibinfo  [0]{\@secondoftwo}%
\providecommand \bibfield  [0]{\@secondoftwo}%
\providecommand \translation [1]{[#1]}%
\providecommand \BibitemOpen [0]{}%
\providecommand \bibitemStop [0]{}%
\providecommand \bibitemNoStop [0]{.\EOS\space}%
\providecommand \EOS [0]{\spacefactor3000\relax}%
\providecommand \BibitemShut  [1]{\csname bibitem#1\endcsname}%
\let\auto@bib@innerbib\@empty
\bibitem [{\citenamefont {Aasi}\ \emph
  {et~al.}(2015{\natexlab{a}})\citenamefont {Aasi} \emph
  {et~al.}}]{LIGOScientific:2014pky}%
  \BibitemOpen
  \bibfield  {author} {\bibinfo {author} {\bibfnamefont {J.}~\bibnamefont
  {Aasi}} \emph {et~al.} (\bibinfo {collaboration} {LIGO Scientific}),\ }\href
  {\doibase 10.1088/0264-9381/32/7/074001} {\bibfield  {journal} {\bibinfo
  {journal} {Class. Quant. Grav.}\ }\textbf {\bibinfo {volume} {32}},\ \bibinfo
  {pages} {074001} (\bibinfo {year} {2015}{\natexlab{a}})},\ \Eprint
  {http://arxiv.org/abs/1411.4547} {arXiv:1411.4547 [gr-qc]} \BibitemShut
  {NoStop}%
\bibitem [{\citenamefont {Acernese}\ \emph {et~al.}(2015)\citenamefont
  {Acernese} \emph {et~al.}}]{VIRGO:2014yos}%
  \BibitemOpen
  \bibfield  {author} {\bibinfo {author} {\bibfnamefont {F.}~\bibnamefont
  {Acernese}} \emph {et~al.} (\bibinfo {collaboration} {VIRGO}),\ }\href
  {\doibase 10.1088/0264-9381/32/2/024001} {\bibfield  {journal} {\bibinfo
  {journal} {Class. Quant. Grav.}\ }\textbf {\bibinfo {volume} {32}},\ \bibinfo
  {pages} {024001} (\bibinfo {year} {2015})},\ \Eprint
  {http://arxiv.org/abs/1408.3978} {arXiv:1408.3978 [gr-qc]} \BibitemShut
  {NoStop}%
\bibitem [{\citenamefont {Akutsu}\ \emph {et~al.}(2019)\citenamefont {Akutsu}
  \emph {et~al.}}]{KAGRA:2018plz}%
  \BibitemOpen
  \bibfield  {author} {\bibinfo {author} {\bibfnamefont {T.}~\bibnamefont
  {Akutsu}} \emph {et~al.} (\bibinfo {collaboration} {KAGRA}),\ }\href
  {\doibase 10.1038/s41550-018-0658-y} {\bibfield  {journal} {\bibinfo
  {journal} {Nature Astron.}\ }\textbf {\bibinfo {volume} {3}},\ \bibinfo
  {pages} {35} (\bibinfo {year} {2019})},\ \Eprint
  {http://arxiv.org/abs/1811.08079} {arXiv:1811.08079 [gr-qc]} \BibitemShut
  {NoStop}%
\bibitem [{\citenamefont {Abbott}\ \emph
  {et~al.}(2021{\natexlab{a}})\citenamefont {Abbott} \emph
  {et~al.}}]{LIGOScientific:2021djp}%
  \BibitemOpen
  \bibfield  {author} {\bibinfo {author} {\bibfnamefont {R.}~\bibnamefont
  {Abbott}} \emph {et~al.} (\bibinfo {collaboration} {LIGO Scientific, VIRGO,
  KAGRA}),\ }\href@noop {} {\  (\bibinfo {year} {2021}{\natexlab{a}})},\
  \Eprint {http://arxiv.org/abs/2111.03606} {arXiv:2111.03606 [gr-qc]}
  \BibitemShut {NoStop}%
\bibitem [{\citenamefont {Hall}\ and\ \citenamefont
  {Evans}(2019)}]{Hall:2019xmm}%
  \BibitemOpen
  \bibfield  {author} {\bibinfo {author} {\bibfnamefont {E.~D.}\ \bibnamefont
  {Hall}}\ and\ \bibinfo {author} {\bibfnamefont {M.}~\bibnamefont {Evans}},\
  }\href {\doibase 10.1088/1361-6382/ab41d6} {\bibfield  {journal} {\bibinfo
  {journal} {Class. Quant. Grav.}\ }\textbf {\bibinfo {volume} {36}},\ \bibinfo
  {pages} {225002} (\bibinfo {year} {2019})},\ \Eprint
  {http://arxiv.org/abs/1902.09485} {arXiv:1902.09485 [astro-ph.IM]}
  \BibitemShut {NoStop}%
\bibitem [{\citenamefont {Schutz}(1986)}]{Schutz:1986gp}%
  \BibitemOpen
  \bibfield  {author} {\bibinfo {author} {\bibfnamefont {B.~F.}\ \bibnamefont
  {Schutz}},\ }\href {\doibase 10.1038/323310a0} {\bibfield  {journal}
  {\bibinfo  {journal} {Nature}\ }\textbf {\bibinfo {volume} {323}},\ \bibinfo
  {pages} {310} (\bibinfo {year} {1986})}\BibitemShut {NoStop}%
\bibitem [{\citenamefont {Holz}\ and\ \citenamefont
  {Hughes}(2005)}]{Holz:2005df}%
  \BibitemOpen
  \bibfield  {author} {\bibinfo {author} {\bibfnamefont {D.~E.}\ \bibnamefont
  {Holz}}\ and\ \bibinfo {author} {\bibfnamefont {S.~A.}\ \bibnamefont
  {Hughes}},\ }\href {\doibase 10.1086/431341} {\bibfield  {journal} {\bibinfo
  {journal} {Astrophys. J.}\ }\textbf {\bibinfo {volume} {629}},\ \bibinfo
  {pages} {15} (\bibinfo {year} {2005})},\ \Eprint
  {http://arxiv.org/abs/astro-ph/0504616} {arXiv:astro-ph/0504616} \BibitemShut
  {NoStop}%
\bibitem [{\citenamefont {Riess}\ \emph {et~al.}(2021)\citenamefont {Riess}
  \emph {et~al.}}]{Riess:2021jrx}%
  \BibitemOpen
  \bibfield  {author} {\bibinfo {author} {\bibfnamefont {A.~G.}\ \bibnamefont
  {Riess}} \emph {et~al.},\ }\href@noop {} {\  (\bibinfo {year} {2021})},\
  \Eprint {http://arxiv.org/abs/2112.04510} {arXiv:2112.04510 [astro-ph.CO]}
  \BibitemShut {NoStop}%
\bibitem [{\citenamefont {Sandage}\ \emph {et~al.}(1996)\citenamefont
  {Sandage}, \citenamefont {Saha}, \citenamefont {Tammann}, \citenamefont
  {Labhardt}, \citenamefont {Panagia},\ and\ \citenamefont
  {Macchetto}}]{Sandage1996CepheidCO}%
  \BibitemOpen
  \bibfield  {author} {\bibinfo {author} {\bibfnamefont {A.~R.}\ \bibnamefont
  {Sandage}}, \bibinfo {author} {\bibfnamefont {A.}~\bibnamefont {Saha}},
  \bibinfo {author} {\bibfnamefont {G.~A.}\ \bibnamefont {Tammann}}, \bibinfo
  {author} {\bibfnamefont {L.}~\bibnamefont {Labhardt}}, \bibinfo {author}
  {\bibfnamefont {N.}~\bibnamefont {Panagia}}, \ and\ \bibinfo {author}
  {\bibfnamefont {F.~D.}\ \bibnamefont {Macchetto}},\ }\href@noop {} {\bibfield
   {journal} {\bibinfo  {journal} {The Astrophysical Journal Letters}\ }\textbf
  {\bibinfo {volume} {460}},\ \bibinfo {pages} {L15 } (\bibinfo {year}
  {1996})}\BibitemShut {NoStop}%
\bibitem [{\citenamefont {Aghanim}\ \emph {et~al.}(2020)\citenamefont {Aghanim}
  \emph {et~al.}}]{Planck:2018vyg}%
  \BibitemOpen
  \bibfield  {author} {\bibinfo {author} {\bibfnamefont {N.}~\bibnamefont
  {Aghanim}} \emph {et~al.} (\bibinfo {collaboration} {Planck}),\ }\href
  {\doibase 10.1051/0004-6361/201833910} {\bibfield  {journal} {\bibinfo
  {journal} {Astron. Astrophys.}\ }\textbf {\bibinfo {volume} {641}},\ \bibinfo
  {pages} {A6} (\bibinfo {year} {2020})},\ \bibinfo {note} {[Erratum:
  Astron.Astrophys. 652, C4 (2021)]},\ \Eprint
  {http://arxiv.org/abs/1807.06209} {arXiv:1807.06209 [astro-ph.CO]}
  \BibitemShut {NoStop}%
\bibitem [{\citenamefont {Di~Valentino}\ \emph {et~al.}(2021)\citenamefont
  {Di~Valentino}, \citenamefont {Mena}, \citenamefont {Pan}, \citenamefont
  {Visinelli}, \citenamefont {Yang}, \citenamefont {Melchiorri}, \citenamefont
  {Mota}, \citenamefont {Riess},\ and\ \citenamefont
  {Silk}}]{DiValentino:2021izs}%
  \BibitemOpen
  \bibfield  {author} {\bibinfo {author} {\bibfnamefont {E.}~\bibnamefont
  {Di~Valentino}}, \bibinfo {author} {\bibfnamefont {O.}~\bibnamefont {Mena}},
  \bibinfo {author} {\bibfnamefont {S.}~\bibnamefont {Pan}}, \bibinfo {author}
  {\bibfnamefont {L.}~\bibnamefont {Visinelli}}, \bibinfo {author}
  {\bibfnamefont {W.}~\bibnamefont {Yang}}, \bibinfo {author} {\bibfnamefont
  {A.}~\bibnamefont {Melchiorri}}, \bibinfo {author} {\bibfnamefont {D.~F.}\
  \bibnamefont {Mota}}, \bibinfo {author} {\bibfnamefont {A.~G.}\ \bibnamefont
  {Riess}}, \ and\ \bibinfo {author} {\bibfnamefont {J.}~\bibnamefont {Silk}},\
  }\href {\doibase 10.1088/1361-6382/ac086d} {\bibfield  {journal} {\bibinfo
  {journal} {Class. Quant. Grav.}\ }\textbf {\bibinfo {volume} {38}},\ \bibinfo
  {pages} {153001} (\bibinfo {year} {2021})},\ \Eprint
  {http://arxiv.org/abs/2103.01183} {arXiv:2103.01183 [astro-ph.CO]}
  \BibitemShut {NoStop}%
\bibitem [{\citenamefont {Abbott}\ \emph
  {et~al.}(2017{\natexlab{a}})\citenamefont {Abbott} \emph
  {et~al.}}]{LIGOScientific:2017vwq}%
  \BibitemOpen
  \bibfield  {author} {\bibinfo {author} {\bibfnamefont {B.~P.}\ \bibnamefont
  {Abbott}} \emph {et~al.} (\bibinfo {collaboration} {LIGO Scientific,
  Virgo}),\ }\href {\doibase 10.1103/PhysRevLett.119.161101} {\bibfield
  {journal} {\bibinfo  {journal} {Phys. Rev. Lett.}\ }\textbf {\bibinfo
  {volume} {119}},\ \bibinfo {pages} {161101} (\bibinfo {year}
  {2017}{\natexlab{a}})},\ \Eprint {http://arxiv.org/abs/1710.05832}
  {arXiv:1710.05832 [gr-qc]} \BibitemShut {NoStop}%
\bibitem [{\citenamefont {Abbott}\ \emph
  {et~al.}(2017{\natexlab{b}})\citenamefont {Abbott} \emph
  {et~al.}}]{LIGOScientific:2017ync}%
  \BibitemOpen
  \bibfield  {author} {\bibinfo {author} {\bibfnamefont {B.~P.}\ \bibnamefont
  {Abbott}} \emph {et~al.} (\bibinfo {collaboration} {LIGO Scientific, Virgo,
  Fermi GBM, INTEGRAL, IceCube, AstroSat Cadmium Zinc Telluride Imager Team,
  IPN, Insight-Hxmt, ANTARES, Swift, AGILE Team, 1M2H Team, Dark Energy Camera
  GW-EM, DES, DLT40, GRAWITA, Fermi-LAT, ATCA, ASKAP, Las Cumbres Observatory
  Group, OzGrav, DWF (Deeper Wider Faster Program), AST3, CAASTRO, VINROUGE,
  MASTER, J-GEM, GROWTH, JAGWAR, CaltechNRAO, TTU-NRAO, NuSTAR, Pan-STARRS,
  MAXI Team, TZAC Consortium, KU, Nordic Optical Telescope, ePESSTO, GROND,
  Texas Tech University, SALT Group, TOROS, BOOTES, MWA, CALET, IKI-GW
  Follow-up, H.E.S.S., LOFAR, LWA, HAWC, Pierre Auger, ALMA, Euro VLBI Team, Pi
  of Sky, Chandra Team at McGill University, DFN, ATLAS Telescopes, High Time
  Resolution Universe Survey, RIMAS, RATIR, SKA South Africa/MeerKAT}),\ }\href
  {\doibase 10.3847/2041-8213/aa91c9} {\bibfield  {journal} {\bibinfo
  {journal} {Astrophys. J. Lett.}\ }\textbf {\bibinfo {volume} {848}},\
  \bibinfo {pages} {L12} (\bibinfo {year} {2017}{\natexlab{b}})},\ \Eprint
  {http://arxiv.org/abs/1710.05833} {arXiv:1710.05833 [astro-ph.HE]}
  \BibitemShut {NoStop}%
\bibitem [{\citenamefont {Abbott}\ \emph
  {et~al.}(2017{\natexlab{c}})\citenamefont {Abbott} \emph
  {et~al.}}]{LIGOScientific:2017adf}%
  \BibitemOpen
  \bibfield  {author} {\bibinfo {author} {\bibfnamefont {B.~P.}\ \bibnamefont
  {Abbott}} \emph {et~al.} (\bibinfo {collaboration} {LIGO Scientific, Virgo,
  1M2H, Dark Energy Camera GW-E, DES, DLT40, Las Cumbres Observatory, VINROUGE,
  MASTER}),\ }\href {\doibase 10.1038/nature24471} {\bibfield  {journal}
  {\bibinfo  {journal} {Nature}\ }\textbf {\bibinfo {volume} {551}},\ \bibinfo
  {pages} {85} (\bibinfo {year} {2017}{\natexlab{c}})},\ \Eprint
  {http://arxiv.org/abs/1710.05835} {arXiv:1710.05835 [astro-ph.CO]}
  \BibitemShut {NoStop}%
\bibitem [{\citenamefont {Borhanian}\ \emph {et~al.}(2020)\citenamefont
  {Borhanian}, \citenamefont {Dhani}, \citenamefont {Gupta}, \citenamefont
  {Arun},\ and\ \citenamefont {Sathyaprakash}}]{Borhanian:2020vyr}%
  \BibitemOpen
  \bibfield  {author} {\bibinfo {author} {\bibfnamefont {S.}~\bibnamefont
  {Borhanian}}, \bibinfo {author} {\bibfnamefont {A.}~\bibnamefont {Dhani}},
  \bibinfo {author} {\bibfnamefont {A.}~\bibnamefont {Gupta}}, \bibinfo
  {author} {\bibfnamefont {K.~G.}\ \bibnamefont {Arun}}, \ and\ \bibinfo
  {author} {\bibfnamefont {B.~S.}\ \bibnamefont {Sathyaprakash}},\ }\href
  {\doibase 10.3847/2041-8213/abcaf5} {\bibfield  {journal} {\bibinfo
  {journal} {Astrophys. J. Lett.}\ }\textbf {\bibinfo {volume} {905}},\
  \bibinfo {pages} {L28} (\bibinfo {year} {2020})},\ \Eprint
  {http://arxiv.org/abs/2007.02883} {arXiv:2007.02883 [astro-ph.CO]}
  \BibitemShut {NoStop}%
\bibitem [{\citenamefont {Nishizawa}(2017)}]{Nishizawa:2016ood}%
  \BibitemOpen
  \bibfield  {author} {\bibinfo {author} {\bibfnamefont {A.}~\bibnamefont
  {Nishizawa}},\ }\href {\doibase 10.1103/PhysRevD.96.101303} {\bibfield
  {journal} {\bibinfo  {journal} {Phys. Rev.}\ }\textbf {\bibinfo {volume}
  {D96}},\ \bibinfo {pages} {101303} (\bibinfo {year} {2017})},\ \Eprint
  {http://arxiv.org/abs/1612.06060} {arXiv:1612.06060 [astro-ph.CO]}
  \BibitemShut {NoStop}%
\bibitem [{\citenamefont {Yu}\ \emph {et~al.}(2020)\citenamefont {Yu},
  \citenamefont {Wang}, \citenamefont {Zhao},\ and\ \citenamefont
  {Lu}}]{Yu:2020vyy}%
  \BibitemOpen
  \bibfield  {author} {\bibinfo {author} {\bibfnamefont {J.}~\bibnamefont
  {Yu}}, \bibinfo {author} {\bibfnamefont {Y.}~\bibnamefont {Wang}}, \bibinfo
  {author} {\bibfnamefont {W.}~\bibnamefont {Zhao}}, \ and\ \bibinfo {author}
  {\bibfnamefont {Y.}~\bibnamefont {Lu}},\ }\href {\doibase
  10.1093/mnras/staa2465} {\bibfield  {journal} {\bibinfo  {journal} {Mon. Not.
  Roy. Astron. Soc.}\ }\textbf {\bibinfo {volume} {498}},\ \bibinfo {pages}
  {1786} (\bibinfo {year} {2020})},\ \Eprint {http://arxiv.org/abs/2003.06586}
  {arXiv:2003.06586 [astro-ph.CO]} \BibitemShut {NoStop}%
\bibitem [{\citenamefont {Del~Pozzo}(2012)}]{DelPozzo:2011vcw}%
  \BibitemOpen
  \bibfield  {author} {\bibinfo {author} {\bibfnamefont {W.}~\bibnamefont
  {Del~Pozzo}},\ }\href {\doibase 10.1103/PhysRevD.86.043011} {\bibfield
  {journal} {\bibinfo  {journal} {Phys. Rev. D}\ }\textbf {\bibinfo {volume}
  {86}},\ \bibinfo {pages} {043011} (\bibinfo {year} {2012})},\ \Eprint
  {http://arxiv.org/abs/1108.1317} {arXiv:1108.1317 [astro-ph.CO]} \BibitemShut
  {NoStop}%
\bibitem [{\citenamefont {Soares-Santos}\ \emph {et~al.}(2019)\citenamefont
  {Soares-Santos} \emph {et~al.}}]{DES:2019ccw}%
  \BibitemOpen
  \bibfield  {author} {\bibinfo {author} {\bibfnamefont {M.}~\bibnamefont
  {Soares-Santos}} \emph {et~al.} (\bibinfo {collaboration} {DES, LIGO
  Scientific, Virgo}),\ }\href {\doibase 10.3847/2041-8213/ab14f1} {\bibfield
  {journal} {\bibinfo  {journal} {Astrophys. J. Lett.}\ }\textbf {\bibinfo
  {volume} {876}},\ \bibinfo {pages} {L7} (\bibinfo {year} {2019})},\ \Eprint
  {http://arxiv.org/abs/1901.01540} {arXiv:1901.01540 [astro-ph.CO]}
  \BibitemShut {NoStop}%
\bibitem [{\citenamefont {Abbott}\ \emph
  {et~al.}(2021{\natexlab{b}})\citenamefont {Abbott} \emph
  {et~al.}}]{LIGOScientific:2019zcs}%
  \BibitemOpen
  \bibfield  {author} {\bibinfo {author} {\bibfnamefont {B.~P.}\ \bibnamefont
  {Abbott}} \emph {et~al.} (\bibinfo {collaboration} {LIGO Scientific, Virgo,
  VIRGO}),\ }\href {\doibase 10.3847/1538-4357/abdcb7} {\bibfield  {journal}
  {\bibinfo  {journal} {Astrophys. J.}\ }\textbf {\bibinfo {volume} {909}},\
  \bibinfo {pages} {218} (\bibinfo {year} {2021}{\natexlab{b}})},\ \bibinfo
  {note} {[Erratum: Astrophys.J. 923, 279 (2021)]},\ \Eprint
  {http://arxiv.org/abs/1908.06060} {arXiv:1908.06060 [astro-ph.CO]}
  \BibitemShut {NoStop}%
\bibitem [{\citenamefont {Palmese}\ \emph {et~al.}(2020)\citenamefont {Palmese}
  \emph {et~al.}}]{DES:2020nay}%
  \BibitemOpen
  \bibfield  {author} {\bibinfo {author} {\bibfnamefont {A.}~\bibnamefont
  {Palmese}} \emph {et~al.} (\bibinfo {collaboration} {DES}),\ }\href {\doibase
  10.3847/2041-8213/abaeff} {\bibfield  {journal} {\bibinfo  {journal}
  {Astrophys. J. Lett.}\ }\textbf {\bibinfo {volume} {900}},\ \bibinfo {pages}
  {L33} (\bibinfo {year} {2020})},\ \Eprint {http://arxiv.org/abs/2006.14961}
  {arXiv:2006.14961 [astro-ph.CO]} \BibitemShut {NoStop}%
\bibitem [{\citenamefont {Oguri}(2016)}]{Oguri:2016dgk}%
  \BibitemOpen
  \bibfield  {author} {\bibinfo {author} {\bibfnamefont {M.}~\bibnamefont
  {Oguri}},\ }\href {\doibase 10.1103/PhysRevD.93.083511} {\bibfield  {journal}
  {\bibinfo  {journal} {Phys. Rev. D}\ }\textbf {\bibinfo {volume} {93}},\
  \bibinfo {pages} {083511} (\bibinfo {year} {2016})},\ \Eprint
  {http://arxiv.org/abs/1603.02356} {arXiv:1603.02356 [astro-ph.CO]}
  \BibitemShut {NoStop}%
\bibitem [{\citenamefont {Mukherjee}\ \emph {et~al.}(2021)\citenamefont
  {Mukherjee}, \citenamefont {Wandelt}, \citenamefont {Nissanke},\ and\
  \citenamefont {Silvestri}}]{Mukherjee:2020hyn}%
  \BibitemOpen
  \bibfield  {author} {\bibinfo {author} {\bibfnamefont {S.}~\bibnamefont
  {Mukherjee}}, \bibinfo {author} {\bibfnamefont {B.~D.}\ \bibnamefont
  {Wandelt}}, \bibinfo {author} {\bibfnamefont {S.~M.}\ \bibnamefont
  {Nissanke}}, \ and\ \bibinfo {author} {\bibfnamefont {A.}~\bibnamefont
  {Silvestri}},\ }\href {\doibase 10.1103/PhysRevD.103.043520} {\bibfield
  {journal} {\bibinfo  {journal} {Phys. Rev. D}\ }\textbf {\bibinfo {volume}
  {103}},\ \bibinfo {pages} {043520} (\bibinfo {year} {2021})},\ \Eprint
  {http://arxiv.org/abs/2007.02943} {arXiv:2007.02943 [astro-ph.CO]}
  \BibitemShut {NoStop}%
\bibitem [{\citenamefont {Mukherjee}\ \emph {et~al.}(2022)\citenamefont
  {Mukherjee}, \citenamefont {Krolewski}, \citenamefont {Wandelt},\ and\
  \citenamefont {Silk}}]{Mukherjee:2022afz}%
  \BibitemOpen
  \bibfield  {author} {\bibinfo {author} {\bibfnamefont {S.}~\bibnamefont
  {Mukherjee}}, \bibinfo {author} {\bibfnamefont {A.}~\bibnamefont
  {Krolewski}}, \bibinfo {author} {\bibfnamefont {B.~D.}\ \bibnamefont
  {Wandelt}}, \ and\ \bibinfo {author} {\bibfnamefont {J.}~\bibnamefont
  {Silk}},\ }\href@noop {} {\  (\bibinfo {year} {2022})},\ \Eprint
  {http://arxiv.org/abs/2203.03643} {arXiv:2203.03643 [astro-ph.CO]}
  \BibitemShut {NoStop}%
\bibitem [{\citenamefont {Aghamousa}\ \emph {et~al.}(2016)\citenamefont
  {Aghamousa} \emph {et~al.}}]{DESI:2016fyo}%
  \BibitemOpen
  \bibfield  {author} {\bibinfo {author} {\bibfnamefont {A.}~\bibnamefont
  {Aghamousa}} \emph {et~al.} (\bibinfo {collaboration} {DESI}),\ }\href@noop
  {} {\  (\bibinfo {year} {2016})},\ \Eprint {http://arxiv.org/abs/1611.00036}
  {arXiv:1611.00036 [astro-ph.IM]} \BibitemShut {NoStop}%
\bibitem [{\citenamefont {Diaz}\ and\ \citenamefont
  {Mukherjee}(2022)}]{Diaz:2021pem}%
  \BibitemOpen
  \bibfield  {author} {\bibinfo {author} {\bibfnamefont {C.~C.}\ \bibnamefont
  {Diaz}}\ and\ \bibinfo {author} {\bibfnamefont {S.}~\bibnamefont
  {Mukherjee}},\ }\href {\doibase 10.1093/mnras/stac208} {\bibfield  {journal}
  {\bibinfo  {journal} {Mon. Not. Roy. Astron. Soc.}\ }\textbf {\bibinfo
  {volume} {511}},\ \bibinfo {pages} {2782} (\bibinfo {year} {2022})},\ \Eprint
  {http://arxiv.org/abs/2107.12787} {arXiv:2107.12787 [astro-ph.CO]}
  \BibitemShut {NoStop}%
\bibitem [{\citenamefont {Dor\'e}\ \emph {et~al.}(2014)\citenamefont {Dor\'e}
  \emph {et~al.}}]{Dore:2014cca}%
  \BibitemOpen
  \bibfield  {author} {\bibinfo {author} {\bibfnamefont {O.}~\bibnamefont
  {Dor\'e}} \emph {et~al.},\ }\href@noop {} {\  (\bibinfo {year} {2014})},\
  \Eprint {http://arxiv.org/abs/1412.4872} {arXiv:1412.4872 [astro-ph.CO]}
  \BibitemShut {NoStop}%
\bibitem [{\citenamefont {Chernoff}\ and\ \citenamefont
  {Finn}(1993)}]{Chernoff:1993th}%
  \BibitemOpen
  \bibfield  {author} {\bibinfo {author} {\bibfnamefont {D.~F.}\ \bibnamefont
  {Chernoff}}\ and\ \bibinfo {author} {\bibfnamefont {L.~S.}\ \bibnamefont
  {Finn}},\ }\href {\doibase 10.1086/186898} {\bibfield  {journal} {\bibinfo
  {journal} {Astrophys. J. Lett.}\ }\textbf {\bibinfo {volume} {411}},\
  \bibinfo {pages} {L5} (\bibinfo {year} {1993})},\ \Eprint
  {http://arxiv.org/abs/gr-qc/9304020} {arXiv:gr-qc/9304020} \BibitemShut
  {NoStop}%
\bibitem [{\citenamefont {Taylor}\ and\ \citenamefont
  {Gair}(2012)}]{Taylor:2012db}%
  \BibitemOpen
  \bibfield  {author} {\bibinfo {author} {\bibfnamefont {S.~R.}\ \bibnamefont
  {Taylor}}\ and\ \bibinfo {author} {\bibfnamefont {J.~R.}\ \bibnamefont
  {Gair}},\ }\href {\doibase 10.1103/PhysRevD.86.023502} {\bibfield  {journal}
  {\bibinfo  {journal} {Phys. Rev. D}\ }\textbf {\bibinfo {volume} {86}},\
  \bibinfo {pages} {023502} (\bibinfo {year} {2012})},\ \Eprint
  {http://arxiv.org/abs/1204.6739} {arXiv:1204.6739 [astro-ph.CO]} \BibitemShut
  {NoStop}%
\bibitem [{\citenamefont {Farr}\ \emph {et~al.}(2019)\citenamefont {Farr},
  \citenamefont {Fishbach}, \citenamefont {Ye},\ and\ \citenamefont
  {Holz}}]{Farr:2019twy}%
  \BibitemOpen
  \bibfield  {author} {\bibinfo {author} {\bibfnamefont {W.~M.}\ \bibnamefont
  {Farr}}, \bibinfo {author} {\bibfnamefont {M.}~\bibnamefont {Fishbach}},
  \bibinfo {author} {\bibfnamefont {J.}~\bibnamefont {Ye}}, \ and\ \bibinfo
  {author} {\bibfnamefont {D.}~\bibnamefont {Holz}},\ }\href {\doibase
  10.3847/2041-8213/ab4284} {\bibfield  {journal} {\bibinfo  {journal}
  {Astrophys. J. Lett.}\ }\textbf {\bibinfo {volume} {883}},\ \bibinfo {pages}
  {L42} (\bibinfo {year} {2019})},\ \Eprint {http://arxiv.org/abs/1908.09084}
  {arXiv:1908.09084 [astro-ph.CO]} \BibitemShut {NoStop}%
\bibitem [{\citenamefont {Ezquiaga}\ and\ \citenamefont
  {Holz}(2022)}]{Ezquiaga:2022zkx}%
  \BibitemOpen
  \bibfield  {author} {\bibinfo {author} {\bibfnamefont {J.~M.}\ \bibnamefont
  {Ezquiaga}}\ and\ \bibinfo {author} {\bibfnamefont {D.~E.}\ \bibnamefont
  {Holz}},\ }\href {\doibase 10.1103/PhysRevLett.129.061102} {\bibfield
  {journal} {\bibinfo  {journal} {Phys. Rev. Lett.}\ }\textbf {\bibinfo
  {volume} {129}},\ \bibinfo {pages} {061102} (\bibinfo {year} {2022})},\
  \Eprint {http://arxiv.org/abs/2202.08240} {arXiv:2202.08240 [astro-ph.CO]}
  \BibitemShut {NoStop}%
\bibitem [{\citenamefont {Mukherjee}(2022)}]{Mukherjee:2021rtw}%
  \BibitemOpen
  \bibfield  {author} {\bibinfo {author} {\bibfnamefont {S.}~\bibnamefont
  {Mukherjee}},\ }\href {\doibase 10.1093/mnras/stac2152} {\bibfield  {journal}
  {\bibinfo  {journal} {Mon. Not. Roy. Astron. Soc.}\ }\textbf {\bibinfo
  {volume} {515}},\ \bibinfo {pages} {5495} (\bibinfo {year} {2022})},\ \Eprint
  {http://arxiv.org/abs/2112.10256} {arXiv:2112.10256 [astro-ph.CO]}
  \BibitemShut {NoStop}%
\bibitem [{\citenamefont {Karathanasis}\ \emph {et~al.}(2022)\citenamefont
  {Karathanasis}, \citenamefont {Mukherjee},\ and\ \citenamefont
  {Mastrogiovanni}}]{Karathanasis:2022rtr}%
  \BibitemOpen
  \bibfield  {author} {\bibinfo {author} {\bibfnamefont {C.}~\bibnamefont
  {Karathanasis}}, \bibinfo {author} {\bibfnamefont {S.}~\bibnamefont
  {Mukherjee}}, \ and\ \bibinfo {author} {\bibfnamefont {S.}~\bibnamefont
  {Mastrogiovanni}},\ }\href@noop {} {\  (\bibinfo {year} {2022})},\ \Eprint
  {http://arxiv.org/abs/2204.13495} {arXiv:2204.13495 [astro-ph.CO]}
  \BibitemShut {NoStop}%
\bibitem [{\citenamefont {Ding}\ \emph {et~al.}(2019)\citenamefont {Ding},
  \citenamefont {Biesiada}, \citenamefont {Zheng}, \citenamefont {Liao},
  \citenamefont {Li},\ and\ \citenamefont {Zhu}}]{Ding:2018zrk}%
  \BibitemOpen
  \bibfield  {author} {\bibinfo {author} {\bibfnamefont {X.}~\bibnamefont
  {Ding}}, \bibinfo {author} {\bibfnamefont {M.}~\bibnamefont {Biesiada}},
  \bibinfo {author} {\bibfnamefont {X.}~\bibnamefont {Zheng}}, \bibinfo
  {author} {\bibfnamefont {K.}~\bibnamefont {Liao}}, \bibinfo {author}
  {\bibfnamefont {Z.}~\bibnamefont {Li}}, \ and\ \bibinfo {author}
  {\bibfnamefont {Z.-H.}\ \bibnamefont {Zhu}},\ }\href {\doibase
  10.1088/1475-7516/2019/04/033} {\bibfield  {journal} {\bibinfo  {journal}
  {JCAP}\ }\textbf {\bibinfo {volume} {04}},\ \bibinfo {pages} {033} (\bibinfo
  {year} {2019})},\ \Eprint {http://arxiv.org/abs/1801.05073} {arXiv:1801.05073
  [astro-ph.CO]} \BibitemShut {NoStop}%
\bibitem [{\citenamefont {Leandro}\ \emph {et~al.}(2022)\citenamefont
  {Leandro}, \citenamefont {Marra},\ and\ \citenamefont
  {Sturani}}]{Leandro:2021qlc}%
  \BibitemOpen
  \bibfield  {author} {\bibinfo {author} {\bibfnamefont {H.}~\bibnamefont
  {Leandro}}, \bibinfo {author} {\bibfnamefont {V.}~\bibnamefont {Marra}}, \
  and\ \bibinfo {author} {\bibfnamefont {R.}~\bibnamefont {Sturani}},\ }\href
  {\doibase 10.1103/PhysRevD.105.023523} {\bibfield  {journal} {\bibinfo
  {journal} {Phys. Rev. D}\ }\textbf {\bibinfo {volume} {105}},\ \bibinfo
  {pages} {023523} (\bibinfo {year} {2022})},\ \Eprint
  {http://arxiv.org/abs/2109.07537} {arXiv:2109.07537 [gr-qc]} \BibitemShut
  {NoStop}%
\bibitem [{\citenamefont {Mastrogiovanni}\ \emph {et~al.}(2022)\citenamefont
  {Mastrogiovanni}, \citenamefont {Leyde}, \citenamefont {Karathanasis},
  \citenamefont {Chassande-Mottin}, \citenamefont {Steer}, \citenamefont
  {Gair}, \citenamefont {Ghosh}, \citenamefont {Gray}, \citenamefont
  {Mukherjee},\ and\ \citenamefont {Rinaldi}}]{Mastrogiovanni:2022hil}%
  \BibitemOpen
  \bibfield  {author} {\bibinfo {author} {\bibfnamefont {S.}~\bibnamefont
  {Mastrogiovanni}}, \bibinfo {author} {\bibfnamefont {K.}~\bibnamefont
  {Leyde}}, \bibinfo {author} {\bibfnamefont {C.}~\bibnamefont {Karathanasis}},
  \bibinfo {author} {\bibfnamefont {E.}~\bibnamefont {Chassande-Mottin}},
  \bibinfo {author} {\bibfnamefont {D.~A.}\ \bibnamefont {Steer}}, \bibinfo
  {author} {\bibfnamefont {J.}~\bibnamefont {Gair}}, \bibinfo {author}
  {\bibfnamefont {A.}~\bibnamefont {Ghosh}}, \bibinfo {author} {\bibfnamefont
  {R.}~\bibnamefont {Gray}}, \bibinfo {author} {\bibfnamefont {S.}~\bibnamefont
  {Mukherjee}}, \ and\ \bibinfo {author} {\bibfnamefont {S.}~\bibnamefont
  {Rinaldi}},\ }\href {\doibase 10.22323/1.398.0098} {\bibfield  {journal}
  {\bibinfo  {journal} {PoS}\ }\textbf {\bibinfo {volume} {EPS-HEP2021}},\
  \bibinfo {pages} {098} (\bibinfo {year} {2022})},\ \Eprint
  {http://arxiv.org/abs/2205.05421} {arXiv:2205.05421 [gr-qc]} \BibitemShut
  {NoStop}%
\bibitem [{\citenamefont {Messenger}\ and\ \citenamefont
  {Read}(2012)}]{Messenger:2011gi}%
  \BibitemOpen
  \bibfield  {author} {\bibinfo {author} {\bibfnamefont {C.}~\bibnamefont
  {Messenger}}\ and\ \bibinfo {author} {\bibfnamefont {J.}~\bibnamefont
  {Read}},\ }\href {\doibase 10.1103/PhysRevLett.108.091101} {\bibfield
  {journal} {\bibinfo  {journal} {Phys. Rev. Lett.}\ }\textbf {\bibinfo
  {volume} {108}},\ \bibinfo {pages} {091101} (\bibinfo {year} {2012})},\
  \Eprint {http://arxiv.org/abs/1107.5725} {arXiv:1107.5725 [gr-qc]}
  \BibitemShut {NoStop}%
\bibitem [{\citenamefont {Messenger}\ \emph {et~al.}(2014)\citenamefont
  {Messenger}, \citenamefont {Takami}, \citenamefont {Gossan}, \citenamefont
  {Rezzolla},\ and\ \citenamefont {Sathyaprakash}}]{Messenger:2013fya}%
  \BibitemOpen
  \bibfield  {author} {\bibinfo {author} {\bibfnamefont {C.}~\bibnamefont
  {Messenger}}, \bibinfo {author} {\bibfnamefont {K.}~\bibnamefont {Takami}},
  \bibinfo {author} {\bibfnamefont {S.}~\bibnamefont {Gossan}}, \bibinfo
  {author} {\bibfnamefont {L.}~\bibnamefont {Rezzolla}}, \ and\ \bibinfo
  {author} {\bibfnamefont {B.~S.}\ \bibnamefont {Sathyaprakash}},\ }\href
  {\doibase 10.1103/PhysRevX.4.041004} {\bibfield  {journal} {\bibinfo
  {journal} {Phys. Rev. X}\ }\textbf {\bibinfo {volume} {4}},\ \bibinfo {pages}
  {041004} (\bibinfo {year} {2014})},\ \Eprint {http://arxiv.org/abs/1312.1862}
  {arXiv:1312.1862 [gr-qc]} \BibitemShut {NoStop}%
\bibitem [{\citenamefont {Li}\ \emph {et~al.}(2015)\citenamefont {Li},
  \citenamefont {Del~Pozzo},\ and\ \citenamefont {Messenger}}]{Li:2013via}%
  \BibitemOpen
  \bibfield  {author} {\bibinfo {author} {\bibfnamefont {T.~G.~F.}\
  \bibnamefont {Li}}, \bibinfo {author} {\bibfnamefont {W.}~\bibnamefont
  {Del~Pozzo}}, \ and\ \bibinfo {author} {\bibfnamefont {C.}~\bibnamefont
  {Messenger}},\ }in\ \href {\doibase 10.1142/9789814623995_0346} {\emph
  {\bibinfo {booktitle} {{13th Marcel Grossmann Meeting on Recent Developments
  in Theoretical and Experimental General Relativity, Astrophysics, and
  Relativistic Field Theories}}}}\ (\bibinfo {year} {2015})\ pp.\ \bibinfo
  {pages} {2019--2021},\ \Eprint {http://arxiv.org/abs/1303.0855}
  {arXiv:1303.0855 [gr-qc]} \BibitemShut {NoStop}%
\bibitem [{\citenamefont {Wang}\ \emph {et~al.}(2020)\citenamefont {Wang},
  \citenamefont {Zhu}, \citenamefont {Li},\ and\ \citenamefont
  {Zhao}}]{Wang:2020xwn}%
  \BibitemOpen
  \bibfield  {author} {\bibinfo {author} {\bibfnamefont {B.}~\bibnamefont
  {Wang}}, \bibinfo {author} {\bibfnamefont {Z.}~\bibnamefont {Zhu}}, \bibinfo
  {author} {\bibfnamefont {A.}~\bibnamefont {Li}}, \ and\ \bibinfo {author}
  {\bibfnamefont {W.}~\bibnamefont {Zhao}},\ }\href {\doibase
  10.3847/1538-4365/aba2f3} {\bibfield  {journal} {\bibinfo  {journal}
  {Astrophys. J. Suppl.}\ }\textbf {\bibinfo {volume} {250}},\ \bibinfo {pages}
  {6} (\bibinfo {year} {2020})},\ \Eprint {http://arxiv.org/abs/2005.12875}
  {arXiv:2005.12875 [gr-qc]} \BibitemShut {NoStop}%
\bibitem [{\citenamefont {Chen}(2020)}]{Chen:2020dyt}%
  \BibitemOpen
  \bibfield  {author} {\bibinfo {author} {\bibfnamefont {H.-Y.}\ \bibnamefont
  {Chen}},\ }\href {\doibase 10.1103/PhysRevLett.125.201301} {\bibfield
  {journal} {\bibinfo  {journal} {Phys. Rev. Lett.}\ }\textbf {\bibinfo
  {volume} {125}},\ \bibinfo {pages} {201301} (\bibinfo {year} {2020})},\
  \Eprint {http://arxiv.org/abs/2006.02779} {arXiv:2006.02779 [astro-ph.HE]}
  \BibitemShut {NoStop}%
\bibitem [{\citenamefont {Belgacem}\ \emph
  {et~al.}(2019{\natexlab{a}})\citenamefont {Belgacem}, \citenamefont {Dirian},
  \citenamefont {Foffa}, \citenamefont {Howell}, \citenamefont {Maggiore},\
  and\ \citenamefont {Regimbau}}]{Belgacem:2019tbw}%
  \BibitemOpen
  \bibfield  {author} {\bibinfo {author} {\bibfnamefont {E.}~\bibnamefont
  {Belgacem}}, \bibinfo {author} {\bibfnamefont {Y.}~\bibnamefont {Dirian}},
  \bibinfo {author} {\bibfnamefont {S.}~\bibnamefont {Foffa}}, \bibinfo
  {author} {\bibfnamefont {E.~J.}\ \bibnamefont {Howell}}, \bibinfo {author}
  {\bibfnamefont {M.}~\bibnamefont {Maggiore}}, \ and\ \bibinfo {author}
  {\bibfnamefont {T.}~\bibnamefont {Regimbau}},\ }\href {\doibase
  10.1088/1475-7516/2019/08/015} {\bibfield  {journal} {\bibinfo  {journal}
  {JCAP}\ }\textbf {\bibinfo {volume} {08}},\ \bibinfo {pages} {015} (\bibinfo
  {year} {2019}{\natexlab{a}})},\ \Eprint {http://arxiv.org/abs/1907.01487}
  {arXiv:1907.01487 [astro-ph.CO]} \BibitemShut {NoStop}%
\bibitem [{\citenamefont {Calder\'on~Bustillo}\ \emph
  {et~al.}(2021)\citenamefont {Calder\'on~Bustillo}, \citenamefont {Leong},
  \citenamefont {Dietrich},\ and\ \citenamefont
  {Lasky}}]{CalderonBustillo:2020kcg}%
  \BibitemOpen
  \bibfield  {author} {\bibinfo {author} {\bibfnamefont {J.}~\bibnamefont
  {Calder\'on~Bustillo}}, \bibinfo {author} {\bibfnamefont {S.~H.~W.}\
  \bibnamefont {Leong}}, \bibinfo {author} {\bibfnamefont {T.}~\bibnamefont
  {Dietrich}}, \ and\ \bibinfo {author} {\bibfnamefont {P.~D.}\ \bibnamefont
  {Lasky}},\ }\href {\doibase 10.3847/2041-8213/abf502} {\bibfield  {journal}
  {\bibinfo  {journal} {Astrophys. J. Lett.}\ }\textbf {\bibinfo {volume}
  {912}},\ \bibinfo {pages} {L10} (\bibinfo {year} {2021})},\ \Eprint
  {http://arxiv.org/abs/2006.11525} {arXiv:2006.11525 [gr-qc]} \BibitemShut
  {NoStop}%
\bibitem [{\citenamefont {Chen}\ \emph {et~al.}(2021)\citenamefont {Chen},
  \citenamefont {Cowperthwaite}, \citenamefont {Metzger},\ and\ \citenamefont
  {Berger}}]{Chen:2020zoq}%
  \BibitemOpen
  \bibfield  {author} {\bibinfo {author} {\bibfnamefont {H.-Y.}\ \bibnamefont
  {Chen}}, \bibinfo {author} {\bibfnamefont {P.~S.}\ \bibnamefont
  {Cowperthwaite}}, \bibinfo {author} {\bibfnamefont {B.~D.}\ \bibnamefont
  {Metzger}}, \ and\ \bibinfo {author} {\bibfnamefont {E.}~\bibnamefont
  {Berger}},\ }\href {\doibase 10.3847/2041-8213/abdab0} {\bibfield  {journal}
  {\bibinfo  {journal} {Astrophys. J. Lett.}\ }\textbf {\bibinfo {volume}
  {908}},\ \bibinfo {pages} {L4} (\bibinfo {year} {2021})},\ \Eprint
  {http://arxiv.org/abs/2011.01211} {arXiv:2011.01211 [astro-ph.CO]}
  \BibitemShut {NoStop}%
\bibitem [{\citenamefont {Bulla}\ \emph {et~al.}(2022)\citenamefont {Bulla},
  \citenamefont {Coughlin}, \citenamefont {Dhawan},\ and\ \citenamefont
  {Dietrich}}]{Bulla:2022ppy}%
  \BibitemOpen
  \bibfield  {author} {\bibinfo {author} {\bibfnamefont {M.}~\bibnamefont
  {Bulla}}, \bibinfo {author} {\bibfnamefont {M.~W.}\ \bibnamefont {Coughlin}},
  \bibinfo {author} {\bibfnamefont {S.}~\bibnamefont {Dhawan}}, \ and\ \bibinfo
  {author} {\bibfnamefont {T.}~\bibnamefont {Dietrich}},\ }\href {\doibase
  10.3390/universe8050289} {\bibfield  {journal} {\bibinfo  {journal}
  {Universe}\ }\textbf {\bibinfo {volume} {8}},\ \bibinfo {pages} {289}
  (\bibinfo {year} {2022})},\ \Eprint {http://arxiv.org/abs/2205.09145}
  {arXiv:2205.09145 [astro-ph.HE]} \BibitemShut {NoStop}%
\bibitem [{\citenamefont {Chatterjee}\ \emph {et~al.}(2021)\citenamefont
  {Chatterjee}, \citenamefont {Hegade K~R}, \citenamefont {Holder},
  \citenamefont {Holz}, \citenamefont {Perkins}, \citenamefont {Yagi},\ and\
  \citenamefont {Yunes}}]{Chatterjee:2021xrm}%
  \BibitemOpen
  \bibfield  {author} {\bibinfo {author} {\bibfnamefont {D.}~\bibnamefont
  {Chatterjee}}, \bibinfo {author} {\bibfnamefont {A.}~\bibnamefont {Hegade
  K~R}}, \bibinfo {author} {\bibfnamefont {G.}~\bibnamefont {Holder}}, \bibinfo
  {author} {\bibfnamefont {D.~E.}\ \bibnamefont {Holz}}, \bibinfo {author}
  {\bibfnamefont {S.}~\bibnamefont {Perkins}}, \bibinfo {author} {\bibfnamefont
  {K.}~\bibnamefont {Yagi}}, \ and\ \bibinfo {author} {\bibfnamefont
  {N.}~\bibnamefont {Yunes}},\ }\href {\doibase 10.1103/PhysRevD.104.083528}
  {\bibfield  {journal} {\bibinfo  {journal} {Phys. Rev. D}\ }\textbf {\bibinfo
  {volume} {104}},\ \bibinfo {pages} {083528} (\bibinfo {year} {2021})},\
  \Eprint {http://arxiv.org/abs/2106.06589} {arXiv:2106.06589 [gr-qc]}
  \BibitemShut {NoStop}%
\bibitem [{\citenamefont {Yagi}(2014)}]{Yagi:2013sva}%
  \BibitemOpen
  \bibfield  {author} {\bibinfo {author} {\bibfnamefont {K.}~\bibnamefont
  {Yagi}},\ }\href {\doibase 10.1103/PhysRevD.89.043011} {\bibfield  {journal}
  {\bibinfo  {journal} {Phys. Rev. D}\ }\textbf {\bibinfo {volume} {89}},\
  \bibinfo {pages} {043011} (\bibinfo {year} {2014})},\ \bibinfo {note}
  {[Erratum: Phys.Rev.D 96, 129904 (2017), Erratum: Phys.Rev.D 97, 129901
  (2018)]},\ \Eprint {http://arxiv.org/abs/1311.0872} {arXiv:1311.0872 [gr-qc]}
  \BibitemShut {NoStop}%
\bibitem [{\citenamefont {Del~Pozzo}\ \emph {et~al.}(2017)\citenamefont
  {Del~Pozzo}, \citenamefont {Li},\ and\ \citenamefont
  {Messenger}}]{DelPozzo:2015bna}%
  \BibitemOpen
  \bibfield  {author} {\bibinfo {author} {\bibfnamefont {W.}~\bibnamefont
  {Del~Pozzo}}, \bibinfo {author} {\bibfnamefont {T.~G.~F.}\ \bibnamefont
  {Li}}, \ and\ \bibinfo {author} {\bibfnamefont {C.}~\bibnamefont
  {Messenger}},\ }\href {\doibase 10.1103/PhysRevD.95.043502} {\bibfield
  {journal} {\bibinfo  {journal} {Phys. Rev. D}\ }\textbf {\bibinfo {volume}
  {95}},\ \bibinfo {pages} {043502} (\bibinfo {year} {2017})},\ \Eprint
  {http://arxiv.org/abs/1506.06590} {arXiv:1506.06590 [gr-qc]} \BibitemShut
  {NoStop}%
\bibitem [{\citenamefont {Jin}\ \emph {et~al.}(2022)\citenamefont {Jin},
  \citenamefont {Li}, \citenamefont {Zhang},\ and\ \citenamefont
  {Zhang}}]{Jin:2022qnj}%
  \BibitemOpen
  \bibfield  {author} {\bibinfo {author} {\bibfnamefont {S.-J.}\ \bibnamefont
  {Jin}}, \bibinfo {author} {\bibfnamefont {T.-N.}\ \bibnamefont {Li}},
  \bibinfo {author} {\bibfnamefont {J.-F.}\ \bibnamefont {Zhang}}, \ and\
  \bibinfo {author} {\bibfnamefont {X.}~\bibnamefont {Zhang}},\ }\href@noop {}
  {\  (\bibinfo {year} {2022})},\ \Eprint {http://arxiv.org/abs/2202.11882}
  {arXiv:2202.11882 [gr-qc]} \BibitemShut {NoStop}%
\bibitem [{\citenamefont {Ghosh}\ \emph {et~al.}(2022)\citenamefont {Ghosh},
  \citenamefont {Biswas},\ and\ \citenamefont {Bose}}]{Ghosh:2022muc}%
  \BibitemOpen
  \bibfield  {author} {\bibinfo {author} {\bibfnamefont {T.}~\bibnamefont
  {Ghosh}}, \bibinfo {author} {\bibfnamefont {B.}~\bibnamefont {Biswas}}, \
  and\ \bibinfo {author} {\bibfnamefont {S.}~\bibnamefont {Bose}},\ }\href@noop
  {} {\  (\bibinfo {year} {2022})},\ \Eprint {http://arxiv.org/abs/2203.11756}
  {arXiv:2203.11756 [astro-ph.CO]} \BibitemShut {NoStop}%
\bibitem [{\citenamefont {Evans}\ \emph {et~al.}(2021)\citenamefont {Evans}
  \emph {et~al.}}]{Evans:2021gyd}%
  \BibitemOpen
  \bibfield  {author} {\bibinfo {author} {\bibfnamefont {M.}~\bibnamefont
  {Evans}} \emph {et~al.},\ }\href@noop {} {\  (\bibinfo {year} {2021})},\
  \Eprint {http://arxiv.org/abs/2109.09882} {arXiv:2109.09882 [astro-ph.IM]}
  \BibitemShut {NoStop}%
\bibitem [{\citenamefont {Borhanian}\ and\ \citenamefont
  {Sathyaprakash}(2022)}]{Borhanian:2022czq}%
  \BibitemOpen
  \bibfield  {author} {\bibinfo {author} {\bibfnamefont {S.}~\bibnamefont
  {Borhanian}}\ and\ \bibinfo {author} {\bibfnamefont {B.~S.}\ \bibnamefont
  {Sathyaprakash}},\ }\href@noop {} {\  (\bibinfo {year} {2022})},\ \Eprint
  {http://arxiv.org/abs/2202.11048} {arXiv:2202.11048 [gr-qc]} \BibitemShut
  {NoStop}%
\bibitem [{\citenamefont {Aasi}\ \emph
  {et~al.}(2015{\natexlab{b}})\citenamefont {Aasi} \emph {et~al.}}]{aLIGO_ref}%
  \BibitemOpen
  \bibfield  {author} {\bibinfo {author} {\bibfnamefont {J.}~\bibnamefont
  {Aasi}} \emph {et~al.} (\bibinfo {collaboration} {LIGO Scientific}),\ }\href
  {\doibase 10.1088/0264-9381/32/7/074001} {\bibfield  {journal} {\bibinfo
  {journal} {Class. Quant. Grav.}\ }\textbf {\bibinfo {volume} {32}},\ \bibinfo
  {pages} {074001} (\bibinfo {year} {2015}{\natexlab{b}})},\ \Eprint
  {http://arxiv.org/abs/1411.4547} {arXiv:1411.4547 [gr-qc]} \BibitemShut
  {NoStop}%
\bibitem [{\citenamefont {Adhikari}\ \emph {et~al.}(2020)\citenamefont
  {Adhikari} \emph {et~al.}}]{LIGO:2020xsf}%
  \BibitemOpen
  \bibfield  {author} {\bibinfo {author} {\bibfnamefont {R.~X.}\ \bibnamefont
  {Adhikari}} \emph {et~al.} (\bibinfo {collaboration} {LIGO}),\ }\href
  {\doibase 10.1088/1361-6382/ab9143} {\bibfield  {journal} {\bibinfo
  {journal} {Class. Quant. Grav.}\ }\textbf {\bibinfo {volume} {37}},\ \bibinfo
  {pages} {165003} (\bibinfo {year} {2020})},\ \Eprint
  {http://arxiv.org/abs/2001.11173} {arXiv:2001.11173 [astro-ph.IM]}
  \BibitemShut {NoStop}%
\bibitem [{\citenamefont {Punturo}\ \emph {et~al.}(2010)\citenamefont {Punturo}
  \emph {et~al.}}]{Punturo:2010zza}%
  \BibitemOpen
  \bibfield  {author} {\bibinfo {author} {\bibfnamefont {M.}~\bibnamefont
  {Punturo}} \emph {et~al.},\ }\href {\doibase 10.1088/0264-9381/27/8/084007}
  {\bibfield  {journal} {\bibinfo  {journal} {Class. Quant. Grav.}\ }\textbf
  {\bibinfo {volume} {27}},\ \bibinfo {pages} {084007} (\bibinfo {year}
  {2010})}\BibitemShut {NoStop}%
\bibitem [{\citenamefont {Reitze}\ \emph {et~al.}(2019)\citenamefont {Reitze}
  \emph {et~al.}}]{Reitze:2019iox}%
  \BibitemOpen
  \bibfield  {author} {\bibinfo {author} {\bibfnamefont {D.}~\bibnamefont
  {Reitze}} \emph {et~al.},\ }\href@noop {} {\bibfield  {journal} {\bibinfo
  {journal} {Bull. Am. Astron. Soc.}\ }\textbf {\bibinfo {volume} {51}},\
  \bibinfo {pages} {035} (\bibinfo {year} {2019})},\ \Eprint
  {http://arxiv.org/abs/1907.04833} {arXiv:1907.04833 [astro-ph.IM]}
  \BibitemShut {NoStop}%
\bibitem [{\citenamefont {Martynov}\ \emph {et~al.}(2019)\citenamefont
  {Martynov} \emph {et~al.}}]{Martynov:2019gvu}%
  \BibitemOpen
  \bibfield  {author} {\bibinfo {author} {\bibfnamefont {D.}~\bibnamefont
  {Martynov}} \emph {et~al.},\ }\href {\doibase 10.1103/PhysRevD.99.102004}
  {\bibfield  {journal} {\bibinfo  {journal} {Phys. Rev. D}\ }\textbf {\bibinfo
  {volume} {99}},\ \bibinfo {pages} {102004} (\bibinfo {year} {2019})},\
  \Eprint {http://arxiv.org/abs/1901.03885} {arXiv:1901.03885 [astro-ph.IM]}
  \BibitemShut {NoStop}%
\bibitem [{\citenamefont {Vangioni}\ \emph {et~al.}(2015)\citenamefont
  {Vangioni}, \citenamefont {Olive}, \citenamefont {Prestegard}, \citenamefont
  {Silk}, \citenamefont {Petitjean},\ and\ \citenamefont
  {Mandic}}]{Vangioni:2014axa}%
  \BibitemOpen
  \bibfield  {author} {\bibinfo {author} {\bibfnamefont {E.}~\bibnamefont
  {Vangioni}}, \bibinfo {author} {\bibfnamefont {K.~A.}\ \bibnamefont {Olive}},
  \bibinfo {author} {\bibfnamefont {T.}~\bibnamefont {Prestegard}}, \bibinfo
  {author} {\bibfnamefont {J.}~\bibnamefont {Silk}}, \bibinfo {author}
  {\bibfnamefont {P.}~\bibnamefont {Petitjean}}, \ and\ \bibinfo {author}
  {\bibfnamefont {V.}~\bibnamefont {Mandic}},\ }\href {\doibase
  10.1093/mnras/stu2600} {\bibfield  {journal} {\bibinfo  {journal} {Mon. Not.
  Roy. Astron. Soc.}\ }\textbf {\bibinfo {volume} {447}},\ \bibinfo {pages}
  {2575} (\bibinfo {year} {2015})},\ \Eprint {http://arxiv.org/abs/1409.2462}
  {arXiv:1409.2462 [astro-ph.GA]} \BibitemShut {NoStop}%
\bibitem [{\citenamefont {Abbott}\ \emph
  {et~al.}(2021{\natexlab{c}})\citenamefont {Abbott} \emph
  {et~al.}}]{LIGOScientific:2020kqk}%
  \BibitemOpen
  \bibfield  {author} {\bibinfo {author} {\bibfnamefont {R.}~\bibnamefont
  {Abbott}} \emph {et~al.} (\bibinfo {collaboration} {LIGO Scientific,
  Virgo}),\ }\href {\doibase 10.3847/2041-8213/abe949} {\bibfield  {journal}
  {\bibinfo  {journal} {Astrophys. J. Lett.}\ }\textbf {\bibinfo {volume}
  {913}},\ \bibinfo {pages} {L7} (\bibinfo {year} {2021}{\natexlab{c}})},\
  \Eprint {http://arxiv.org/abs/2010.14533} {arXiv:2010.14533 [astro-ph.HE]}
  \BibitemShut {NoStop}%
\bibitem [{\citenamefont {M\"uther}\ \emph {et~al.}(1987)\citenamefont
  {M\"uther}, \citenamefont {Prakash},\ and\ \citenamefont
  {Ainsworth}}]{Muther:1987xaa}%
  \BibitemOpen
  \bibfield  {author} {\bibinfo {author} {\bibfnamefont {H.}~\bibnamefont
  {M\"uther}}, \bibinfo {author} {\bibfnamefont {M.}~\bibnamefont {Prakash}}, \
  and\ \bibinfo {author} {\bibfnamefont {T.~L.}\ \bibnamefont {Ainsworth}},\
  }\href {\doibase 10.1016/0370-2693(87)91611-X} {\bibfield  {journal}
  {\bibinfo  {journal} {Phys. Lett. B}\ }\textbf {\bibinfo {volume} {199}},\
  \bibinfo {pages} {469} (\bibinfo {year} {1987})}\BibitemShut {NoStop}%
\bibitem [{\citenamefont {Flanagan}\ and\ \citenamefont
  {Hinderer}(2008)}]{Flanagan:2007ix}%
  \BibitemOpen
  \bibfield  {author} {\bibinfo {author} {\bibfnamefont {E.~E.}\ \bibnamefont
  {Flanagan}}\ and\ \bibinfo {author} {\bibfnamefont {T.}~\bibnamefont
  {Hinderer}},\ }\href {\doibase 10.1103/PhysRevD.77.021502} {\bibfield
  {journal} {\bibinfo  {journal} {Phys. Rev. D}\ }\textbf {\bibinfo {volume}
  {77}},\ \bibinfo {pages} {021502} (\bibinfo {year} {2008})},\ \Eprint
  {http://arxiv.org/abs/0709.1915} {arXiv:0709.1915 [astro-ph]} \BibitemShut
  {NoStop}%
\bibitem [{\citenamefont {Favata}(2014)}]{Favata:2013rwa}%
  \BibitemOpen
  \bibfield  {author} {\bibinfo {author} {\bibfnamefont {M.}~\bibnamefont
  {Favata}},\ }\href {\doibase 10.1103/PhysRevLett.112.101101} {\bibfield
  {journal} {\bibinfo  {journal} {Phys. Rev. Lett.}\ }\textbf {\bibinfo
  {volume} {112}},\ \bibinfo {pages} {101101} (\bibinfo {year} {2014})},\
  \Eprint {http://arxiv.org/abs/1310.8288} {arXiv:1310.8288 [gr-qc]}
  \BibitemShut {NoStop}%
\bibitem [{\citenamefont {Buonanno}\ \emph {et~al.}(2009)\citenamefont
  {Buonanno}, \citenamefont {Iyer}, \citenamefont {Ochsner}, \citenamefont
  {Pan},\ and\ \citenamefont {Sathyaprakash}}]{Buonanno:2009zt}%
  \BibitemOpen
  \bibfield  {author} {\bibinfo {author} {\bibfnamefont {A.}~\bibnamefont
  {Buonanno}}, \bibinfo {author} {\bibfnamefont {B.}~\bibnamefont {Iyer}},
  \bibinfo {author} {\bibfnamefont {E.}~\bibnamefont {Ochsner}}, \bibinfo
  {author} {\bibfnamefont {Y.}~\bibnamefont {Pan}}, \ and\ \bibinfo {author}
  {\bibfnamefont {B.~S.}\ \bibnamefont {Sathyaprakash}},\ }\href {\doibase
  10.1103/PhysRevD.80.084043} {\bibfield  {journal} {\bibinfo  {journal} {Phys.
  Rev. D}\ }\textbf {\bibinfo {volume} {80}},\ \bibinfo {pages} {084043}
  (\bibinfo {year} {2009})},\ \Eprint {http://arxiv.org/abs/0907.0700}
  {arXiv:0907.0700 [gr-qc]} \BibitemShut {NoStop}%
\bibitem [{\citenamefont {Soultanis}\ \emph {et~al.}(2022)\citenamefont
  {Soultanis}, \citenamefont {Bauswein},\ and\ \citenamefont
  {Stergioulas}}]{Soultanis:2021oia}%
  \BibitemOpen
  \bibfield  {author} {\bibinfo {author} {\bibfnamefont {T.}~\bibnamefont
  {Soultanis}}, \bibinfo {author} {\bibfnamefont {A.}~\bibnamefont {Bauswein}},
  \ and\ \bibinfo {author} {\bibfnamefont {N.}~\bibnamefont {Stergioulas}},\
  }\href {\doibase 10.1103/PhysRevD.105.043020} {\bibfield  {journal} {\bibinfo
   {journal} {Phys. Rev. D}\ }\textbf {\bibinfo {volume} {105}},\ \bibinfo
  {pages} {043020} (\bibinfo {year} {2022})},\ \Eprint
  {http://arxiv.org/abs/2111.08353} {arXiv:2111.08353 [astro-ph.HE]}
  \BibitemShut {NoStop}%
\bibitem [{\citenamefont {Berti}\ \emph {et~al.}(2006)\citenamefont {Berti},
  \citenamefont {Cardoso},\ and\ \citenamefont {Will}}]{Berti:2005ys}%
  \BibitemOpen
  \bibfield  {author} {\bibinfo {author} {\bibfnamefont {E.}~\bibnamefont
  {Berti}}, \bibinfo {author} {\bibfnamefont {V.}~\bibnamefont {Cardoso}}, \
  and\ \bibinfo {author} {\bibfnamefont {C.~M.}\ \bibnamefont {Will}},\ }\href
  {\doibase 10.1103/PhysRevD.73.064030} {\bibfield  {journal} {\bibinfo
  {journal} {Phys. Rev. D}\ }\textbf {\bibinfo {volume} {73}},\ \bibinfo
  {pages} {064030} (\bibinfo {year} {2006})},\ \Eprint
  {http://arxiv.org/abs/gr-qc/0512160} {arXiv:gr-qc/0512160} \BibitemShut
  {NoStop}%
\bibitem [{\citenamefont {Borhanian}(2021)}]{Borhanian:2020ypi}%
  \BibitemOpen
  \bibfield  {author} {\bibinfo {author} {\bibfnamefont {S.}~\bibnamefont
  {Borhanian}},\ }\href {\doibase 10.1088/1361-6382/ac1618} {\bibfield
  {journal} {\bibinfo  {journal} {Class. Quant. Grav.}\ }\textbf {\bibinfo
  {volume} {38}},\ \bibinfo {pages} {175014} (\bibinfo {year} {2021})},\
  \Eprint {http://arxiv.org/abs/2010.15202} {arXiv:2010.15202 [gr-qc]}
  \BibitemShut {NoStop}%
\bibitem [{\citenamefont {Dietrich}\ \emph {et~al.}(2021)\citenamefont
  {Dietrich}, \citenamefont {Hinderer},\ and\ \citenamefont
  {Samajdar}}]{Dietrich:2020eud}%
  \BibitemOpen
  \bibfield  {author} {\bibinfo {author} {\bibfnamefont {T.}~\bibnamefont
  {Dietrich}}, \bibinfo {author} {\bibfnamefont {T.}~\bibnamefont {Hinderer}},
  \ and\ \bibinfo {author} {\bibfnamefont {A.}~\bibnamefont {Samajdar}},\
  }\href {\doibase 10.1007/s10714-020-02751-6} {\bibfield  {journal} {\bibinfo
  {journal} {Gen. Rel. Grav.}\ }\textbf {\bibinfo {volume} {53}},\ \bibinfo
  {pages} {27} (\bibinfo {year} {2021})},\ \Eprint
  {http://arxiv.org/abs/2004.02527} {arXiv:2004.02527 [gr-qc]} \BibitemShut
  {NoStop}%
\bibitem [{\citenamefont {Gupta}\ \emph {et~al.}(2020)\citenamefont {Gupta},
  \citenamefont {Datta}, \citenamefont {Kastha}, \citenamefont {Borhanian},
  \citenamefont {Arun},\ and\ \citenamefont {Sathyaprakash}}]{Gupta:2020lxa}%
  \BibitemOpen
  \bibfield  {author} {\bibinfo {author} {\bibfnamefont {A.}~\bibnamefont
  {Gupta}}, \bibinfo {author} {\bibfnamefont {S.}~\bibnamefont {Datta}},
  \bibinfo {author} {\bibfnamefont {S.}~\bibnamefont {Kastha}}, \bibinfo
  {author} {\bibfnamefont {S.}~\bibnamefont {Borhanian}}, \bibinfo {author}
  {\bibfnamefont {K.~G.}\ \bibnamefont {Arun}}, \ and\ \bibinfo {author}
  {\bibfnamefont {B.~S.}\ \bibnamefont {Sathyaprakash}},\ }\href {\doibase
  10.1103/PhysRevLett.125.201101} {\bibfield  {journal} {\bibinfo  {journal}
  {Phys. Rev. Lett.}\ }\textbf {\bibinfo {volume} {125}},\ \bibinfo {pages}
  {201101} (\bibinfo {year} {2020})},\ \Eprint
  {http://arxiv.org/abs/2005.09607} {arXiv:2005.09607 [gr-qc]} \BibitemShut
  {NoStop}%
\bibitem [{\citenamefont {Datta}\ \emph {et~al.}(2021)\citenamefont {Datta},
  \citenamefont {Gupta}, \citenamefont {Kastha}, \citenamefont {Arun},\ and\
  \citenamefont {Sathyaprakash}}]{Datta:2020vcj}%
  \BibitemOpen
  \bibfield  {author} {\bibinfo {author} {\bibfnamefont {S.}~\bibnamefont
  {Datta}}, \bibinfo {author} {\bibfnamefont {A.}~\bibnamefont {Gupta}},
  \bibinfo {author} {\bibfnamefont {S.}~\bibnamefont {Kastha}}, \bibinfo
  {author} {\bibfnamefont {K.~G.}\ \bibnamefont {Arun}}, \ and\ \bibinfo
  {author} {\bibfnamefont {B.~S.}\ \bibnamefont {Sathyaprakash}},\ }\href
  {\doibase 10.1103/PhysRevD.103.024036} {\bibfield  {journal} {\bibinfo
  {journal} {Phys. Rev. D}\ }\textbf {\bibinfo {volume} {103}},\ \bibinfo
  {pages} {024036} (\bibinfo {year} {2021})},\ \Eprint
  {http://arxiv.org/abs/2006.12137} {arXiv:2006.12137 [gr-qc]} \BibitemShut
  {NoStop}%
\bibitem [{\citenamefont {Mandel}\ \emph {et~al.}(2014)\citenamefont {Mandel},
  \citenamefont {Berry}, \citenamefont {Ohme}, \citenamefont {Fairhurst},\ and\
  \citenamefont {Farr}}]{Mandel:2014tca}%
  \BibitemOpen
  \bibfield  {author} {\bibinfo {author} {\bibfnamefont {I.}~\bibnamefont
  {Mandel}}, \bibinfo {author} {\bibfnamefont {C.~P.~L.}\ \bibnamefont
  {Berry}}, \bibinfo {author} {\bibfnamefont {F.}~\bibnamefont {Ohme}},
  \bibinfo {author} {\bibfnamefont {S.}~\bibnamefont {Fairhurst}}, \ and\
  \bibinfo {author} {\bibfnamefont {W.~M.}\ \bibnamefont {Farr}},\ }\href
  {\doibase 10.1088/0264-9381/31/15/155005} {\bibfield  {journal} {\bibinfo
  {journal} {Class. Quant. Grav.}\ }\textbf {\bibinfo {volume} {31}},\ \bibinfo
  {pages} {155005} (\bibinfo {year} {2014})},\ \Eprint
  {http://arxiv.org/abs/1404.2382} {arXiv:1404.2382 [gr-qc]} \BibitemShut
  {NoStop}%
\bibitem [{\citenamefont {Damour}\ \emph {et~al.}(2001)\citenamefont {Damour},
  \citenamefont {Iyer},\ and\ \citenamefont {Sathyaprakash}}]{Damour:2000zb}%
  \BibitemOpen
  \bibfield  {author} {\bibinfo {author} {\bibfnamefont {T.}~\bibnamefont
  {Damour}}, \bibinfo {author} {\bibfnamefont {B.~R.}\ \bibnamefont {Iyer}}, \
  and\ \bibinfo {author} {\bibfnamefont {B.~S.}\ \bibnamefont
  {Sathyaprakash}},\ }\href {\doibase 10.1103/PhysRevD.63.044023} {\bibfield
  {journal} {\bibinfo  {journal} {Phys. Rev. D}\ }\textbf {\bibinfo {volume}
  {63}},\ \bibinfo {pages} {044023} (\bibinfo {year} {2001})},\ \bibinfo {note}
  {[Erratum: Phys.Rev.D 72, 029902 (2005)]},\ \Eprint
  {http://arxiv.org/abs/gr-qc/0010009} {arXiv:gr-qc/0010009} \BibitemShut
  {NoStop}%
\bibitem [{\citenamefont {Sathyaprakash}\ \emph {et~al.}(2010)\citenamefont
  {Sathyaprakash}, \citenamefont {Schutz},\ and\ \citenamefont {Van
  Den~Broeck}}]{Sathyaprakash:2009xt}%
  \BibitemOpen
  \bibfield  {author} {\bibinfo {author} {\bibfnamefont {B.~S.}\ \bibnamefont
  {Sathyaprakash}}, \bibinfo {author} {\bibfnamefont {B.~F.}\ \bibnamefont
  {Schutz}}, \ and\ \bibinfo {author} {\bibfnamefont {C.}~\bibnamefont {Van
  Den~Broeck}},\ }\href {\doibase 10.1088/0264-9381/27/21/215006} {\bibfield
  {journal} {\bibinfo  {journal} {Class. Quant. Grav.}\ }\textbf {\bibinfo
  {volume} {27}},\ \bibinfo {pages} {215006} (\bibinfo {year} {2010})},\
  \Eprint {http://arxiv.org/abs/0906.4151} {arXiv:0906.4151 [astro-ph.CO]}
  \BibitemShut {NoStop}%
\bibitem [{\citenamefont {Arun}\ \emph {et~al.}(2014)\citenamefont {Arun},
  \citenamefont {Tagoshi}, \citenamefont {Mishra},\ and\ \citenamefont
  {Pai}}]{Arun:2014ysa}%
  \BibitemOpen
  \bibfield  {author} {\bibinfo {author} {\bibfnamefont {K.~G.}\ \bibnamefont
  {Arun}}, \bibinfo {author} {\bibfnamefont {H.}~\bibnamefont {Tagoshi}},
  \bibinfo {author} {\bibfnamefont {C.~K.}\ \bibnamefont {Mishra}}, \ and\
  \bibinfo {author} {\bibfnamefont {A.}~\bibnamefont {Pai}},\ }\href {\doibase
  10.1103/PhysRevD.90.024060} {\bibfield  {journal} {\bibinfo  {journal} {Phys.
  Rev. D}\ }\textbf {\bibinfo {volume} {90}},\ \bibinfo {pages} {024060}
  (\bibinfo {year} {2014})},\ \Eprint {http://arxiv.org/abs/1403.6917}
  {arXiv:1403.6917 [astro-ph.HE]} \BibitemShut {NoStop}%
\bibitem [{\citenamefont {Tolman}(1939)}]{Tolman:1939jz}%
  \BibitemOpen
  \bibfield  {author} {\bibinfo {author} {\bibfnamefont {R.~C.}\ \bibnamefont
  {Tolman}},\ }\href {\doibase 10.1103/PhysRev.55.364} {\bibfield  {journal}
  {\bibinfo  {journal} {Phys. Rev.}\ }\textbf {\bibinfo {volume} {55}},\
  \bibinfo {pages} {364} (\bibinfo {year} {1939})}\BibitemShut {NoStop}%
\bibitem [{\citenamefont {Oppenheimer}\ and\ \citenamefont
  {Volkoff}(1939)}]{Oppenheimer:1939ne}%
  \BibitemOpen
  \bibfield  {author} {\bibinfo {author} {\bibfnamefont {J.~R.}\ \bibnamefont
  {Oppenheimer}}\ and\ \bibinfo {author} {\bibfnamefont {G.~M.}\ \bibnamefont
  {Volkoff}},\ }\href {\doibase 10.1103/PhysRev.55.374} {\bibfield  {journal}
  {\bibinfo  {journal} {Phys. Rev.}\ }\textbf {\bibinfo {volume} {55}},\
  \bibinfo {pages} {374} (\bibinfo {year} {1939})}\BibitemShut {NoStop}%
\bibitem [{\citenamefont {Heavens}\ \emph {et~al.}(2014)\citenamefont
  {Heavens}, \citenamefont {Seikel}, \citenamefont {Nord}, \citenamefont
  {Aich}, \citenamefont {Bouffanais}, \citenamefont {Bassett},\ and\
  \citenamefont {Hobson}}]{Heavens:2014xba}%
  \BibitemOpen
  \bibfield  {author} {\bibinfo {author} {\bibfnamefont {A.~F.}\ \bibnamefont
  {Heavens}}, \bibinfo {author} {\bibfnamefont {M.}~\bibnamefont {Seikel}},
  \bibinfo {author} {\bibfnamefont {B.~D.}\ \bibnamefont {Nord}}, \bibinfo
  {author} {\bibfnamefont {M.}~\bibnamefont {Aich}}, \bibinfo {author}
  {\bibfnamefont {Y.}~\bibnamefont {Bouffanais}}, \bibinfo {author}
  {\bibfnamefont {B.~A.}\ \bibnamefont {Bassett}}, \ and\ \bibinfo {author}
  {\bibfnamefont {M.~P.}\ \bibnamefont {Hobson}},\ }\href {\doibase
  10.1093/mnras/stu1866} {\bibfield  {journal} {\bibinfo  {journal} {Mon. Not.
  Roy. Astron. Soc.}\ }\textbf {\bibinfo {volume} {445}},\ \bibinfo {pages}
  {1687} (\bibinfo {year} {2014})},\ \Eprint {http://arxiv.org/abs/1404.2854}
  {arXiv:1404.2854 [astro-ph.CO]} \BibitemShut {NoStop}%
\bibitem [{\citenamefont {Abbott}\ \emph
  {et~al.}(2017{\natexlab{d}})\citenamefont {Abbott} \emph
  {et~al.}}]{LIGOScientific:2017zic}%
  \BibitemOpen
  \bibfield  {author} {\bibinfo {author} {\bibfnamefont {B.~P.}\ \bibnamefont
  {Abbott}} \emph {et~al.} (\bibinfo {collaboration} {LIGO Scientific, Virgo,
  Fermi-GBM, INTEGRAL}),\ }\href {\doibase 10.3847/2041-8213/aa920c} {\bibfield
   {journal} {\bibinfo  {journal} {Astrophys. J. Lett.}\ }\textbf {\bibinfo
  {volume} {848}},\ \bibinfo {pages} {L13} (\bibinfo {year}
  {2017}{\natexlab{d}})},\ \Eprint {http://arxiv.org/abs/1710.05834}
  {arXiv:1710.05834 [astro-ph.HE]} \BibitemShut {NoStop}%
\bibitem [{\citenamefont {Coulter}\ \emph {et~al.}(2017)\citenamefont {Coulter}
  \emph {et~al.}}]{Coulter:2017wya}%
  \BibitemOpen
  \bibfield  {author} {\bibinfo {author} {\bibfnamefont {D.~A.}\ \bibnamefont
  {Coulter}} \emph {et~al.},\ }\href {\doibase 10.1126/science.aap9811}
  {\bibfield  {journal} {\bibinfo  {journal} {Science}\ }\textbf {\bibinfo
  {volume} {358}},\ \bibinfo {pages} {1556} (\bibinfo {year} {2017})},\ \Eprint
  {http://arxiv.org/abs/1710.05452} {arXiv:1710.05452 [astro-ph.HE]}
  \BibitemShut {NoStop}%
\bibitem [{\citenamefont {Cowperthwaite}\ \emph {et~al.}(2017)\citenamefont
  {Cowperthwaite} \emph {et~al.}}]{Cowperthwaite:2017dyu}%
  \BibitemOpen
  \bibfield  {author} {\bibinfo {author} {\bibfnamefont {P.~S.}\ \bibnamefont
  {Cowperthwaite}} \emph {et~al.},\ }\href {\doibase 10.3847/2041-8213/aa8fc7}
  {\bibfield  {journal} {\bibinfo  {journal} {Astrophys. J. Lett.}\ }\textbf
  {\bibinfo {volume} {848}},\ \bibinfo {pages} {L17} (\bibinfo {year}
  {2017})},\ \Eprint {http://arxiv.org/abs/1710.05840} {arXiv:1710.05840
  [astro-ph.HE]} \BibitemShut {NoStop}%
\bibitem [{\citenamefont {Kasen}\ \emph {et~al.}(2017)\citenamefont {Kasen},
  \citenamefont {Metzger}, \citenamefont {Barnes}, \citenamefont {Quataert},\
  and\ \citenamefont {Ramirez-Ruiz}}]{Kasen:2017sxr}%
  \BibitemOpen
  \bibfield  {author} {\bibinfo {author} {\bibfnamefont {D.}~\bibnamefont
  {Kasen}}, \bibinfo {author} {\bibfnamefont {B.}~\bibnamefont {Metzger}},
  \bibinfo {author} {\bibfnamefont {J.}~\bibnamefont {Barnes}}, \bibinfo
  {author} {\bibfnamefont {E.}~\bibnamefont {Quataert}}, \ and\ \bibinfo
  {author} {\bibfnamefont {E.}~\bibnamefont {Ramirez-Ruiz}},\ }\href {\doibase
  10.1038/nature24453} {\bibfield  {journal} {\bibinfo  {journal} {Nature}\
  }\textbf {\bibinfo {volume} {551}},\ \bibinfo {pages} {80} (\bibinfo {year}
  {2017})},\ \Eprint {http://arxiv.org/abs/1710.05463} {arXiv:1710.05463
  [astro-ph.HE]} \BibitemShut {NoStop}%
\bibitem [{\citenamefont {Soares-Santos}\ \emph {et~al.}(2017)\citenamefont
  {Soares-Santos} \emph {et~al.}}]{DES:2017kbs}%
  \BibitemOpen
  \bibfield  {author} {\bibinfo {author} {\bibfnamefont {M.}~\bibnamefont
  {Soares-Santos}} \emph {et~al.} (\bibinfo {collaboration} {DES, Dark Energy
  Camera GW-EM}),\ }\href {\doibase 10.3847/2041-8213/aa9059} {\bibfield
  {journal} {\bibinfo  {journal} {Astrophys. J. Lett.}\ }\textbf {\bibinfo
  {volume} {848}},\ \bibinfo {pages} {L16} (\bibinfo {year} {2017})},\ \Eprint
  {http://arxiv.org/abs/1710.05459} {arXiv:1710.05459 [astro-ph.HE]}
  \BibitemShut {NoStop}%
\bibitem [{\citenamefont {Valenti}\ \emph {et~al.}(2017)\citenamefont
  {Valenti}, \citenamefont {Sand}, \citenamefont {Yang}, \citenamefont
  {Cappellaro}, \citenamefont {Tartaglia}, \citenamefont {Corsi}, \citenamefont
  {Jha}, \citenamefont {Reichart}, \citenamefont {Haislip},\ and\ \citenamefont
  {Kouprianov}}]{Valenti:2017ngx}%
  \BibitemOpen
  \bibfield  {author} {\bibinfo {author} {\bibfnamefont {S.}~\bibnamefont
  {Valenti}}, \bibinfo {author} {\bibfnamefont {D.~J.}\ \bibnamefont {Sand}},
  \bibinfo {author} {\bibfnamefont {S.}~\bibnamefont {Yang}}, \bibinfo {author}
  {\bibfnamefont {E.}~\bibnamefont {Cappellaro}}, \bibinfo {author}
  {\bibfnamefont {L.}~\bibnamefont {Tartaglia}}, \bibinfo {author}
  {\bibfnamefont {A.}~\bibnamefont {Corsi}}, \bibinfo {author} {\bibfnamefont
  {S.~W.}\ \bibnamefont {Jha}}, \bibinfo {author} {\bibfnamefont {D.~E.}\
  \bibnamefont {Reichart}}, \bibinfo {author} {\bibfnamefont {J.}~\bibnamefont
  {Haislip}}, \ and\ \bibinfo {author} {\bibfnamefont {V.}~\bibnamefont
  {Kouprianov}},\ }\href {\doibase 10.3847/2041-8213/aa8edf} {\bibfield
  {journal} {\bibinfo  {journal} {Astrophys. J. Lett.}\ }\textbf {\bibinfo
  {volume} {848}},\ \bibinfo {pages} {L24} (\bibinfo {year} {2017})},\ \Eprint
  {http://arxiv.org/abs/1710.05854} {arXiv:1710.05854 [astro-ph.HE]}
  \BibitemShut {NoStop}%
\bibitem [{\citenamefont {Arcavi}\ \emph {et~al.}(2017)\citenamefont {Arcavi}
  \emph {et~al.}}]{Arcavi:2017xiz}%
  \BibitemOpen
  \bibfield  {author} {\bibinfo {author} {\bibfnamefont {I.}~\bibnamefont
  {Arcavi}} \emph {et~al.},\ }\href {\doibase 10.1038/nature24291} {\bibfield
  {journal} {\bibinfo  {journal} {Nature}\ }\textbf {\bibinfo {volume} {551}},\
  \bibinfo {pages} {64} (\bibinfo {year} {2017})},\ \Eprint
  {http://arxiv.org/abs/1710.05843} {arXiv:1710.05843 [astro-ph.HE]}
  \BibitemShut {NoStop}%
\bibitem [{\citenamefont {Tanvir}\ \emph {et~al.}(2017)\citenamefont {Tanvir}
  \emph {et~al.}}]{Tanvir:2017pws}%
  \BibitemOpen
  \bibfield  {author} {\bibinfo {author} {\bibfnamefont {N.~R.}\ \bibnamefont
  {Tanvir}} \emph {et~al.},\ }\href {\doibase 10.3847/2041-8213/aa90b6}
  {\bibfield  {journal} {\bibinfo  {journal} {Astrophys. J. Lett.}\ }\textbf
  {\bibinfo {volume} {848}},\ \bibinfo {pages} {L27} (\bibinfo {year}
  {2017})},\ \Eprint {http://arxiv.org/abs/1710.05455} {arXiv:1710.05455
  [astro-ph.HE]} \BibitemShut {NoStop}%
\bibitem [{\citenamefont {Lipunov}\ \emph {et~al.}(2017)\citenamefont {Lipunov}
  \emph {et~al.}}]{Lipunov:2017dwd}%
  \BibitemOpen
  \bibfield  {author} {\bibinfo {author} {\bibfnamefont {V.~M.}\ \bibnamefont
  {Lipunov}} \emph {et~al.},\ }\href {\doibase 10.3847/2041-8213/aa92c0}
  {\bibfield  {journal} {\bibinfo  {journal} {Astrophys. J. Lett.}\ }\textbf
  {\bibinfo {volume} {850}},\ \bibinfo {pages} {L1} (\bibinfo {year} {2017})},\
  \Eprint {http://arxiv.org/abs/1710.05461} {arXiv:1710.05461 [astro-ph.HE]}
  \BibitemShut {NoStop}%
\bibitem [{\citenamefont {Evans}\ \emph {et~al.}(2017)\citenamefont {Evans}
  \emph {et~al.}}]{Evans:2017mmy}%
  \BibitemOpen
  \bibfield  {author} {\bibinfo {author} {\bibfnamefont {P.~A.}\ \bibnamefont
  {Evans}} \emph {et~al.},\ }\href {\doibase 10.1126/science.aap9580}
  {\bibfield  {journal} {\bibinfo  {journal} {Science}\ }\textbf {\bibinfo
  {volume} {358}},\ \bibinfo {pages} {1565} (\bibinfo {year} {2017})},\ \Eprint
  {http://arxiv.org/abs/1710.05437} {arXiv:1710.05437 [astro-ph.HE]}
  \BibitemShut {NoStop}%
\bibitem [{\citenamefont {Margutti}\ \emph {et~al.}(2017)\citenamefont
  {Margutti} \emph {et~al.}}]{Margutti:2017cjl}%
  \BibitemOpen
  \bibfield  {author} {\bibinfo {author} {\bibfnamefont {R.}~\bibnamefont
  {Margutti}} \emph {et~al.},\ }\href {\doibase 10.3847/2041-8213/aa9057}
  {\bibfield  {journal} {\bibinfo  {journal} {Astrophys. J. Lett.}\ }\textbf
  {\bibinfo {volume} {848}},\ \bibinfo {pages} {L20} (\bibinfo {year}
  {2017})},\ \Eprint {http://arxiv.org/abs/1710.05431} {arXiv:1710.05431
  [astro-ph.HE]} \BibitemShut {NoStop}%
\bibitem [{\citenamefont {Hajela}\ \emph {et~al.}(2022)\citenamefont {Hajela}
  \emph {et~al.}}]{Hajela:2021faz}%
  \BibitemOpen
  \bibfield  {author} {\bibinfo {author} {\bibfnamefont {A.}~\bibnamefont
  {Hajela}} \emph {et~al.},\ }\href {\doibase 10.3847/2041-8213/ac504a}
  {\bibfield  {journal} {\bibinfo  {journal} {Astrophys. J. Lett.}\ }\textbf
  {\bibinfo {volume} {927}},\ \bibinfo {pages} {L17} (\bibinfo {year}
  {2022})},\ \Eprint {http://arxiv.org/abs/2104.02070} {arXiv:2104.02070
  [astro-ph.HE]} \BibitemShut {NoStop}%
\bibitem [{\citenamefont {Hallinan}\ \emph {et~al.}(2017)\citenamefont
  {Hallinan} \emph {et~al.}}]{Hallinan:2017woc}%
  \BibitemOpen
  \bibfield  {author} {\bibinfo {author} {\bibfnamefont {G.}~\bibnamefont
  {Hallinan}} \emph {et~al.},\ }\href {\doibase 10.1126/science.aap9855}
  {\bibfield  {journal} {\bibinfo  {journal} {Science}\ }\textbf {\bibinfo
  {volume} {358}},\ \bibinfo {pages} {1579} (\bibinfo {year} {2017})},\ \Eprint
  {http://arxiv.org/abs/1710.05435} {arXiv:1710.05435 [astro-ph.HE]}
  \BibitemShut {NoStop}%
\bibitem [{\citenamefont {Balasubramanian}\ \emph {et~al.}(2022)\citenamefont
  {Balasubramanian}, \citenamefont {Corsi}, \citenamefont {Mooley},
  \citenamefont {Hotokezaka}, \citenamefont {Kaplan}, \citenamefont {Frail},
  \citenamefont {Hallinan}, \citenamefont {Lazzati},\ and\ \citenamefont
  {Murphy}}]{Balasubramanian:2022sie}%
  \BibitemOpen
  \bibfield  {author} {\bibinfo {author} {\bibfnamefont {A.}~\bibnamefont
  {Balasubramanian}}, \bibinfo {author} {\bibfnamefont {A.}~\bibnamefont
  {Corsi}}, \bibinfo {author} {\bibfnamefont {K.~P.}\ \bibnamefont {Mooley}},
  \bibinfo {author} {\bibfnamefont {K.}~\bibnamefont {Hotokezaka}}, \bibinfo
  {author} {\bibfnamefont {D.~L.}\ \bibnamefont {Kaplan}}, \bibinfo {author}
  {\bibfnamefont {D.~A.}\ \bibnamefont {Frail}}, \bibinfo {author}
  {\bibfnamefont {G.}~\bibnamefont {Hallinan}}, \bibinfo {author}
  {\bibfnamefont {D.}~\bibnamefont {Lazzati}}, \ and\ \bibinfo {author}
  {\bibfnamefont {E.~J.}\ \bibnamefont {Murphy}},\ }\href {\doibase
  10.3847/1538-4357/ac9133} {\bibfield  {journal} {\bibinfo  {journal}
  {Astrophys. J.}\ }\textbf {\bibinfo {volume} {938}},\ \bibinfo {pages} {12}
  (\bibinfo {year} {2022})},\ \Eprint {http://arxiv.org/abs/2205.14788}
  {arXiv:2205.14788 [astro-ph.HE]} \BibitemShut {NoStop}%
\bibitem [{Swi()}]{Swift:2022}%
  \BibitemOpen
  \href {https://swift.gsfc.nasa.gov/about_swift/} {\enquote {\bibinfo {title}
  {Swift: About swift},}\ }\BibitemShut {NoStop}%
\bibitem [{\citenamefont {Metzger}(2020)}]{Metzger:2019zeh}%
  \BibitemOpen
  \bibfield  {author} {\bibinfo {author} {\bibfnamefont {B.~D.}\ \bibnamefont
  {Metzger}},\ }\href {\doibase 10.1007/s41114-019-0024-0} {\bibfield
  {journal} {\bibinfo  {journal} {Living Rev. Rel.}\ }\textbf {\bibinfo
  {volume} {23}},\ \bibinfo {pages} {1} (\bibinfo {year} {2020})},\ \Eprint
  {http://arxiv.org/abs/1910.01617} {arXiv:1910.01617 [astro-ph.HE]}
  \BibitemShut {NoStop}%
\bibitem [{\citenamefont {Kalogera}\ \emph {et~al.}(2021)\citenamefont
  {Kalogera} \emph {et~al.}}]{Kalogera:2021bya}%
  \BibitemOpen
  \bibfield  {author} {\bibinfo {author} {\bibfnamefont {V.}~\bibnamefont
  {Kalogera}} \emph {et~al.},\ }\href@noop {} {\  (\bibinfo {year} {2021})},\
  \Eprint {http://arxiv.org/abs/2111.06990} {arXiv:2111.06990 [gr-qc]}
  \BibitemShut {NoStop}%
\bibitem [{\citenamefont {Howell}\ \emph {et~al.}(2018)\citenamefont {Howell},
  \citenamefont {Ackley}, \citenamefont {Rowlinson},\ and\ \citenamefont
  {Coward}}]{Howell:2018nhu}%
  \BibitemOpen
  \bibfield  {author} {\bibinfo {author} {\bibfnamefont {E.~J.}\ \bibnamefont
  {Howell}}, \bibinfo {author} {\bibfnamefont {K.}~\bibnamefont {Ackley}},
  \bibinfo {author} {\bibfnamefont {A.}~\bibnamefont {Rowlinson}}, \ and\
  \bibinfo {author} {\bibfnamefont {D.}~\bibnamefont {Coward}},\ }\href
  {\doibase 10.1093/mnras/stz455} {\  (\bibinfo {year} {2018}),\
  10.1093/mnras/stz455},\ \Eprint {http://arxiv.org/abs/1811.09168}
  {arXiv:1811.09168 [astro-ph.HE]} \BibitemShut {NoStop}%
\bibitem [{\citenamefont {Wanderman}\ and\ \citenamefont
  {Piran}(2015)}]{Wanderman:2014eza}%
  \BibitemOpen
  \bibfield  {author} {\bibinfo {author} {\bibfnamefont {D.}~\bibnamefont
  {Wanderman}}\ and\ \bibinfo {author} {\bibfnamefont {T.}~\bibnamefont
  {Piran}},\ }\href {\doibase 10.1093/mnras/stv123} {\bibfield  {journal}
  {\bibinfo  {journal} {Mon. Not. Roy. Astron. Soc.}\ }\textbf {\bibinfo
  {volume} {448}},\ \bibinfo {pages} {3026} (\bibinfo {year} {2015})},\ \Eprint
  {http://arxiv.org/abs/1405.5878} {arXiv:1405.5878 [astro-ph.HE]} \BibitemShut
  {NoStop}%
\bibitem [{\citenamefont {Burns}\ \emph {et~al.}(2016)\citenamefont {Burns},
  \citenamefont {Connaughton}, \citenamefont {Zhang}, \citenamefont {Lien},
  \citenamefont {Briggs}, \citenamefont {Goldstein}, \citenamefont {Pelassa},\
  and\ \citenamefont {Troja}}]{Burns:2015fol}%
  \BibitemOpen
  \bibfield  {author} {\bibinfo {author} {\bibfnamefont {E.}~\bibnamefont
  {Burns}}, \bibinfo {author} {\bibfnamefont {V.}~\bibnamefont {Connaughton}},
  \bibinfo {author} {\bibfnamefont {B.-B.}\ \bibnamefont {Zhang}}, \bibinfo
  {author} {\bibfnamefont {A.}~\bibnamefont {Lien}}, \bibinfo {author}
  {\bibfnamefont {M.~S.}\ \bibnamefont {Briggs}}, \bibinfo {author}
  {\bibfnamefont {A.}~\bibnamefont {Goldstein}}, \bibinfo {author}
  {\bibfnamefont {V.}~\bibnamefont {Pelassa}}, \ and\ \bibinfo {author}
  {\bibfnamefont {E.}~\bibnamefont {Troja}},\ }\href {\doibase
  10.3847/0004-637X/818/2/110} {\bibfield  {journal} {\bibinfo  {journal}
  {Astrophys. J.}\ }\textbf {\bibinfo {volume} {818}},\ \bibinfo {pages} {110}
  (\bibinfo {year} {2016})},\ \Eprint {http://arxiv.org/abs/1512.00923}
  {arXiv:1512.00923 [astro-ph.HE]} \BibitemShut {NoStop}%
\bibitem [{\citenamefont {Maggiore}(2018)}]{Maggiore:2018sht}%
  \BibitemOpen
  \bibfield  {author} {\bibinfo {author} {\bibfnamefont {M.}~\bibnamefont
  {Maggiore}},\ }\href@noop {} {\emph {\bibinfo {title} {{Gravitational Waves.
  Vol. 2: Astrophysics and Cosmology}}}}\ (\bibinfo  {publisher} {Oxford
  University Press},\ \bibinfo {year} {2018})\BibitemShut {NoStop}%
\bibitem [{\citenamefont {Belgacem}\ \emph {et~al.}(2018)\citenamefont
  {Belgacem}, \citenamefont {Dirian}, \citenamefont {Foffa},\ and\
  \citenamefont {Maggiore}}]{Belgacem:2018lbp}%
  \BibitemOpen
  \bibfield  {author} {\bibinfo {author} {\bibfnamefont {E.}~\bibnamefont
  {Belgacem}}, \bibinfo {author} {\bibfnamefont {Y.}~\bibnamefont {Dirian}},
  \bibinfo {author} {\bibfnamefont {S.}~\bibnamefont {Foffa}}, \ and\ \bibinfo
  {author} {\bibfnamefont {M.}~\bibnamefont {Maggiore}},\ }\href {\doibase
  10.1103/PhysRevD.98.023510} {\bibfield  {journal} {\bibinfo  {journal} {Phys.
  Rev. D}\ }\textbf {\bibinfo {volume} {98}},\ \bibinfo {pages} {023510}
  (\bibinfo {year} {2018})},\ \Eprint {http://arxiv.org/abs/1805.08731}
  {arXiv:1805.08731 [gr-qc]} \BibitemShut {NoStop}%
\bibitem [{\citenamefont {Belgacem}\ \emph
  {et~al.}(2019{\natexlab{b}})\citenamefont {Belgacem}, \citenamefont {Dirian},
  \citenamefont {Finke}, \citenamefont {Foffa},\ and\ \citenamefont
  {Maggiore}}]{Belgacem:2019lwx}%
  \BibitemOpen
  \bibfield  {author} {\bibinfo {author} {\bibfnamefont {E.}~\bibnamefont
  {Belgacem}}, \bibinfo {author} {\bibfnamefont {Y.}~\bibnamefont {Dirian}},
  \bibinfo {author} {\bibfnamefont {A.}~\bibnamefont {Finke}}, \bibinfo
  {author} {\bibfnamefont {S.}~\bibnamefont {Foffa}}, \ and\ \bibinfo {author}
  {\bibfnamefont {M.}~\bibnamefont {Maggiore}},\ }\href {\doibase
  10.1088/1475-7516/2019/11/022} {\bibfield  {journal} {\bibinfo  {journal}
  {JCAP}\ }\textbf {\bibinfo {volume} {11}},\ \bibinfo {pages} {022} (\bibinfo
  {year} {2019}{\natexlab{b}})},\ \Eprint {http://arxiv.org/abs/1907.02047}
  {arXiv:1907.02047 [astro-ph.CO]} \BibitemShut {NoStop}%
\bibitem [{\citenamefont {Pratten}\ \emph {et~al.}(2022)\citenamefont
  {Pratten}, \citenamefont {Schmidt},\ and\ \citenamefont
  {Williams}}]{Pratten:2021pro}%
  \BibitemOpen
  \bibfield  {author} {\bibinfo {author} {\bibfnamefont {G.}~\bibnamefont
  {Pratten}}, \bibinfo {author} {\bibfnamefont {P.}~\bibnamefont {Schmidt}}, \
  and\ \bibinfo {author} {\bibfnamefont {N.}~\bibnamefont {Williams}},\ }\href
  {\doibase 10.1103/PhysRevLett.129.081102} {\bibfield  {journal} {\bibinfo
  {journal} {Phys. Rev. Lett.}\ }\textbf {\bibinfo {volume} {129}},\ \bibinfo
  {pages} {081102} (\bibinfo {year} {2022})},\ \Eprint
  {http://arxiv.org/abs/2109.07566} {arXiv:2109.07566 [astro-ph.HE]}
  \BibitemShut {NoStop}%
\bibitem [{\citenamefont {Bauswein}\ and\ \citenamefont
  {Stergioulas}(2015)}]{Bauswein:2015yca}%
  \BibitemOpen
  \bibfield  {author} {\bibinfo {author} {\bibfnamefont {A.}~\bibnamefont
  {Bauswein}}\ and\ \bibinfo {author} {\bibfnamefont {N.}~\bibnamefont
  {Stergioulas}},\ }\href {\doibase 10.1103/PhysRevD.91.124056} {\bibfield
  {journal} {\bibinfo  {journal} {Phys. Rev. D}\ }\textbf {\bibinfo {volume}
  {91}},\ \bibinfo {pages} {124056} (\bibinfo {year} {2015})},\ \Eprint
  {http://arxiv.org/abs/1502.03176} {arXiv:1502.03176 [astro-ph.SR]}
  \BibitemShut {NoStop}%
\end{thebibliography}%

\end{document}